\documentclass[twocolumn, prb, showpacs]{revtex4-1}

\usepackage{graphicx}
\usepackage{dcolumn}
\usepackage{bm}
\usepackage{physics}
\usepackage{verbatim}
\usepackage{amsmath,mathrsfs}
\usepackage{mathtools}
\usepackage{placeins}
\usepackage{afterpage}
\usepackage{float}
\usepackage[usenames,dvipsnames]{color}

\newcommand{\sg}{$\mathcal{G}\;$}
\newcommand{\mg}{\mathcal{G}}

\newcommand{\mh}{\mathcal{H}}

\begin{document}

\title{Nonlocality, entropy creation, and entanglement in quantum many-body systems}

\author{Marc Dvorak}
\affiliation{Department of Applied Physics, Aalto University School of Science, 00076-Aalto, Finland}
\email{marc.dvorak@aalto.fi}

\date{\today}

\begin{abstract}
We propose a reinterpretation and reformulation of the single-particle Green's function in nonrelativistic quantum many-body theory with an emphasis on normalization. By downfolding a correlation function covering all of Fock space into the observable portion, we derive a nonlocal Dyson equation which depends on an unknown downfolding frequency. The downfolding frequency is determined by solving the inverse problem so that the spectral function of the single-particle propagator is a Dirac-$\delta$ function. Upon measurement, the system collapses stochastically onto one of these normalized solutions. This collapse has a nonlocal effect on the path the particle takes, in agreement with quantum entanglement. We postulate that the multiplicity of each quantized solution is directly related to the ensemble averaged spectrum and the entropy created by measurement of the particle. We numerically compare the multiplicity spectrum from our theory with a local approximation for a two-level system and nicely reproduce results of the local theory for weak correlation.

In the final part, we outline a new picture of dynamics in quantum many-body systems. As a function of the coupling strength, the multiplicity for collapse has a complicated form due to the shape of the quantization condition. This structure creates an entropic force from counting quantized solutions which is predominantly attractive but likely also has a narrow repulsive regime at weak coupling. Upon collapse, an internal spacetime forms between the two points in order to carry the information gained from the reduction of the probabilistic many-body state. The repeated creation of these spacetime bridges defines an internal spacetime with a complicated shape and history. We treat the quantum system as a finite informational resource that holds information about possible normalized outcomes, collapses the wave function to reset after encountering a conflict, and creates an internal spacetime to carry the information gained with every collapse. We use a simple model to show that, in our informational phenomenology, the curvature of the multiplicity surface is related to the dilation of the internal spacetime. Finally, we consider a special limit of the theory in which no quantized solutions, and therefore no internal spacetime, exist.

\end{abstract}

\maketitle

\section{Introduction}
  The current understanding of the single-particle Green's function ($G$) in quantum many-body theory is not consistent. $G$ is defined in a way that loses norm in a many-body system, but the working equations guarantee a particle conserving $G$. This inconsistency could lead to practical issues or foundational problems for understanding quantum many-body systems. Particle conservation is essential for a complete theory describing a closed system. For this reason, we seek a new theory which emphasizes normalization and is internally consistent.

The topic is of great importance, even if we limit the scope to condensed matter. We are primarily interested in the theory of strongly-correlated materials, which we roughly define as non-Fermi liquids or materials with excitations that have zero quasiparticle residue. There are considerable theoretical and computational research efforts to understand these systems. Considering the issues with the definition of the single-particle Green's function and norm conservation, however, the theoretical tools to understand these materials may not be in place. There is a need for a particle/hole theory which conserves norm in a way that is consistent with the Schr\"odinger equation.

In this work, we approach the problem from a different perspective. We go all the way back to the beginning of the many-body problem and search for a different, \textit{ab-initio}, and universal theory. Our approach begins as a brute force solution to the problem. Eventually, mathematical manipulations will reduce the problem to a compact and familiar form. The meaning and solution method for this new problem, however, are vastly different than the standard theory. Our revised picture sets up deep connections among normalization, entanglement, and statistical mechanics.

We can now set up our brute force approach. We follow a strict adherence to the Schr\"odinger equation to start the problem. In order to conserve norm, we must correlate every degree of freedom with all others. This quantity could be called the total Green's function of the many-body Hamiltonian. Because time evolution mixes all possible configurations, there is no other choice which conserves particle number. This approach could be equivalent to knowing the many-body wave function. Generally, effective field theories try to avoid the full solution of the many-body problem. However, we consider the equivalence (if it should appear) between the wave function and any new theory a desirable result since we know that the wave function fully characterizes a many-body system. The new theory, then, would also fully characterize the system.

Such a full solution based on the total Green's function is generally considered not necessary, or even incorrect, to describe low-energy excitations. We do consider it necessary, however, in order to conserve norm. Instead of adhering to the Schr\"odinger equation, one can impose local continuity in the particle/hole basis to describe addition/removal of normalized particles. Requiring local continuity, no matter how well-reasoned it may be, is not necessarily the same as obeying time evolution according to the Schr\"odinger equation. In Ref.~[arXiv:Dvorak], we argue that these ideas are not equivalent and represent two distinct starting points, even though $G$ is often presented as if it satisfies both. Both choices make sense in their own ways, and they most likely quantitatively agree with each other in most cases. Neither choice, however, is completely satisfactory. Our motivation for this work is to separate these concepts from each other and approach the problem from a new perspective to find a consistent solution.

Our understanding of the situation is that information about microscopic, virtual degrees of freedom that may be lost when constructing the effective field theory based on local continuity is usually not considered relevant. It is believed that the structure of the effective theory agrees with that of the actual physics by respecting the proper spacetime symmetries and conservation laws, that this agreement is most important, and the full solution is not necessary to describe what is observable. However, we consider aspects of this transformation from exact to effective theories to be inconsistent or only roughly defined. What remains to be understood is how the full solution and rigorous normalization of the theory may change the \textit{structure} of the physics compared to the effective theory and if the microscopic information in the total Green's function really is only an unimportant detail. Of course, to know this, one needs the rigorously normalized theory.

If we leave the Green's function picture, we can instead compute the many-body wave function, but this has limitations. The wave function is not directly related to experimental spectroscopies (not by itself, at least). Because the wave function is not observable, it is most useful for calculating expectation values, including correlation functions, that describe experiments and observable physics. We want to develop a theory for the ``exact," particle conserving correlation function which describes particle addition or removal. We are interested in the most general case in which the system has no low-energy quasiparticle theory. We leave the quasiparticle picture entirely. A formulation for a correlation function, as opposed to some other quantity like the wave function, is essential since correlation functions most closely describe experimental spectroscopies and observable physics. We need a way to compute spectra from the wave functions. Furthermore, we want the \textit{exact} spectrum instead of the spectral function of $G$.

We advise the reader to read Ref.~[arXiv:Dvorak] before going any further in this manuscript. In Sec.~\ref{sec:theory}, we present the theory for our new approach. In Sec.~\ref{sec:numerics}, we present numerical results for a two-level system. In Sec.~\ref{sec:analysis}, we interpret the result and outline a revised picture of quantum dynamics based on the statistical mechanics that appear.

  \label{sec:introduction}

\section{Theory}
  Subsections \ref{sec:setup} and \ref{sec:sing_ref} are mostly a review of theory to set up the remaining derivation. For completeness, we present a thorough derivation, but the downfolding method is familiar to experts. In subsection~\ref{sec:downfolding}, the ideas begin to deviate from the standard formalism. Subsection~\ref{sec:sol_method} describing the solution method is the major feature of our reformulation, and subsection~\ref{sec:nsr} is a summary of our total concept.

  \label{sec:theory}
  
\subsection{Correlating all degrees of freedom}
  In this work, the relevant Hamiltonian is the nonrelativistic interacting Hamiltonian \cite{Fetter/Walecka,helgaker_molecular},
\begin{equation}
\mh = \sum_{ij} t_{ij} a_{i}^{\dagger} a_j + \frac{1}{2} \sum_{ijkl} v_{ijkl} a_{i}^{\dagger} a_{j}^{\dagger} a_l a_k  \label{secondquant} 
\end{equation}
for one-body matrix elements $t_{ij}$, two-body interaction matrix elements $v_{ijkl}$, and fermionic mode creation (annihilation) operators $a_i^{\dagger}$ ($a_i$). We neglect spin throughout and consider only the zero temperature problem. We will mostly assume a $1/r$ two-body interaction, though this is not strictly necessary.

The time-ordered, single-particle Green's function at zero temperature is \cite{Fetter/Walecka,cederbaum_domcke}
\begin{equation}
\label{def:G}
  G(1,2)= (-i) \frac{  \bra{\Psi_0} \hat{T} \, [ \, \hat{\psi}(1) \, \hat{\psi}^\dagger(2) \, ] \ket{\Psi_0}}{\langle \Psi_0 | \Psi_0 \rangle}, 
\end{equation}
where $\hat{T}$ is the time-ordering operator, $\ket{\Psi_0}$ is the interacting ground state, and $\hat{\psi}^{\dagger}(2)$ ($\hat{\psi}(1)$) is a Heisenberg creation (annihilation) field operator. Here, numbers represent points in space and time, $1 = (\mathbf{r}_1, t_1)$.

The Green's function is meant to describe particle creation/annihilation at a single point. However, as we have shown [Ref.~arXiv:Dvorak], a single field-operator neither creates a normalized particle above the ground state nor forms a complete basis to recover the full norm at the final time. The physical field operators that describe normalized particles must instead be complicated superpositions of many bare field operators. These localized, point-like operators must include virtual particle-hole pairs that dress and normalize the observed particle. This is the crux of the problem: the required virtual excitations are not included in the definition of $G$ in Eq.~\ref{def:G} \textit{even though} $G$ does cover the full space that is experimentally accessible. We see no way to consistently and generally define such ``physical'' field operators that create normalized particles. The second quantization procedure is defined in the basis of noninteracting bare particles, as are the field operators which enter $G$. Our goal is to formally incorporate the high-dimensional, virtual degrees of freedom into the single-particle problem to rigorously normalize the theory.

\begin{figure}
\begin{centering}
\includegraphics[width=\columnwidth]{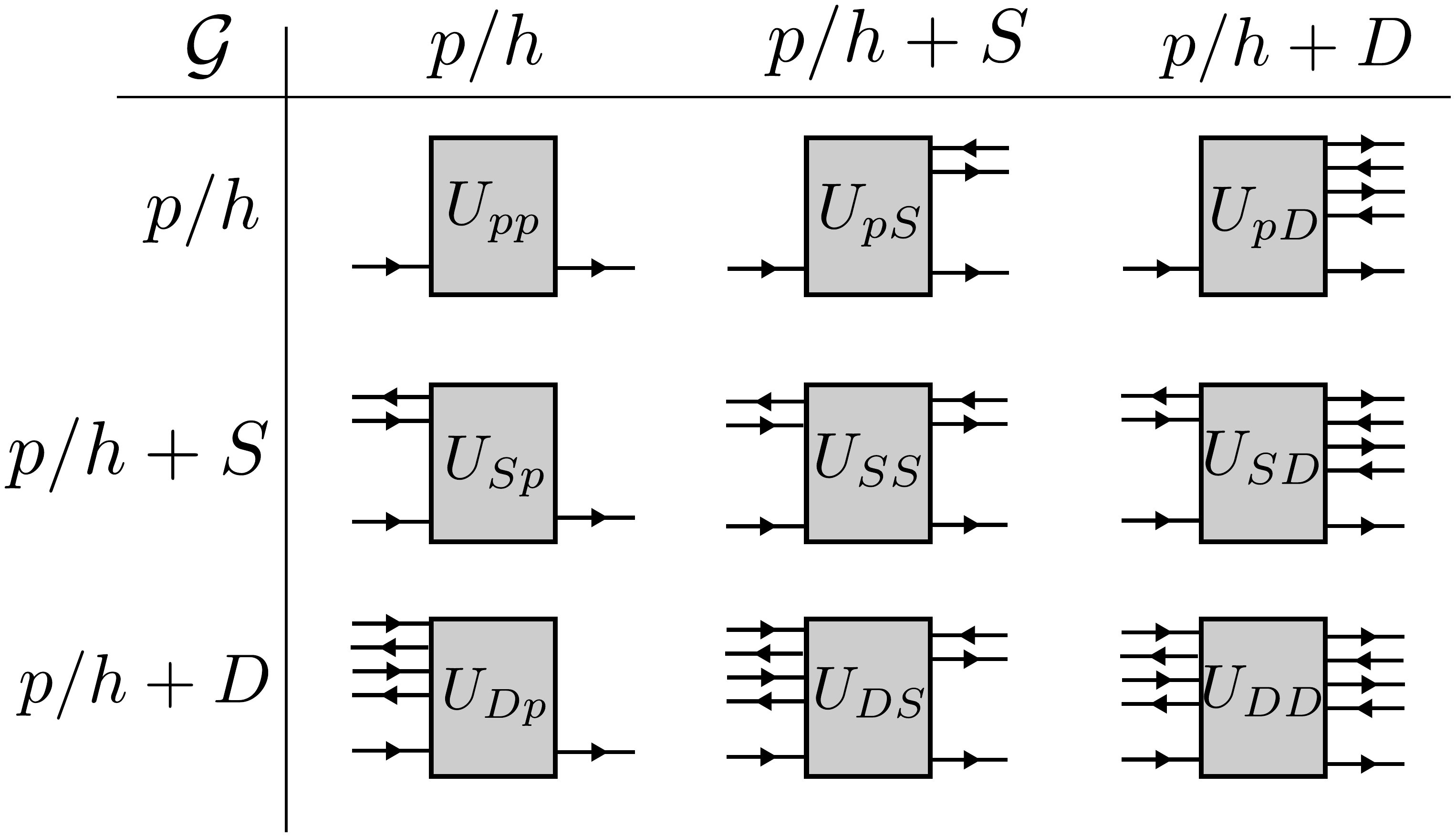}
\caption{Diagrammatic representation of $\mathcal{G}$. Here, $p$ or $h$ refers to a single, bare particle or hole, while $S$, $D$, etc. refer to a single, double, or higher neutral excitation. $p/h + X$ denotes neutral excitation $X$ created or annihilated with a single, bare particle or hole. $U$ is the exact time evolution operator. \label{big_g_fig}}
\end{centering}
\end{figure}

To conserve particle number, we must correlate every channel of Fock space with every other (we restrict ourselves to the $N \pm 1$ portion). With this approach, every possible initial state is allowed to decay into every possible final state. This criterion defines a correlation function which we denote $\mathcal{G}$ for its similarity to $G$. \sg is depicted diagramatically in Fig.~\ref{big_g_fig}. $U_{ij}$ represents the exact time evolution operator which evolves the initial state ($i$) into the final state ($j$). Lines with arrows are $G^0$ lines that represent both possible time orderings of bare particles.

To rigorously define $\mg$, we first define neutral excitation operators $\Omega^{\dagger}_{J}$ meant to act on the interacting ground state which generate all possible neutral excitations.
\begin{equation}
\Omega_J^{\dagger} = \prod_{\alpha,i \in J} a^{\dagger}_{\alpha} a_{i}  
\end{equation}
where $J$ is a composite index covering $\alpha$ and $i$. In a discrete basis, $\alpha$ and $i$ can take all possible noninteracting single-particle states. In real space, $\alpha,i$ can take any position coordinate. There can be any number of operators in the string for $\Omega_J^{\dagger}$:
\begin{eqnarray}
\Omega_{S}^{\dagger}  &=& a^{\dagger}_{\alpha} a_{i}        \\
\Omega_{D}^{\dagger}  &=& a^{\dagger}_{\alpha} a^{\dagger}_{\beta} a_{j} a_{i}   \nonumber \\
\Omega_{T}^{\dagger}  &=& a^{\dagger}_{\alpha} a^{\dagger}_{\beta} a^{\dagger}_{\gamma} a_{k} a_{j} a_{i} \nonumber  \\ 
... \; &=& \; ... \nonumber
\end{eqnarray}
to create single, double, and triple neutral excitations, respectively (and all possible higher excitations). We create the entire $N \pm 1$ portion of Fock space by acting on the ground state with composite operators with an odd number of field operators,
\begin{eqnarray}
\Lambda_K^{\dagger} &=& a_{p}^{\dagger} \,  \Omega_J^{\dagger} 
\end{eqnarray}
or its adjoint. Excitations should not be double-counted. We also include single field operators $a_p^{\dagger}$ or $a_p$ without any neutral excitation $\Omega_J^{\dagger}$ (bare particles and holes) to complete the basis. The set of all $\Lambda_K^{\dagger}$ and $\Lambda_K$ creates every single-particle state as both a particle and hole, combined with every possible neutral excitation (or no neutral excitation). The action of these operators on the ground state is quite complicated. The ground state is not a single reference configuration, meaning that every $\Lambda_K^{\dagger}$ acting on the ground state can potentially return a nonzero value. The idea for $\mg$ as the total Green's function is simple, in principle, but very complicated in practice.

We define $\mg$ in the Heisenberg picture as
\begin{equation}
\mathcal{G}_{KK'}(t , t') = (-i) \frac{ \bra{\Psi_0} \hat{T} \, [ \,  \Lambda_{H,K}(t)   \,  \Lambda_{H,K'}^{\dagger}(t') \,  ] \ket{\Psi_0} } { \langle \Psi_0 | \Psi_0 \rangle }  \label{bigg}
\end{equation}
for Heisenberg operators $\Lambda_{H,K'}^{\dagger}(t')$. $\hat{T}$ is the time-ordering operator to sum over the cases $t' > t$ and $t > t'$. The individual field operators contributing to $\Lambda_{H,K'}^{\dagger}(t')$ are all taken to act at the same time. The time dependence of the individual Heisenberg operators is
\begin{eqnarray}
a_{H,j}(t) &=& e^{i \mh t / \hbar} \,  a_j  \,  e^{-i \mh t / \hbar}   \\
a_{H,j}^{\dagger}(t) &=& e^{i \mh t / \hbar} \, a_j^{\dagger} \, e^{-i \mh t / \hbar} \; . \nonumber
\end{eqnarray}
The time dependence of the composite operators depends only on the outermost operators. For example (we now set $\hbar$ to 1),
\begin{eqnarray}
\Lambda_{H,K} (t) &=& e^{i \mh t } \,  a_i^{\dagger}  \,  e^{-i \mh t }  \, \times \\
& & e^{i \mh t } \,  a_{\alpha}  \,  e^{-i \mh t}  \,  e^{i \mh t  } \,  a_p  \,  e^{-i \mh t } \nonumber  \\
&=&  e^{i \mh t } \,  a_i^{\dagger}  \, e^{-i \mh (t-t) } a_{\alpha} e^{-i \mh (t-t)} \,  a_p e^{-i \mh t}  \nonumber \\
&=&  e^{i \mh t} \,  a_i^{\dagger}  \, a_{\alpha}  \,  a_p \, e^{-i \mh t }  \nonumber \; .
 \end{eqnarray}

We insert a complete set of many-body eigenstates $\ket{\Psi_{N \pm 1}}$ to find the Lehmann representation for $\mg$.
\begin{eqnarray}
i \mathcal{G}_{KK'}(t,t') &=& \sum_{N+1} \bra{\Psi_0} e^{i \mh t } \, \Lambda_K \, e^{-i \mh t / \hbar} \ket{\Psi_{N + 1} } \times   \\
& & \bra{\Psi_{N + 1}} e^{i \mh t' } \, \Lambda_{K'}^{\dagger} \,  e^{-i \mh t' } \ket{\Psi_0} \Theta(t-t')  \nonumber   \\
&-& \sum_{N - 1} \bra{\Psi_0} e^{i \mh t' } \, \Lambda_{K'}^{\dagger} \, e^{-i \mh t' } \ket{\Psi_{N - 1}} \times \nonumber  \\
& & \bra{\Psi_{N - 1}} e^{i \mh t } \, \Lambda_{K} \,  e^{-i \mh t } \ket{\Psi_0}  \Theta(t'-t) \nonumber
\end{eqnarray}
The action of $e^{-i \mh t }$ on the eigenstates returns the scalar $e^{-i E_{N \pm 1} t}$ so that \sg Fourier transforms in the standard way as
\begin{eqnarray}
\mathcal{G}_{KK'}(\omega) &=& \sum_{N+1} \frac{ \bra{\Psi_0} \Lambda_K \ket{\Psi_{N + 1}} \bra{\Psi_{N + 1}} \Lambda_{K'}^{\dagger} \ket{\Psi_0} }{ \omega - (E_{N + 1} - E_0) + i \eta}  \label{lehmann_big_g} \\ 
&+& \sum_{N - 1} \frac{ \bra{\Psi_0} \Lambda_{K'}^{\dagger} \ket{\Psi_{N - 1}} \bra{\Psi_{N - 1}} \Lambda_{K} \ket{\Psi_0} }{ \omega - ( E_0 - E_{N - 1} ) - i \eta}  \; . \nonumber
\end{eqnarray}
$\eta$ is a positive infinitesimal and we have set the chemical potential $\mu$ to zero. The amplitudes in the numerator of Eq.~\ref{lehmann_big_g} are the Lehmann amplitudes, and the poles of $\mathcal{G}_{KK'}(\omega)$ are at the exact particle addition/removal energies of the system.

$n$-particle Green's functions ($G^n$) in the literature can be defined with $n$ initial and $n$ final field operators. The usual interpretation is that the initial process in $G^n$ changes particle number from $N$ to $N \pm n$, and the final process changes back to $N$. This correlation function could be interpreted in terms of a two-electron spectroscopy if $n=2$, for example, or some other multi-electron spectroscopy. Different versions and time orderings of $G^n$ could also be interpreted as electron plus photon processes.

Modeling such electron plus photon processes is not our goal with $\mg$. \sg is meant to describe all channels of \textit{only} the electron addition/removal problem which is ordinarily described by the single-particle $G$. The offdiagonal matrix elements of \sg do not vanish, as depicted in Fig.~\ref{big_g_fig}, even for matrix elements with different numbers of initial/final field operators. Even though these offdiagonal matrix elements have different numbers of initial/final field operators, they create the same particle number ($N\pm1$) and their Lehmann amplitudes are not zero. These offdiagonal matrix elements are crucial to recovering the norm lost to high excitation levels during time evolution and the motivation for this work.

\sg is an enormous and complicated object. We intend it to be the largest possible correlation function in the $N\pm1$ particle portions of Fock space, a criterion which can be taken to define $\mg$. Fortunately, we will develop a simplified theory for which the minimally necessary portion is reduced to the three-particle level.

  \label{sec:setup}
  
\subsection{Single-reference theory}
  Given the complexity of the exact $\mg$, we want to immediately switch to a simpler single-reference theory. The single-reference perspective is simpler and aligns with diagrammatic perturbation theory. We are not attempting to adiabatically connect the exact $\mg$ to a single reference configuration. In fact, we do not believe this is possible for our theory. Instead, we are completely replacing the exact problem with a similar one that is easier to work with and has certain features built in to it. Additionally, we will not use the interaction picture for our reconstruction. We switch to the Schr\"odinger picture and let all the operators in $\Lambda_K$ act at the same time. The kernel of the propagator ($\mathcal{H}$) is also time independent. Time ordering is then only meant to separate the outermost times (initial and final).

We start from the Schr\"odinger propagator for the many-body system,
\begin{equation}
U_S(t,t') = (-i) \, e^{ -i \mh ( t-t' ) } \; .
\end{equation}
We impose causality on the propagator and converge the Fourier transform by deflecting each pole off of the real $\omega$-axis by $\pm i \eta$. We combine the two time-orderings,
\begin{eqnarray}
U_S^{T} (\omega) &=& \frac{1}{ \omega - \bra{ \phi_{N+1} } \mh \ket{ \phi_{N+1} } + i \eta}  \\
& & + \frac{1}{ \omega + \bra{\phi_{N-1}} \mh \ket{\phi_{N-1}} - i \eta }   \\
&=& \frac{1}{ \omega \mp \bra{ \phi_{N \pm 1} } \mh \ket{ \phi_{N\pm1} } \pm i \eta }  \\
&=& \frac{1}{ \omega \mp \mh \pm i \eta} 
\end{eqnarray}
where $U_S^T$ indicates the time-ordered Schr\"odinger propagator. We have selected the $N+1$ ($N-1$) sector of the problem for $+ i \eta$ ($-i\eta$). The basis for $U_S^T$ is the set of all Slater determinants $\phi_{N \pm 1}$ generated from the orbitals of the noninteracting Hamiltonian. This choice of basis is equivalent to letting all possible $\Lambda_K$ and $\Lambda_{K'}^{\dagger}$ excitation operators act on the reference configuration. As we have defined the $\Lambda_K$, certain $\Lambda_K$ or $\Lambda_{K'}^{\dagger}$ acting on the reference return zero. This decouples the particle and hole channels of our single-reference theory, a point which we return to later.

To fully connect to the particle/hole excitation picture and diagrammatic perturbation theory, we want to describe particle addition/removal above or below a reference configuration. For this, we subtract the matrix element of the reference configuration, $\ket{\phi}$. With this subtraction, we have an approximation to $\mg$ instead of the propagator of the full many-body system. To avoid introducing additional notation, we denote this object $\mg$, but it should be understood to be only approximate from here on.
\begin{equation}
\mg_{KK'}(\omega) = \frac{1}{ \omega \mp \left( \mh - \bra{\phi} \mh \ket{\phi} \right)  \pm i \eta } 
\end{equation}
The residual Hamiltonian $\mh - \bra{\phi} \mh \ket{\phi}$ can be rewritten in particle-hole notation as
\begin{eqnarray}
\mh - \bra{\phi} \mh \ket{\phi}  &=&  \sum_{ij \in \mathrm{p} } t_{ij} a_{i}^{\dagger} a_j  - \sum_{ij \in \mathrm{h} } t_{ij} b_i^{\dagger} b_j  \label{residual_h} \\
&& + \frac{1}{2} \sum_{ijkl} v_{ijkl} a_{i}^{\dagger} a_{j}^{\dagger} a_l a_k  - V^{\mathrm{ref}}  \nonumber
\end{eqnarray}
where the new operators $b_i^{\dagger}$ create holes defined relative to our reference configuration $\ket{\phi}$. The sums over the filled Fermi sea in $\mh$ and $\bra{\phi} \mh \ket{\phi}$ cancel. $V^{\mathrm{ref}}$ is only a number, the matrix element of the two-body interaction for the reference configuration. The two-body part of Eq.~\ref{residual_h} is complicated. While a clear set of rules for this two-body term exists, we simply replace it with an operator $\mathcal{E}$.
\begin{equation}
\mathcal{E} =  \pm \left( \frac{1}{2} \sum_{ijkl} v_{ijkl} a_{i}^{\dagger} a_{j}^{\dagger} a_l a_k  - V^{\mathrm{ref}} \right) \label{kernel_e} \end{equation}
We only need to understand the basic structure of the simplest matrix elements of $\mathcal{E}$. The $\pm$ sign in Eq.~\ref{kernel_e} goes with the particle (hole) channel of $\mg$.

We make a final adjustment in order to match our $\Lambda_{K'}^{\dagger}$ excitation basis. We perform the one-body sums over the particles and holes in each many-particle excitation to calculate the noninteracting excitation energy for each $\Lambda_K$.
\begin{eqnarray}
\lambda_K = \pm \left( \sum_{p \in K} t_{pp} - \sum_{h \in K} t_{hh} \right)  \label{lambda_sum}
\end{eqnarray}
where the sums run over particles ($p$) and holes ($h$) in the excitation $\Lambda_K^{\dagger}$ or $\Lambda_K$. $t_{ij}$ is diagonal in this basis, and $\pm$ again indicates the particle or hole channel. This gives a final compact form for our approximate, single reference $\mg$,
\begin{equation}
\mg_{KK'}(\omega) = \frac{1}{ \omega - \left( \lambda + \mathcal{E} \right)  \pm i \eta }  \label{big_g_single}
\end{equation}
which is understood to be a matrix equation in the $K,K'$ basis. As with the exact $\mg$, the concept for the approximate $\mg$ is straightforward: the Hamiltonian in the denominator of $U_S^T$ is rewritten in particle-hole notation and we discard contributions from the filled Fermi sea.

The form in Eq.~\ref{big_g_single} is deceptively simple. Eq.~\ref{big_g_single} can be a shorthand notation for $G$ or any other correlation function. However, $G$ would properly be accompanied by single field operators on the outside, representing the external lines. For the case of $G$, the size of the propagator or kernel at internal times does not match the overall size of the correlation function. In our case for the single-reference $\mathcal{G}$, however, the basis of the correlation function (external lines) \textit{does} match the basis of the kernel, which gives a deceptively simple expression for our single reference theory. In our case, it makes sense to omit the external $\Lambda_K$, $\Lambda_{K'}^{\dagger}$ since the two basis sets agree. Understanding our intent for Eq.~\ref{big_g_single} compared to a shorthand for $G$ is a very important distinction. Eq.~\ref{big_g_single} describes the correlations among all possible \textit{external} excitations, not just the propagator at internal times.

\begin{figure}[ht]
\begin{centering}
\includegraphics[width=0.75\columnwidth]{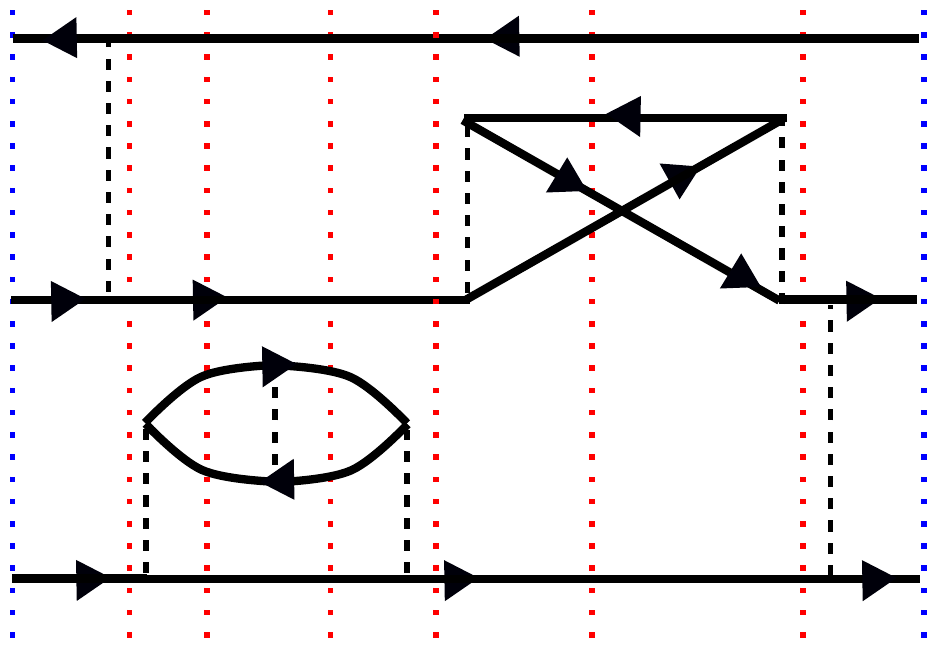}
\caption{An example diagram contributing to the full $\mg$. No matter how complicated the diagram becomes, at any internal time, a cut (red dotted lines) can be made leaving a set of $G^0$ lines. By construction, all possible time orderings of $G^0$ propagators are included in $\mg^0$, so the full \sg is reducible by $\mg^0$. \label{reducible_g}}
\end{centering}
\end{figure}

We group the one-body part of $\mg$ with $\omega$ to rewrite $\mg$ as
\begin{eqnarray}
\mg(\omega) &=& \frac{1}{ \omega - \left( \lambda + \mathcal{E} \right)  \pm i \eta }  \label{big_g_denom}  \\
&=& \frac{1}{ \left( \omega - \lambda \pm i \eta \right) - \mathcal{E} }  \\
&=& \frac{1}{ \mg^0(\omega)^{-1} - \mathcal{E}  }
\end{eqnarray}
which can be further rearranged into the Dyson series
\begin{equation}
\mg(\omega) = \mg^0(\omega) + \mg^0(\omega) \, \mathcal{E} \, \mg(\omega) \; .  \label{big_g_dyson}
\end{equation}
$\mg^0$ is the correlation function of the noninteracting problem. The scattering processes described by Eq.~\ref{big_g_dyson} are shown in Fig.~\ref{reducible_g}. Eq.~\ref{big_g_dyson} shows that the exact $\mg$ is reducible by $\mg^0$. The kernel of the Dyson equation is $\mathcal{E}$, it is frequency independent, and it exists in the full Fock space. Given this reducibility property of $\mg$, it is not clear how the exact $G$ can be reducible by $G^0$. $\mg$ and $G$ obey the same time evolution operator at internal times. It is only by including all possible channels at the internal \textit{and} external times that $\mg$ can be reduced. It is the character of the time evolution operator and $\mathcal{E}$ to connect all of these channels. This is precisely the way $\mg$ is defined and allows us to write the Dyson series. Reducibility is linked to norm conservation and unitarity $-$ the internal evolution can be properly sliced only for the total propagator or the exactly norm conserving correlation function. Our construction so far is exact (within our single-reference model) and assumes nothing about the Hamiltonian like coupling strength.

$G$ is only the projection of $\mg$ into the particle-hole basis at the outer times. This projection at the outermost times cannot alter the structure of the time evolution at internal times, where it is not reducible by $G^0$. This suggests that derivations of Dyson's equation are not entirely consistent with the definition of $G$ and the exact time evolution operator.

  \label{sec:sing_ref}
  
\subsection{Downfolding the kernel}
  To condense the high-dimensional problem into the observable space, we will downfold with the L\"owdin technique.\cite{lowdin_jmp_3} This downfolding approach is well understood and applied extensively. The energy dependence of Dyson's equation, for example, is commonly attributed to such a downfolding. This contradicts the idea, however, that the original hierarchy contained in the definition of $G$ is reducible. \textit{If} this exact, original self-energy hierarchy for $G$ is reducible and obeys Dyson's equation, no downfolding is necessary to achieve its energy dependence since this is simply part of the definition of $G$. Dyson's equation cannot come from both the definition of $G$ \textit{and} downfolding a larger object; the latter implies that the original hierarchy was not reducible by $G^0$. The overall picture is inconsistent.

In our case, we can already say that downfolding $\mg$ will produce something other than the hierarchy of correlation functions contained in $G$ simply because they are defined differently. The effects of initial/final virtual excitations are not contained in the hierarchy which appears in $G$ at \textit{internal} times. However, the effects of initial/final virtual excitations, which originate from external lines instead of internal, should be contained in our final downfolded object. Our downfolding is in the \textit{external} basis. We expect any new hierarchy which may appear in our derivation to be \textit{qualitatively different} than that of $G$ or the structure of the time evolution operator, neither of which involve downfolding. Because $G$ and $U(t,t')$ already include all local scattering processes, and our derivation cannot possibly be just a reproduction of $G$, it is reasonable to expect \textit{nonlocal} effects to appear.  Additionally, downfolding the time evolution operator is a choice that represents an abrupt and drastic change to the dynamics that is not part of any definition. The downfolding must also be interpreted as a physical process, and it evidently must be something beyond normal time evolution.

Certain features of the exact $\mg$ must carry over to the minimal single-reference version of our theory which we are now constructing. Without yet giving a complete justification, we need to include these adjustments at the current stage of the theory. The exact $\mg$ in Eq.~\ref{lehmann_big_g} has a property that is very important for our reformulation: not a single pole of $\mg$ anywhere on the $\omega$-axis has residue 1. In a many-body system, the Lehmann amplitudes in Eq.~\ref{lehmann_big_g} are all strictly $\le 1$, and the equality can only occur as an absurd accident or for the noninteracting system. To be consistent with the exact $\mg$, our single-reference theory cannot have any such ``perfect" poles. While it will not be obvious why this is important until later, it is necessary to introduce the idea and build this feature into our single-reference theory at this stage.

All the diagonal poles of our single-reference $\mg$ are such perfect poles $-$ simple poles of residue 1, shifted off the real $\omega$-axis by $i \eta$. We know this behavior is incorrect and consider it a failure our single-reference construction. Nonetheless, the physics of the single reference is so much more intuitive that we want a way to simulate the effects of the multiconfigurational theory without actually solving the exact problem. Without the ability to compute the Lehmann amplitudes in our single-reference theory, we remove the perfect poles by applying an additional broadening $i \epsilon$. The overbroadened $\mg$ is
\begin{equation}
\mg(\omega) = \frac{1}{ \omega - \left( \lambda + \mathcal{E} \right) \pm i ( \eta + \epsilon) }  \label{overbroadened}
\end{equation}
for real parameter $\epsilon$, $\epsilon > 0$. Generally, interactions broaden spectral peaks away from their bare value, giving the quasiparticle a finite lifetime. This effect would appear in the exact spectral function of $\mg$, which is exactly what we cannot compute with our single-reference theory since we set all the Lehmann amplitudes to 1. To compensate and recover a realistic behavior, we apply a universal extra broadening $i \epsilon$ to every pole. Eq.~\ref{overbroadened} is meant to approximate an \textit{actual} correlation function with a spectrum instead of only the propagator. We can always set $\epsilon = 0$ and recover the original problem.

\begin{figure}[ht]
\begin{centering}
\includegraphics[width=0.95\columnwidth]{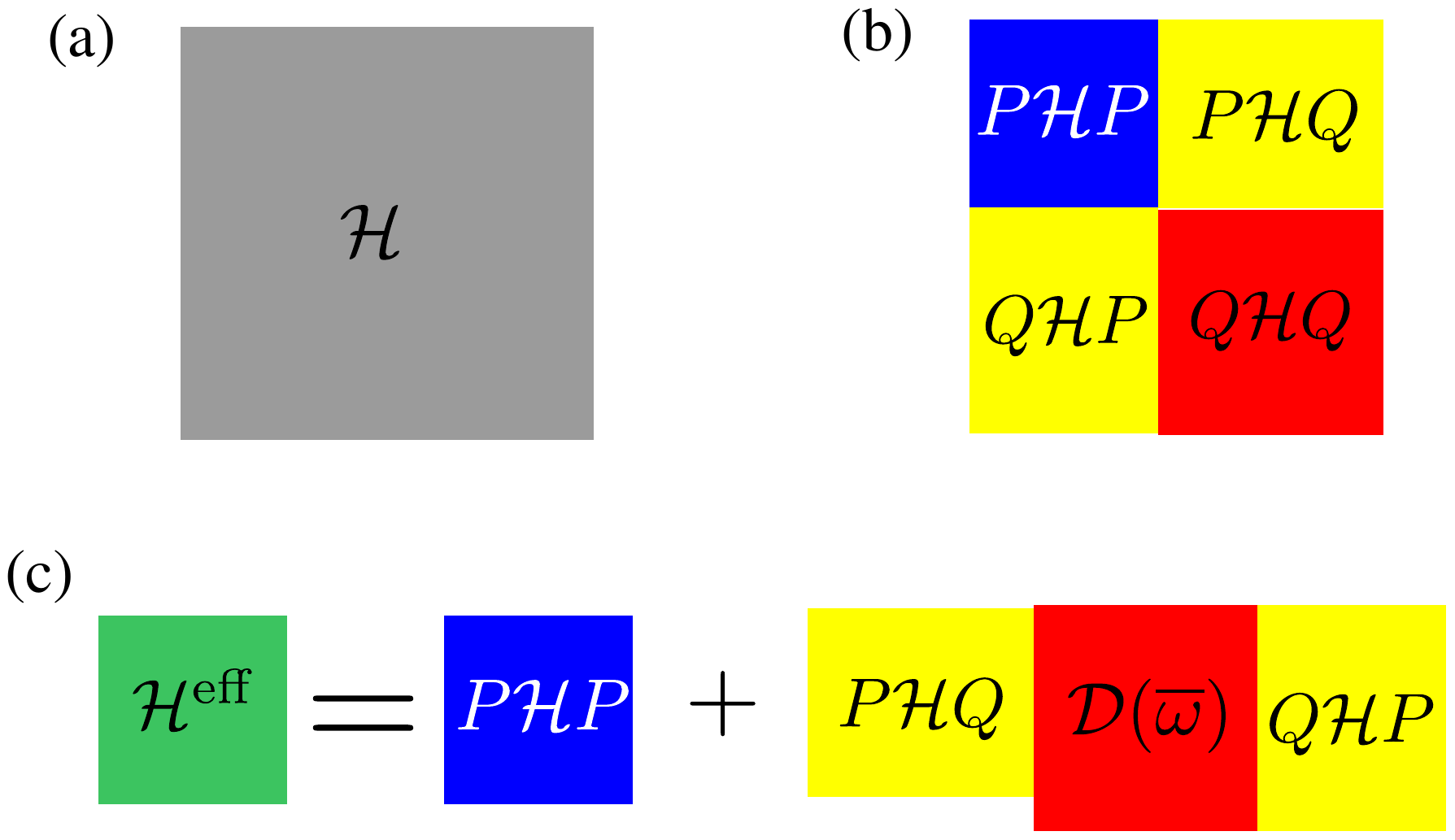}
\caption{L\"owdin downfolding of the particle-hole Hamiltonian. The full Hamiltonian (a) is partitioned into blocks (b) and downfolded (c) \label{downfolding}}
\end{centering}
\end{figure}

In the denominator of $\mg$ is a Hermitian operator, the particle/hole Hamiltonian ($\mh^{\mathrm{ph}}$), rigidly shifted off of the real axis by the amount $i ( \eta + \epsilon )$. We set up the downfolding with the shifted Hermitian eigenvalue problem from the denominator of $\mg$:
\begin{eqnarray}
\mg(\omega) &=& \frac{1}{ \omega - \mathcal{H}^{\mathrm{ph}} } \\
\mh^{\mathrm{ph}} \xi &=& E \xi  \\
\left[ \left( \lambda + \mathcal{E} \right) \mp i (\eta + \epsilon) \right] \xi &=&  \left[ \overline{\omega} \mp i \eta \right] \xi \; . \label{ph_hamiltonian}
\end{eqnarray}
for complex parameter $E$ and real parameter $\overline{\omega}$. Even though we have artificially broadened $\mg$ by the extra amount $i \epsilon$ on the LHS of Eq.~\ref{ph_hamiltonian}, we still force the poles of $\mg$ to appear at the physical value $i \eta$ on the RHS. The effect of this overbroadening is critical, and we will discuss its role more in Sec.~\ref{sec:nsr}. We partition $\mh^{\mathrm{ph}}$ into two blocks: a block containing all bare particles and holes, with a projection operator labeled $P$, and the remaining multiple excitations, with a projector labeled $Q$. The eigenvalue problem is rewritten in block form as
\begin{equation}
\begin{bmatrix}
P \, \mh^{\mathrm{ph}} \, P & P \, \mh^{\mathrm{ph}} \, Q  \\
Q \, \mh^{\mathrm{ph}} \, P & Q \, \mh^{\mathrm{ph}} \, Q 
\end{bmatrix}
\begin{bmatrix}
\psi \\
\chi
\end{bmatrix}
= E
\begin{bmatrix}
\psi \\
\chi
\end{bmatrix} \, . \label{block_eigenvalue}
\end{equation}
Here, $\psi$ is the portion of the full eigenvector $\xi$ in the $P$ space and $\chi$ the portion in $Q$. The second line of the eigenvalue problem, using the real parameter $\overline{\omega}$, is
\begin{equation}
\left[ Q \, \mh^{\mathrm{ph}} \, P \right] \psi + \left[ Q \, \mh^{\mathrm{ph}} \, Q \right] \chi = \left( \overline{\omega} \mp i \eta \right) \chi.
\end{equation}
We solve this equation for $\chi$ 
\begin{eqnarray}
\mathcal{D}(\overline{\omega}) &=& \frac{1}{ \left( \overline{\omega} \mp i \eta \right) - Q \, \mh^{\mathrm{ph}} \, Q}   \label{resolvent}  \\
\chi &=& \mathcal{D}(\overline{\omega}) \left[ Q \, \mh^{\mathrm{ph}} \, P \right] \psi    \nonumber 
\end{eqnarray}
which defines $\mathcal{D}(\overline{\omega})$, the resolvent. We insert $\mh^{\mathrm{ph}}$ into $\mathcal{D}$ as
\begin{eqnarray}
\mathcal{D}(\overline{\omega}) &=& \frac{1}{ \left( \overline{\omega} \mp i \eta \right) - Q \left( \left( \lambda + \mathcal{E} \right)  \mp i \left( \eta + \epsilon \right) \right) Q }   \\
&=& \frac{1}{ \left( \overline{\omega} \mp i \eta \pm i \left( \eta + \epsilon \right) \right) - Q \left( \lambda + \mathcal{E} \right) Q }  \\
&=& \frac{1}{ \left( \overline{\omega}  \pm i \epsilon \right) - Q \left( \lambda + \mathcal{E} \right) Q }  \\
&=& \frac{1}{ \overline{\omega}  - Q \left( \lambda + \mathcal{E} \right) Q \pm i \epsilon }
\end{eqnarray}
Replacing $\chi$ in the first line of Eq.~\ref{block_eigenvalue} with Eq.~\ref{resolvent} gives
\begin{eqnarray}
M(\overline{\omega}) &=& \left[ P \, \mh^{\mathrm{ph}} Q \, \right] \mathcal{D}(\overline{\omega}) \left[ Q \, \mh^{\mathrm{ph}} \, P \right]   \label{effective}  \\
\mh^{\mathrm{eff}}(\overline{\omega}) \psi &=& \left[ P \, \mh^{\mathrm{ph}} \, P + M(\overline{\omega}) \right] \psi = \left( \overline{\omega} \mp i \eta \right) \psi.  \label{heff}  
\end{eqnarray}
The $P \, \mh^{\mathrm{ph}} \, P$ matrix includes the free propagation of and interaction between bare particles and holes, and the infinitesimal $i (\eta+\epsilon)$. The interaction within the $P$ block is the Hartree-Fock diagrams, represented by the Fock matrix $\mathcal{F}$ (we exclude the one-body part from our definition of $\mathcal{F}$).
\begin{equation}
P \, \mh^{\mathrm{ph}} \, P = \lambda_P + \mathcal{F} \mp i \left( \eta + \epsilon \right) 
\end{equation}
The downfolding is shown schematically in Fig.~\ref{downfolding}.

We pause to discuss the appearance of the frequency $\overline{\omega}$ in detail since it is the most important element of the derivation. We find that, unlike the energy dependence of Dyson's equation,\cite{dyson_pr} the downfolding frequency $\overline{\omega}$ is not the physical test frequency $\omega$. We already showed that $\mh$ is frequency independent after the Fourier transform. Any downfolding frequency of Eq.~\ref{block_eigenvalue} must be a different frequency since $\omega$ is already defined for the Fourier transform. The downfolding frequency can be related to the eigenvalues of the Hamiltonian, but these eigenvalues are not related to $\omega$. Even when treating $\overline{\omega}$ as a continuous parameter, as we will do, its meaning as an eigenvalue and downfolding frequency is different than that of $\omega$, the Fourier transform variable. From this point on, we treat $\overline{\omega}$ as a parameter. Determining this parameter will be the subject of the solution method in Sec.~\ref{sec:sol_method}.

By construction, the downfolded propagator must be able to reproduce the $P$ projection of the original, total propagator with the proper choice of downfolding frequency. For this reason, there must be a free parameter to choose in order to meet the projection matching condition. $\omega$ is not a free parameter and cannot be used for this purpose. The downfolding procedure is only algebraic manipulations of an eigenvalue problem that does not depend on time (or frequency). The final downfolded Hamiltonian is then also time independent. The eigenvalue $\overline{\omega}$ has no conjugate variable as $\omega$ does ($\omega \leftrightarrow t$). If we were working in the time domain, we would replace $\mh$ with $\mh^{\mathrm{eff}}(\overline{\omega})$ in the propagator $e^{\mp i \mh (t-t') }$. In the time domain, the meaning of $\overline{\omega}$ as an eigenvalue is more obvious and the only choice which makes sense.

If we choose a value for $\omega$, $\mg$ has many different correlations at that fixed frequency that cover the entire Fock space. This just depends on the size of $\mg$, which happens to be very large, and not the value of $\omega$ we choose. After the downfolding \textit{and at the same fixed value of} $\omega$, we must be able to recover the particle/hole projection of every original $\mg$ correlation. For our simplified single-reference $\mg$, these different downfolded correlations are different projections of $\mh$. This multiple projection behavior requires a free parameter, $\overline{\omega}$, in order to choose and reproduce the multiple different projections. We still have not changed $\omega$, which is not a free parameter to use for any self-consistency or projection matching condition. These projections are not related to $\omega$ but instead allow us to choose the pole position on the $\omega$-axis $-$ $\overline{\omega}$ is the \textit{eigenvalue}, not the frequency axis. If $\overline{\omega}$ is self-consistently fixed to some eigenstate, there must still be a frequency coordinate, $\omega$, on which to plot the spectrum and show the peak of the propagating stationary state, just as it appeared before downfolding.

\begin{figure}[ht]
\begin{centering}
\includegraphics[width=1.0\columnwidth]{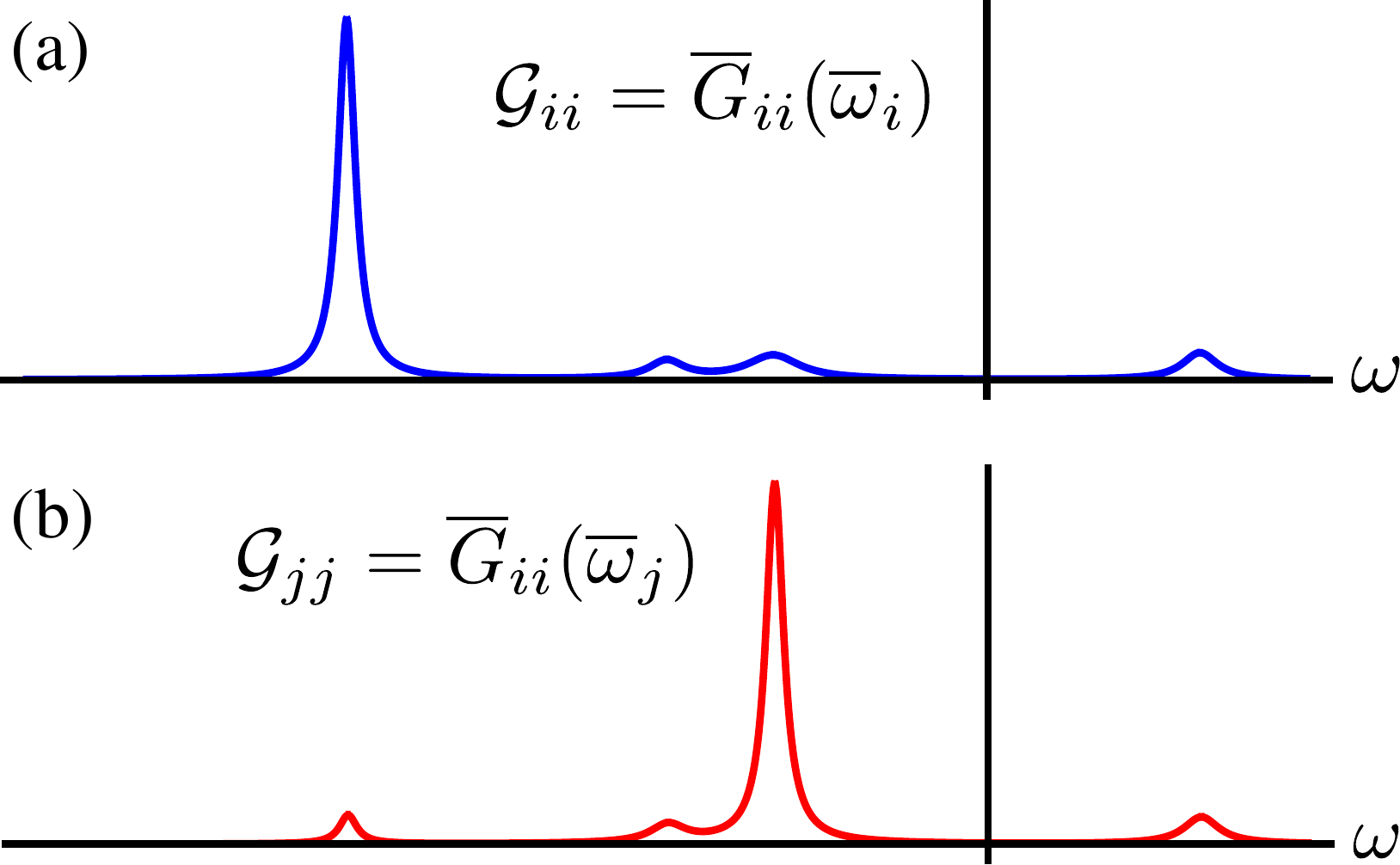}
\caption{Schematic showing the layered spectra contained in the downfolded version of $\mg$, denoted $\overline{G}$. For a $P$ dominated excitation ($i \in P$), the $i^{\mathrm{th}}$ diagonal component of $\mg$ or $\overline{G}$ shows the $i^{\mathrm{th}}$ spectrum, related to the imaginary part of $\mg$ or $\overline{G}$. For a $Q$ excitation ($j \in Q$), its spectrum appears in the $j^{\mathrm{th}}$ element of $\mg$ but in the $i^{\mathrm{th}}$ element of $\overline{G}$ $-$ the $j^{\mathrm{th}}$ component is eliminated with the downfolding. $\overline{G}_{ii}$ contains two spectra at different downfolding frequencies, $\overline{\omega}_i$ or $\overline{\omega}_j$, which can be chosen self-consistently to place a pole at the eigenvalue for the peak in the original $\mg_{ii}$ or $\mg_{jj}$. While the figure illustrates the structure of the downfolding, choosing the projections to reproduce the original spectra of $\mg$ is \textit{not} our solution method. \label{layers}}
\end{centering}
\end{figure}

We can think of first calculating the full spectra on the $\omega$-axis for all $P$ and $Q$ excitations before the downfolding. After downfolding, we must be able to reproduce the spectrum on the $\omega$-axis of any $P$ or $Q$ excitation in \textit{only} the $P$ subspace, which requires some extra layer to produce multiple spectra in $P$. These layered spectra are shown in Fig.~\ref{layers} and can roughly be thought of as the separate spectra showing all the self-consistent, stationary states of the time independent, downfolded Hamiltonian. In contrast, information about the total system is lost when projecting $\mg$ onto $G$, which has no layers, instead of downfolding $\mg$ onto $\overline{G}$. A key observation is that the projections of any stationary state can only be reproduced as they appear in the total system by a time independent subspace Hamiltonian.

The significance of our finding is that $\overline{\omega}$ gives us freedom to \textit{choose} how the system propagates in the particle/hole space. We claim this is sensible. We have not changed the time evolution operator, but instead can only see a small portion of any eigenstate as it propagates, as if it has been cut off from the rest. There are still the same number of eigenstates, and the portion of any eigenvector we see is truly stationary (infinite lifetime). The reduction in size while maintaining the same number of eigenstates creates a non-unique or overcomplete time evolution operator in the observable subspace. Indeed, the eigenvectors of $\mh^{\mathrm{eff}}$ are not naively orthogonal. The time evolution operator in the subspace really would appear different depending on the projection. There are many more possible combinations of the total configuration basis, and an equivalent number of the ``tails" of any one vector that are seen in the subspace, than can be created by working only in the subspace. This concept is not new, but we find that it is an actual physical effect that ruins the uniqueness of the subspace time propagation.

These \textit{are} all projections of eigenvectors of the total system and they \textit{do} all exist as stationary states in the subspace: we \textit{must} be able to reproduce them if the system is prepared in that state. Compressing them into a subspace creates the layered spectrum. There just is not a diagonal representation in the subspace which can show all the eigenstates or, equivalently, show all the information about the total time evolution. This layering effect is necessary in order to hold all information about the total system, contained in the eigenstates, in only the subspace in order to create the correct projection. With any choice of $\overline{\omega}$, the observable subspace still propagates on a time or frequency coordinate according to the Schr\"odinger equation as if it is attached to the total eigenvector and time evolution. However, the non-uniqueness related to the choice of the projection creates inherent uncertainty about the time evolution in the subspace.

We consider it accidental that our procedure resembles an internal downfolding or a possible derivation of Dyson's equation. With the additional broadening by $i \epsilon$, we obtain a \textit{spectrum}, which is then subject to an external downfolding, instead of downfolding $\mh$ by itself on the internal time. Of course, we take advantage of the very simple external downfolding of our model correlation function, but we have built in this simple form. We want to maintain the internal time evolution as it is. What we want, in principle, is an external downfolding of the entire correlation function $\mg$. A downfolding of $\mh$ on the internal time of $\mg$ or $G$ may also be possible and useful for other reasons, although we claim the internal downfolding frequency is not $\omega$ either. The internal downfolding procedure would very closely resemble our derivation here and follow the same arguments in order to exactly reproduce the spectra of the original $\mg$ or $G$. Internal downfolding does not serve our purpose of particle conservation, however.

Briefly, we propose choosing $\overline{\omega}$ to match the observed physics, which has a nonlocal effect on the dynamics. On a heuristic level, we argue that constructing an object which violates local time evolution is a plausible route to normalizing the problem. It is the nature of the local time evolution operator to distribute probability amplitude across high-dimensional configurations, which directly results in probability amplitude lost to virtual excitations. Given this property of the time evolution operator, a normalized and three-dimensional object which obeys local time evolution seems impossible. Local time evolution and norm conservation in the particle/hole basis appear incompatible \textit{if we follow the Schr\"odinger equation exactly.} There is also experimental and theoretical evidence that nonlocality exists in quantum mechanics. Specifically, we have in mind quantum eraser\cite{jacques_science_315,kim_prl_84,kwiat_pra_45,bienfait_prx,ionicioiu_prl_107,scully_pra_25,wheeler} experiments, Bell's inequality\cite{bell}, and the Einstein-Podolsky-Rosen\cite{einstein_pr_47} (EPR) paradox. This discussion is not proof of anything but is meant to motivate the role of a second frequency in the equation.

In this work, we have found a free parameter from quantum mechanical first principles which we can use to normalize the theory nonlocally. We can use $\overline{\omega}$ to find a particle/hole theory which remains consistent with the Schr\"odinger equation. In contrast, the meaning of Dyson's equation ($G^0$-reducible time evolution and local continuity) actually disagrees with the Schr\"odinger equation, which says that local norm conservation is only possible in the full configuration basis. We conclude that the field theory framework to which Dyson's equation belongs is a local physics formalism separate from the Schr\"odinger equation.

We return to the single-reference construction to analyze the $\overline{\omega}$ dependent correction $M(\overline{\omega})$. While we are not enumerating matrix elements of $\mathcal{E}$, we can safely say that offdiagonal elements of $\mathcal{E}$ only couple excitations which are one particle-hole excitation away. We have defined $P$ as the space of bare excitations. Therefore, matrix elements of $P \, \mh^{\mathrm{ph}} \, Q$ and $Q \, \mh^{\mathrm{ph}} \, P$ are a two-body interaction between bare particles and bare particles plus a particle-hole pair. All other matrix elements of $P \mh Q$ are zero. The one-body term has no offdiagonal contribution. We generically replace $P\mh Q$ and $Q \mh P$ with two-body matrix elements $v_{ijkl}$ or simply $v$. This generic two-body interaction $v$ is understood to contain direct and exchange terms. At this point, the effective Hamiltonian in the particle-hole space is (without imaginary infinitesimals)
\begin{equation}
\mh^{\mathrm{eff}} (\overline{\omega}) = \lambda_P  + \mathcal{F}  + v \, \mathcal{D} (\overline{\omega}) \, v 
\end{equation}
where a contraction of the object $v_{ijkl} \, D_{jkl,pqr}(\overline{\omega}) \, v_{pqrs}$ is implied. $v$ is the interaction strength between the particle (hole) and p+ph (h+ph) excitations.

Since we downfolded the kernel, the entire matrix for \sg is reduced in size. We no longer have an equation for \sg but for its downfolded counterpart which we label $\overline{G}$. $\overline{G}$ exists in the same particle/hole basis as $G$. We use the overbar to denote downfolded quantities or frequencies. This gives
\begin{equation}
\overline{G}(\omega) = \frac{1}{\omega - \left( \lambda_P + \mathcal{F}  + v \, \mathcal{D}( \overline{\omega} ) \, v \right)  \pm i \left( \eta + \epsilon \right) } \; .  \label{big_g_equation}
\end{equation}
We do not mean to overcomplicate a standard method (downfolding). The main point of this subsection is that the downfolding frequency is not $\omega$. To proceed further, we examine the resolvent $\mathcal{D}$.

  \label{sec:downfolding}
  
\subsection{The six-point object $D$}
  
The block $Q \, \mh^{\mathrm{ph}} \, Q$ needed for the resolvent $\mathcal{D}$ also has a free part and an interaction kernel,
\begin{equation}
Q \, \mh^{\mathrm{ph}} \, Q = \lambda_Q + \mathcal{E}_Q  \mp i \left( \eta + \epsilon \right)  \; .
\end{equation}
As demonstrated above, the imaginary parts of $\mh^{\mathrm{ph}}$ and $E$ partially cancel so that the resolvent is
\begin{equation}
\mathcal{D}(\overline{\omega}) = \frac{1}{\overline{\omega} - (\lambda_Q + \mathcal{E}_Q) \pm i \epsilon} \; \label{d_function} .
\end{equation}
By comparison with Eq.~\ref{big_g_denom}, $\mathcal{D}(\overline{\omega})$ can be interpreted as a correlation function on the $\overline{\omega}$-axis with broadening $i \epsilon$. We can unfold Eq.~\ref{d_function} into a Dyson series as
\begin{eqnarray}
\mathcal{D}^0(\overline{\omega}) &=& \frac{1}{\overline{\omega} - \lambda_Q  \pm i \eta } \label{d_dyson} \\
\mathcal{D}(\overline{\omega}) &=& \frac{1}{\overline{\omega} - (\lambda_Q +  \mathcal{E}_Q ) \pm i \epsilon }  \\
 &=& \frac{1}{ \mathcal{D}^0(\overline{\omega})^{-1} - \mathcal{E}_Q }  \\
 &=& \mathcal{D}^0(\overline{\omega}) + \mathcal{D}^0(\overline{\omega}) \, \mathcal{E}_Q \, \mathcal{D}(\overline{\omega}) \; . 
\end{eqnarray}
$\mathcal{D}$ describes an initial $Q$ excitation, time evolution which mixes \textit{only} $Q$ excitations at internal times, and a final $Q$ annihilation. All external and internal lines of $\mathcal{D}$ must belong to the $Q$ space. By definition, $\mathcal{E}_Q$ is the kernel $\mathcal{E}$ subject to $Q$ projectors on both sides. By construction, $\mathcal{D}$ is reducible by $\mathcal{D}^0$ $-$ the kernel $\mathcal{E}_Q$ connects all possible $Q$ excitations, and every arrangement of $Q$ excitations is contained in $\mathcal{D}^0$. Here, we see the effect of applying the lifetime broadening to $\mg$. $\epsilon$ is the broadening of $\mathcal{D}$, the correlation function contained in the virtual space.

To compute $\overline{G}$, which is our final goal, we only need the portion of $\mathcal{D}$ with three external lines on each side. We label this three-particle, six-point correlation function $D$. $D$ is not reducible by $D^0$, $G^0$, or $\mg^0$. Reducibility by $G^0$ is already forbidden for $\mathcal{D}$. To go from $\mathcal{D}$ to $D$, we restrict the external lines of $\mathcal{D}$ to the three-particle level. However, we still allow higher excitations at intermediate times, which destroys the reducibility of $D$. The structure of $D$ is a topological similarity to $\Sigma^*$. In our eventual Dyson series for $\overline{G}$, the contraction of $v \, D(\overline{\omega}) \, v$ will play the role of $\Sigma^*(\omega)$. Just as $\Sigma^*$ is the irreducible block of Dyson's equation for $G$, $v \, D \, v$ is the irreducible building block for $\overline{G}$.

\begin{figure}[ht]
\begin{centering}
\includegraphics[width=\columnwidth]{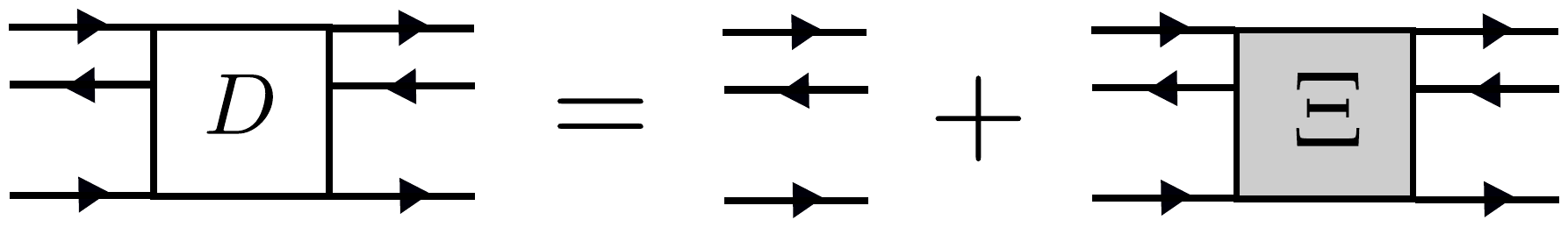}
\caption{The irreducible object $D$. The kernel $\Xi$ is irreducible by $D^0$ and can be computed order-by-order with perturbation theory. \label{d3_dyson}}
\end{centering}
\end{figure}

$D$ is the object which can be computed order-by-order with perturbation theory, although we have the exact $D$ in mind here. There are a clear and well-defined set of rules for the matrix elements of $\Xi$, the sum of all possible intra-$Q$ scattering processes, though we do not show them here. Given the scope of this manuscript, the precise form of these matrix elements is unimportant. We are focused on the form of the equations, their frequency structure, and solution method. Free and interaction parts of $D$ or $\Xi$ are always in the $Q$ basis. We know that, by virtue of the Slater-Condon rules for $\mathcal{H}$ and the configuration basis, $D$ can have no self-interaction effects like self-screening or any $D$ diagrams which violate the Pauli exclusion principle.

If we simply contract $v \, D \, v$ and set $\overline{\omega} = \omega$, we get a diagrammatic construction similar to $\Sigma^*$, the proper self-energy of Dyson's equation. The approximation $\overline{\omega} = \omega$ gives an $\omega$ dependent Dyson series for $\overline{G}$ that is of the same structure to the normal Dyson series for $G$. The meaning of this replacement ($\overline{\omega} = \omega$) is not obvious. It resembles Dyson's equation, which has its own meaning altogether, and, furthermore, this approximation does not necessarily include the same diagrams as the exact $G$. However, it is convenient for us to assign some meaning and make use of the $\overline{\omega} = \omega$ approximation in our test calculations.

We group $D$ with its contracted interactions into a new object, $\overline{\Sigma}^* = v \, D \, v$, to follow the notation of other downfolded quantities and parallel to Dyson's equation.
\begin{figure}[htb]
\begin{centering}
\includegraphics[width=\columnwidth]{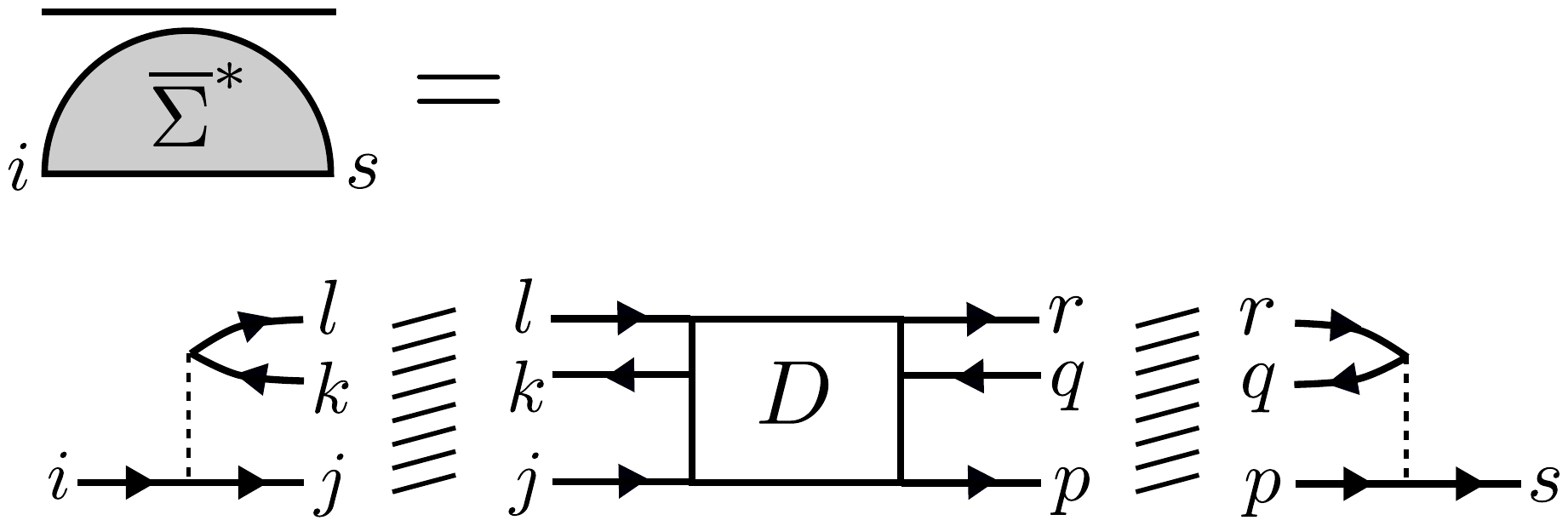}
\caption{Convention for contracting the internal indices of $v \, D \, v$. The cross hatching indicates breaks where the nonlocal jump from $\omega$ to $\overline{\omega}$ occurs. \label{contraction_diagram}}
\end{centering}
\end{figure}
The interpretation of $\overline{\Sigma}^*$ must be done carefully. Before inserting $\overline{\Sigma}^*$ into a diagram for $\overline{G}$, we must make the nonlocality clear. We place breaks in Fig.~\ref{contraction_diagram} between $D$ and $v_{ijkl}$ to make the nonlocality obvious.

The equation for $\overline{G}$, which we name the nonlocal single-particle Green's function, is
\begin{eqnarray}
\overline{G}(\omega) &=& \frac{1}{\omega - ( \lambda_P + \mathcal{F}  + v \, D(\overline{\omega}) \, v ) \pm i \left( \eta + \epsilon \right)  }     \\
\overline{G}(\omega) &=& \frac{1}{\omega - ( \lambda_P + \mathcal{F}  + v \frac{1}{\overline{\omega} - ( \lambda_Q + \mathcal{E}_Q ) \pm i \epsilon}  v ) \pm i \left( \eta + \epsilon \right)  } \; .   \nonumber
\end{eqnarray}

\begin{figure}[htb]
\begin{centering}
\includegraphics[width=\columnwidth]{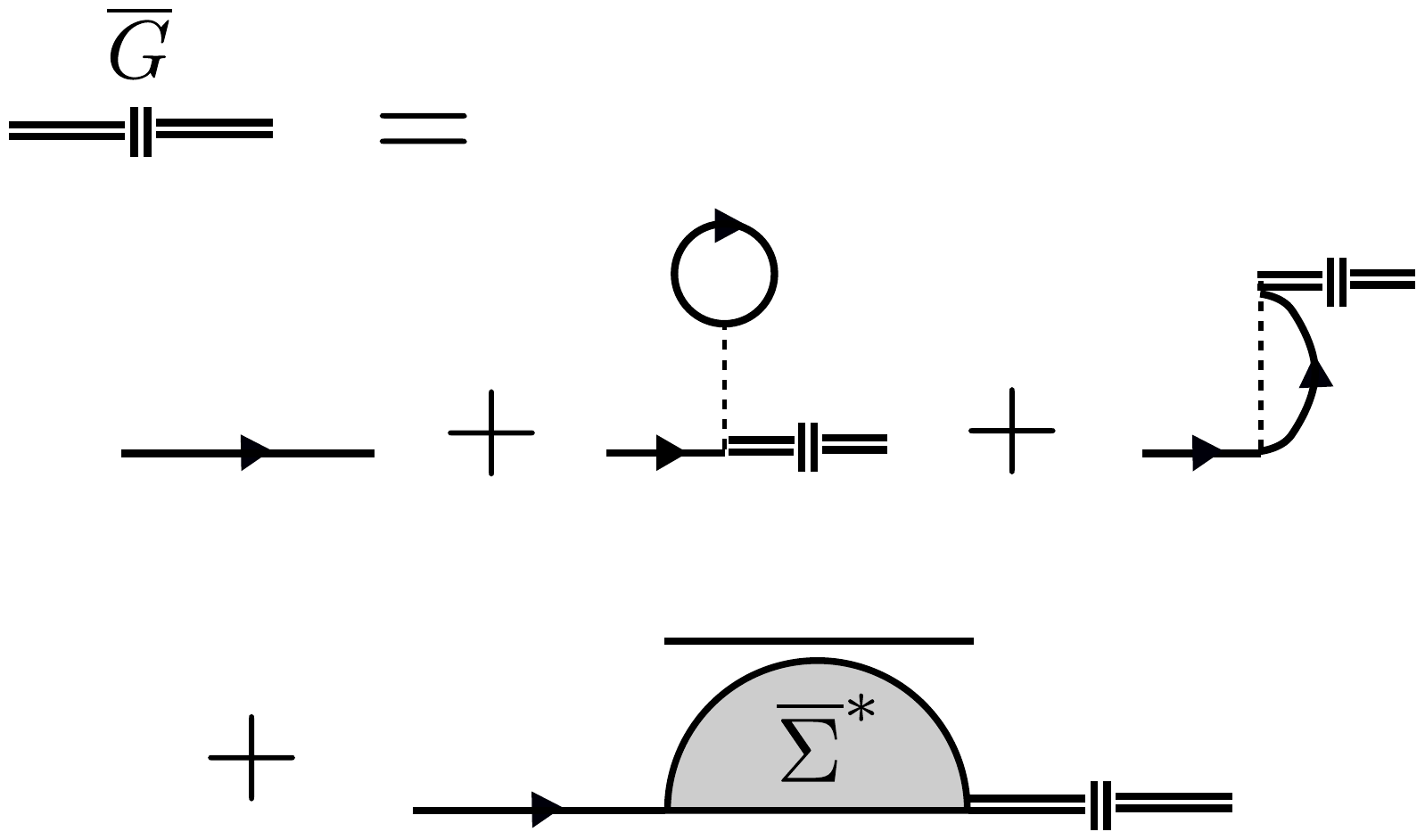}
\caption{Nonlocal Dyson equation obeyed by $\overline{G}$, represented by the bold line with a vertical break in the middle. \label{dyson_equation_gbar}}
\end{centering}
\end{figure}

$\overline{G}(\omega)$ therefore obeys the nonlocal Dyson equation
\begin{equation}
\overline{G}(\omega) = G^0(\omega) + G^0(\omega) ( \mathcal{F} +    \overline{\Sigma}^*(\overline{\omega}) \mp i \epsilon ) \overline{G}(\omega)  \label{g_bar_dyson}
\end{equation}
($\overline{\Sigma}^*$ can also be defined to include $\mp i \epsilon$). The nonlocal Dyson equation is shown diagramatically in Fig.~\ref{dyson_equation_gbar}. Our Dyson equation is similar in structure to Brillouin-Wigner perturbation theory, but the perturbation expansion for the irreducible kernel is on a different frequency coordinate than $\omega$. This diagram must be interpreted carefully $-$ it is not a Feynman diagram. The propagation of the particle does not run continuously through the time coordinate, but goes through a nonlocal jump. It is not a Dyson equation at all, and we will not apply perturbation theory in the usual sense. The diagram is only meant as a guide to give a connection to the standard theory.
We have a Dyson-like equation in place and must now address how to solve it.

  \label{sec:six_point}

\subsection{Solution method}
  We must find a way to determine the parameter $\overline{\omega}$. One possibility is to self-consistently diagonalize the effective Hamiltonian in the denominator of $\overline{G}$,
\begin{equation}
\mh^{\mathrm{eff}}(\overline{\omega}) \psi = [ \lambda_P + \mathcal{F} + \overline{\Sigma}^*(\overline{\omega}) ] \psi = \overline{\omega} \, \psi \; .
\end{equation}
Diagonalizing $\mh^{\mathrm{eff}}(\overline{\omega})$ this way would give the downfolded eigenstates and excitation energies, by construction, but this does not test particle conservation. Despite their prevalence in the discussion in Sec.~\ref{sec:downfolding}, we are not actually concerned with eigenstates of $\mh$. Without an $\omega$ dependence, there is also no way to generate an $\omega$ dependent spectrum. We do not pursue an effective Hamiltonian formalism.

Up to this point in the theory, we have downfolded the many-body probability amplitude into the observable space with the quantity $\overline{G}$. The actual behavior of measurements implies a much more strict condition on the probability amplitude, though, than just being condensed to observable space. Spectroscopic measurements are a type of collapse, localized onto a single eigenstate of the measured quantity. In real space, we are interested in the problem of a normalized particle created at a \textit{single} point and later annihilated at a \textit{single} point. The observed probability of this event is 1, and the probability of all other events is 0. To be consistent with collapse of the system onto a single, quantized event, the full probability amplitude must be downfolded onto a a pair of points.

This is a significantly more restrictive condition than how we initially set up the downfolding. In an experiment, though, the only \textit{observed} probabilities at any point are 0 or 1. Probability amplitudes or spectra are then deduced from an ensemble of quantized events. The single-particle $G$ is interpreted as the probability amplitude of a quantized event, but it is not provably built as a distribution of events which can each be measured \textit{individually}. In contrast, the experimental distribution is measured as a set of discrete pings against a detector, all with the same normalization and identical signal. We use the extra degree of freedom $\overline{\omega}$ to directly search for these discrete events. We search the downfolded state for individual quantized events to match what experiments measure. Once we have a condition for the collapse, we will use the multiplicity of each solution to build a distribution for comparison with experimental spectra.

We seek a condition in the energy domain which enforces the collapse and normalization criteria. In the energy domain, spectroscopic measurements on many-body systems yield a spectrum of events as a function of frequency. A single event, however, gives a sharp and normalized peak. Normalization requires that, for any initial condition, the sum of probabilities over all possible outcomes equals one. Collapse of the measurement requires that the end points are only single points in space. These two conditions set up a spectrum for an individual event realized from $x$ to $y$ at frequency $\omega_t$ with the form
\begin{equation}
\overline{A}_{ij}(\omega)=
\begin{cases}
  \delta(\omega - \omega_t), &   i,j=x,y \\
    \mathrm{arb.}, & \mathrm{otherwise}\label{single_spectrum}
\end{cases}
\end{equation}
Here, we introduce $\overline{A}(\omega)$ as the spectrum of $\overline{G}(\omega)$. Eq.~\ref{single_spectrum} is the critical piece to determine $\overline{\omega}$: an observed event is represented by a single, normalized Dirac-$\delta$ function. Normalization is guaranteed by choosing the amplitude of the Dirac-$\delta$ function to be 1. Collapse is represented by a single matrix element with a $\delta$ function in the spectrum. For any initial point, given by $x$, there can only be one final observable point, given by $y$. Normalization and collapse only need to be enforced among the \textit{observable} events, which means that we allow other matrix elements to take arbitrary values. This appears to transfer norm away from our collapsed solution and ruin the normalization condition. However, we argue that as long as other matrix elements do not give a $\delta$ function, their effect is not observable and normalization and collapse are both preserved.

\begin{figure}[htb]
\begin{centering}
\includegraphics[width=\columnwidth]{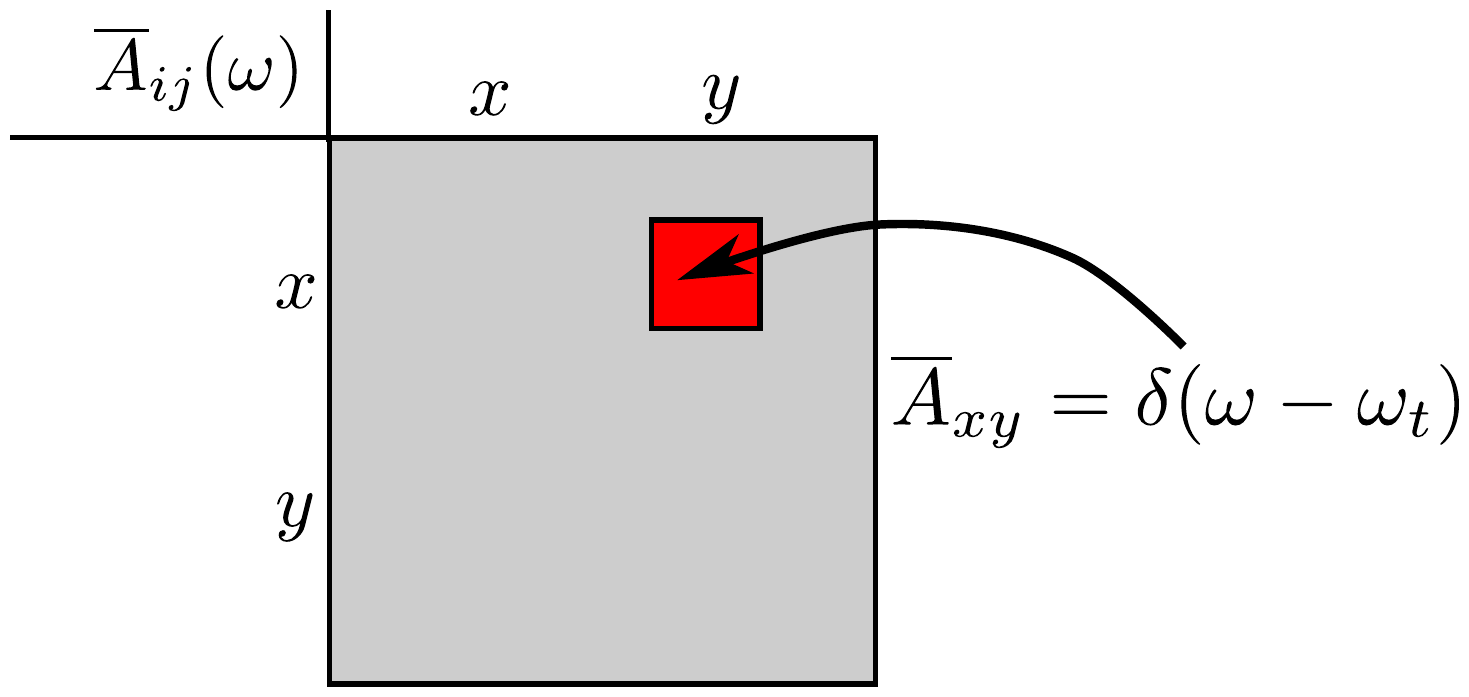}
\caption{Condition for a quantized and collapsed event in the single event spectrum $\overline{A}$. The quantization condition requires one matrix element of $\overline{A}$ to be a Dirac-$\delta$ function. Grey matrix elements are unimportant in the search for a selected $x,y$ event. The sign of $\overline{A}$ indicates the particle or hole character of the event. \label{collapse_spectrum}}
\end{centering}
\end{figure}

We relate the spectrum to $\overline{G}$ using the identity
\begin{equation}
\frac{1}{x \pm i \eta} = \mathcal{P}\frac{1}{x}  \mp i \pi \delta(x) 
\end{equation}
for the infinitesimal $\eta$ and Cauchy principle value $\mathcal{P}$. We take the imaginary  part of $\overline{G}$ to define $\overline{A}_{ij}$ as
\begin{equation}
\overline{A}_{ij}(\omega) = \frac{1}{\pi} | \, \mathrm{Im} \, \overline{G}_{ij}(\omega)  |
\end{equation}
Like $\overline{G}$, $\overline{A}_{ij}(\omega)$ also has an $\overline{\omega}$ dependence. To exactly model the observed event, $\overline{\omega}$ is chosen to match the condition in Eq.~\ref{single_spectrum}. For every possible $x,y$ event, we search for $\overline{\omega}$ which satisfy
\begin{equation}
\frac{1}{\pi} \, | \, \mathrm{Im} \, \overline{G}_{ij}(\omega) | =
\begin{cases}
  \delta(\omega - \omega_t), &   i,j=x,y \\
    \mathrm{arb.}, & \mathrm{otherwise}\label{im_g}
\end{cases}
\end{equation}
where the $\delta$ function is centered on some frequency $\omega_t$. Guided by the Fourier transform of the total propagator, we choose the broadening of the collapsed spectral function to also be $\eta$. Solving Eq.~\ref{im_g} is an inverse problem. We know the desired result $-$ an $\eta$-broadened $\delta$ function in the single event spectrum $-$ and search for all $\overline{\omega}$ projections into the particle/hole space which give this solution. We use $\overline{\omega}$ to quantize the spectral function.

\begin{figure}[htb]
\begin{centering}
\includegraphics[width=0.9\columnwidth]{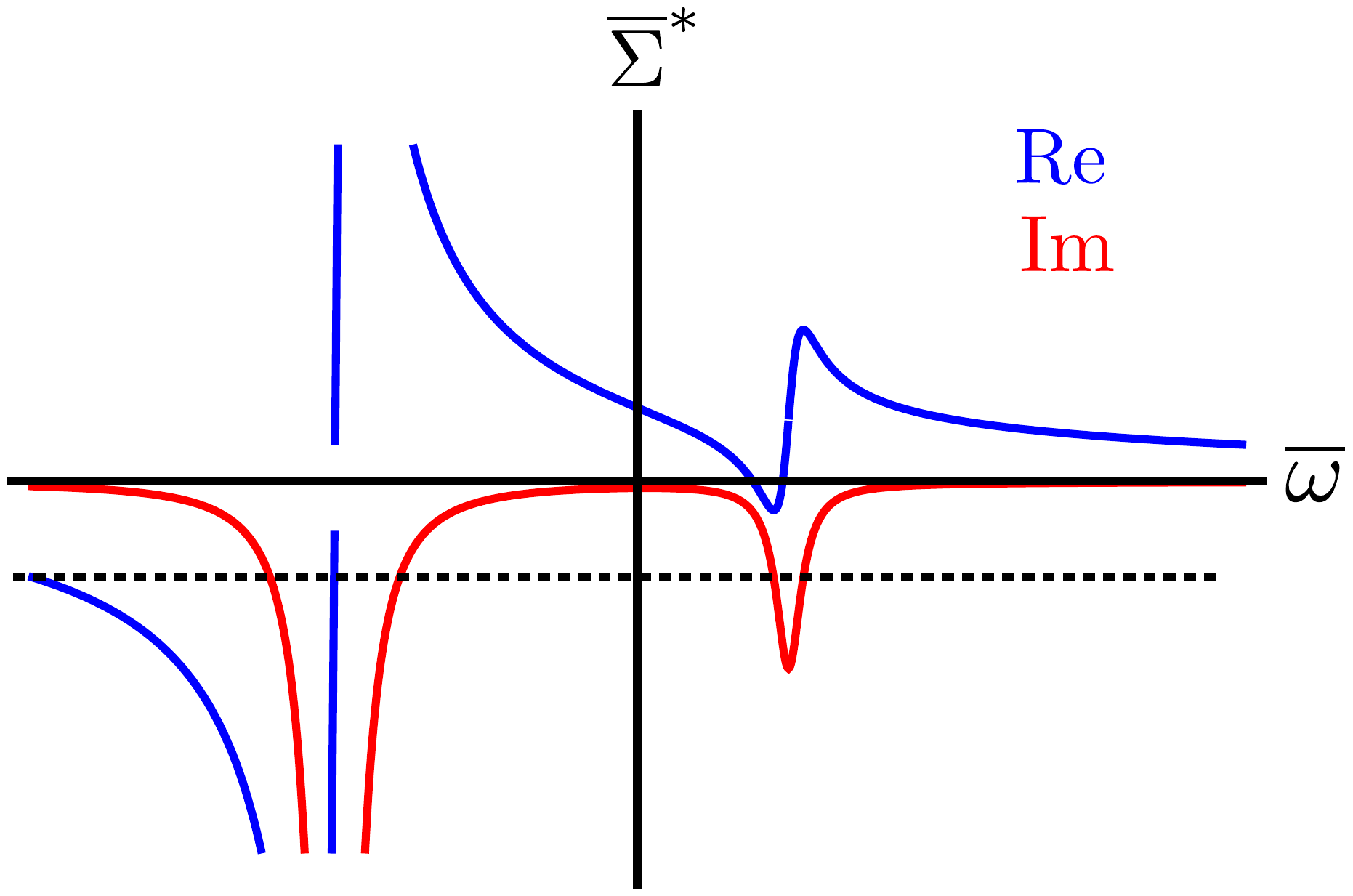}
\caption{Crude schematic showing the structure of $\overline{\Sigma}^*$ and the existence of multiple solutions for a given condition, shown by the dotted line. One can search the entire $\overline{\omega}$ axis looking for solutions. The true condition to find a quantized solution is not what is shown here. Because $D$ has multiple poles of different strengths, the solution region can pass through different regions of the $\overline{\omega}$-axis, adding solutions with every pole in $D$. \label{multiple_solutions}}
\end{centering}
\end{figure}

The spectrum $\overline{A}$ is not useful for comparison with an ensemble average. By construction, $\overline{A}$ is a $\delta$ function to model a single event exactly as it is observed. We can generate a meaningful spectrum from the multiplicity of each solution. For a given $x,y$, there may be quantized events at different energies $\omega_t$. At a given $x,y,$ \textit{and} $\omega_t$, there may be multiple $\overline{\omega}$ which give the same observable solution. We actually expect there to be multiple solutions given the structure of $D$, which has multiple poles and multiple crossings of the $\overline{\omega}$ axis. A rough schematic of the structure of $\overline{\Sigma}^*$ demonstrating the possibility of multiple solutions for some condition is shown in Fig.~\ref{multiple_solutions}. The multiplicity is determined by the amount of the $\overline{\omega}$-axis which matches the quantization condition.

We define a multiplicity function $W_{xy}(\omega_t)$ which simply counts the number of quantized solutions to Eq.~\ref{im_g} at each $x$, $y$, and $\omega_t$. Defining the multiplicity function allows us to reinterpret the propagation of the particle in an entropic way: we count all the ways that a normalized particle created at $x$ with energy $\omega_t$ can be annihilated at $y$.

The multiplicity is the key connection to experiment. If we assume that every quantized $\overline{\omega}$ solution is equally likely, the parameters with high multiplicity will produce a peak in a spectrum generated as an ensemble average of measurements based on the same initial conditions. We do require that the realized $\overline{\omega}$ gives an observable solution. Many other values of $\overline{\omega}$ will give not give a quantized, collapsed solution and not be observable. $D$ may even be poorly behaved on portions of the $\overline{\omega}$-axis. We do not need to artificially rule these values of $\overline{\omega}$ out because they will not contribute to the multiplicity, which is generated only from observable, normalized, and collapsed events. We have assumed nothing about properties of the Hamiltonian like the coupling strength. Any poor behavior or divergences in $\overline{\Sigma}^*$ which do not give a quantized solution can simply be ignored because they do not give physical solutions. This applies to any finite order order expansion of $D$ or the exact $D$. The search for solutions should cover the entire $\overline{\omega}$-axis.

Given an initial point $x$ and a final point $y$, we do not know which $\overline{\omega}$ solution the particle takes from $x$ to $y$. All we know is that the particle created at $x$ arrived at $y$ $-$ we are ignorant of the exact microstate. The collapse of the system creates entropy which we define as
\begin{equation}
S_{xy}(\omega_t) = \mathrm{log}_b \, W_{xy}(\omega_t)
\end{equation}
for base $b$ and we assign no dimension. We term this the collapse entropy. The collapse entropy is based on simple counting arguments. The observed event is a macrostate which can be connected to a number, $W$, of different microstates which correspond to different $\overline{\omega}$. The different $\overline{\omega}$ solutions come from different downfoldings of the many-body wave function onto $x,y$ that we cannot detect. This microstate-macrostate correspondence defines the collapse entropy.

The condition in Eqs.~\ref{single_spectrum} or \ref{im_g} is not a trivial one. It allows for interference among the different phases and amplitudes in $\overline{G}$, of which there are quite many because of the definition of $D$ at the three-particle level and its scattering processes at the beyond-three-particle level (really, beyond three field operators). $D$ has many poles, which could lead to many solutions at different regions of the $\overline{\omega}$-axis. We postulate that any conditions which meet Eq.~\ref{im_g} give an observable event, regardless of the microscopic details of the interference. Normalized solutions are possible, in principle, for any pair of points, even far from a quasiparticle solution where the residue of the bare field operator may be weak. Our insistence on this condition is explained by the Kramers-Kronig relation. We use the relation
\begin{equation}
\overline{G}_{ij}(\omega) =  \int_{- \infty}^{0} d\omega' \frac{\overline{A}_{ij}(\omega')}{\omega - \omega' - i\eta} + \int_{0}^{\infty} d\omega' \frac{\overline{A}_{ij}(\omega')}{\omega - \omega' + i\eta}  \nonumber
\end{equation}
to write the relevant element of $\overline{G}$ for a quantized event as
\begin{equation}
\overline{G}_{xy}(\omega) =  \int d\omega' \frac{\delta(\omega'-\omega_t)}{\omega - \omega' \pm i\eta} = \frac{1}{\omega - \omega_t \pm i \eta} \; .
\end{equation}
Another relation shows the self-consistency of the criterion
\begin{eqnarray}
\overline{G}_{xy}(\omega) &=& \frac{1}{\omega - \omega_t \pm i\eta} \\
&=&  \mathcal{P}\frac{1}{\omega-\omega_t} \mp i \pi \delta_{\eta}(\omega - \omega_t)  \\
\rightarrow \frac{1}{\pi} | \, \mathrm{Im} \, \overline{G}_{xy}(\omega) \, | &=& \delta_{\eta}(\omega - \omega_t) .
\end{eqnarray}
We can use a final equation
\begin{eqnarray}
\frac{1}{ \omega - \omega_t \pm i \eta}  &=& (\mp i) \int d \Delta t \; \Theta(\pm \Delta t) \, e^{i \omega \Delta t \mp \eta \Delta t } \, e^{ -i \omega_t \Delta t }  \nonumber  \\
&=& (\mp i) \, \mathscr{F} \left( \Theta(\pm \Delta t) \, e^{-i  \omega_t  \Delta t} \right) 
\end{eqnarray}
to recognize the Fourier transform of the time domain quantity
\begin{equation}
\overline{G}_{xy}( \Delta t) = (\mp i) \, \Theta(\pm \Delta  t) \, e^{- i  \omega_t  \Delta t} \; . \label{gbar_time}
\end{equation}
where the $-$ ($+$) sign refers to particle (hole) solutions with $\Delta t > 0$ ($\Delta t <0$). Eq.~\ref{gbar_time} describes the observed two-point process of particle creation/annihilation. A single, normalized particle with frequency $\omega_t$ is created at $x$, propagates one way in time, and is annihilated at $y$. This matches well with the most microscopic single-particle quantum mechanical process or idealized experiment without being overly restrictive. The spectral peak always has the same broadening, likely set by the type of particle, and normalization. For our approximate $\overline{G}$, only the imaginary part of the pole needs to be corrected, hence its appearance in Eq.~\ref{im_g}.

While the quantization condition is determined only by the imaginary part of $\overline{G}$, the real part of $\overline{\Sigma}^*$ sets the frequency at which the peak appears. The real part of $\overline{\Sigma}^*$ in Fig.~\ref{multiple_solutions} also has a nontrivial shape. $\mathrm{Re} \, \overline{\Sigma}^*$ determines the shift of the peak away from the Hartree-Fock solution. At a given point on the $\overline{\omega}$-axis, the number of nearby solutions depends on how rapidly the real and imaginary parts of $\overline{\Sigma}^*$ change away from this point.

In order to have complete information about the system, we couple the forward and backward processes to each other, as we have already done. The final $\overline{G}$ describes particles and holes, with the two microscopic channels mixing in the downfolding. The final observed event is of only particle or hole character, however, which can be determined by checking the sign of $\overline{A}(\omega)$
\begin{figure}[htb]
\begin{centering}
\includegraphics[width=0.65\columnwidth]{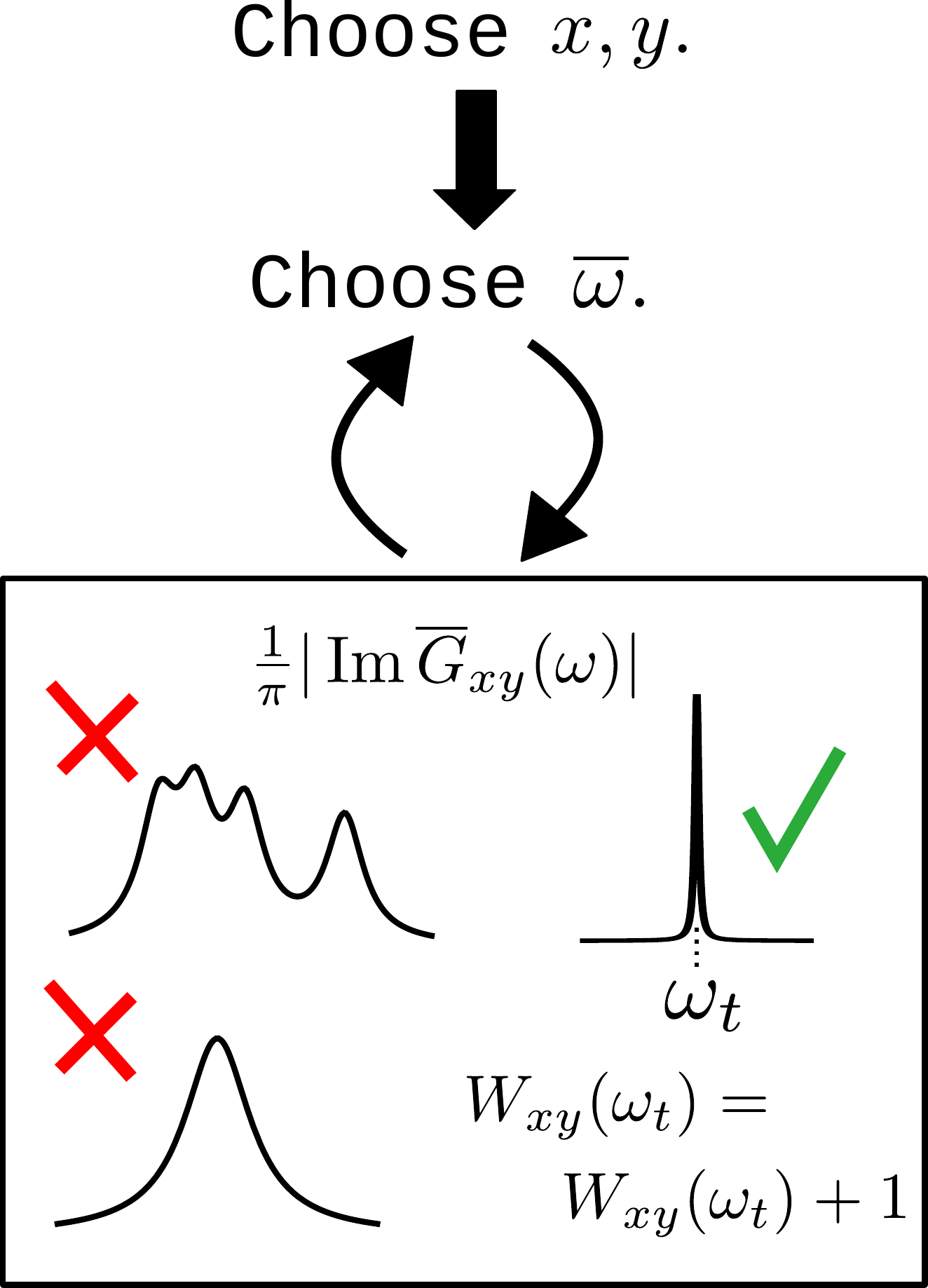}
\caption{Schematic showing the solution method for the inverse problem in Eq.~\ref{im_g}. The search for quantized solutions should cover the entire frequency axis. At every solution, the multiplicity increases by one. \label{solution}}
\end{centering}
\end{figure}

The solution method is summarized at the conceptual level in Fig.~\ref{solution}. For a given $x,y$ event, search the $\overline{\omega}$ axis for $\delta$ function peaks in $| \, \mathrm{Im} \, \overline{G} (\omega)  | / \pi $. When this condition is met and centered at energy $\omega_t$, raise the multiplicity $W_{xy}(\omega_t)$ by 1. Importantly, these spectral peaks at $\omega_t$ are not at the eigenvalues of $\mh$ or the energy differences $ \pm ( E_{\mathrm{N \pm 1}} - E_0)$. For small broadening values $\eta$ and $\epsilon$ and/or weak interactions, we expect the quantized spectral peaks to usually be very close to these energy differences. There could be noticeable deviations from this trend for strongly interacting systems, however.

Because of the emphasis on individual events, the multiplicity spectrum generated by solving our inverse problem is conceptually quite different than the ordinary spectral function of $G$. We believe the solution method based on searching for $\delta$ functions more closely matches experimental spectroscopies than the probability amplitude interpretation of the Born rule or inspecting the spectral function of $G$.

The purpose of the downfolding was to incorporate the effects of external lines. This is precisely the meaning of our inverse problem, which gives us some control of the behavior at the \textit{external} points while preserving the definition of the field operators and the internal time evolution based on the Schr\"odinger equation. The normalization of the external particle/hole lines does not happen until we perform the downfolding, which would appear to \textit{necessarily} have a nonlocal effect on the the initial point.

\begin{figure}[ht]
\begin{centering}
\includegraphics[width=\columnwidth]{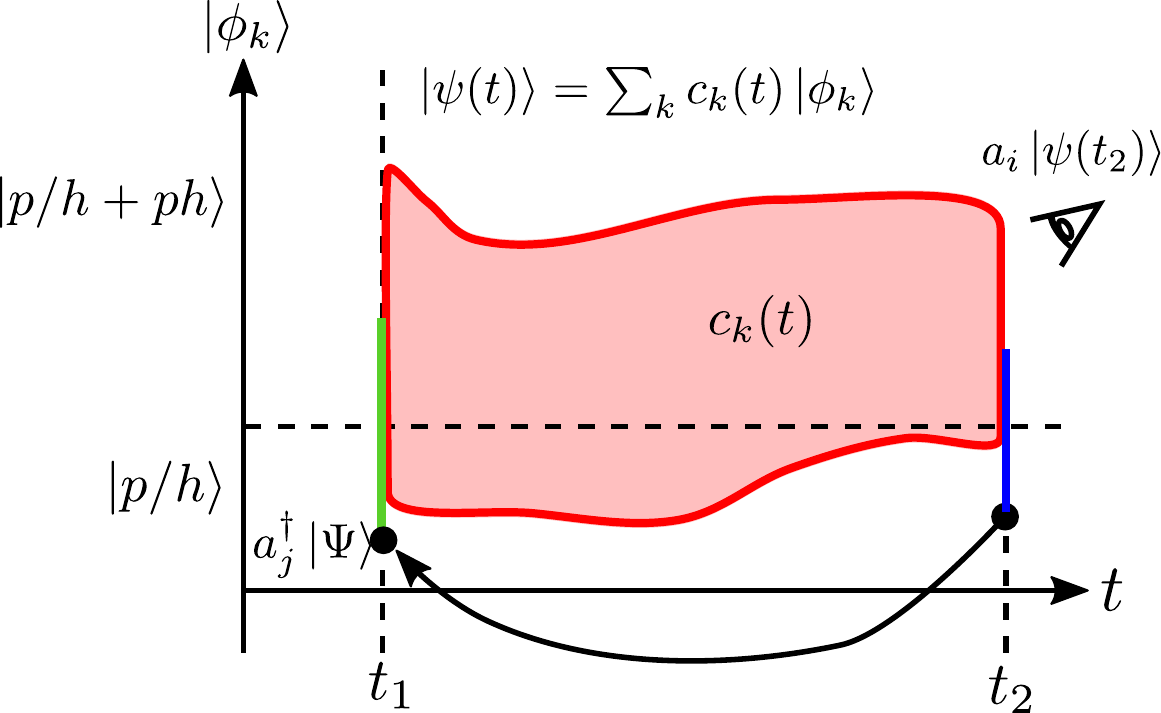}
\caption{The creation process (green), unitary time evolution (red), and annihilation (blue). It is not until the collapse at the time of measurement that the time evolution is downfolded. The nonlocal effect the collapse has on the initial state is indicated by the arrow from $t_2$ to $t_1$. \label{time_evolution}}
\end{centering}
\end{figure}

Despite the strangeness of our proposal, we believe our nonlocal Dyson equation accurately models quantum entanglement in an \textit{ab-initio} framework. To enforce normalization of the initial and final states, $\overline{\omega}$ is determined instantaneously with the measurement and has a nonlocal effect on the particle's initial point, as demonstrated in Fig.~\ref{time_evolution}. Until the measurement, the system remains in a high-dimensional superposition of configurations according to the Schr\"odinger equation. Then, the time evolution operator instantly collapses and the system's timeline is revised. The normalization of the initial state that happens with the collapse, and the very \textit{choice} of initial state, does not occur until the measurement is made. The probability amplitude does not pass continuously from one point to the other. Instead, initial and final states are both determined at the same instant and treated equally as a shared property. Our theory is a two-point collapse, which is consistent with entanglement, instead of a single-point collapse that could instead describe, for example, the density. If collapse onto every $\overline{\omega}$ solution is equally likely, the observable spectrum follows the multiplicity function.

Even though the nonlocal Dyson equation presented here deviates significantly from established many-body theory, there are remarkably few new postulates that went into its derivation. We first constructed an \textit{exactly} norm conserving correlation function based on unitary time evolution and the structure of Fock space. We downfolded the correlation function into the particle/hole space with the well-known L\"owdin technique. The downfolding necessarily leaves us with a choice of how the subspace propagates. The interpretation of the downfolding frequency as a stochastic collapse parameter closely models quantized events, matches the observed experimental signal, and agrees with the collapse postulate of quantum theory. We \textit{choose} $\overline{\omega}$ so that the observable event has the analytic properties we want, avoiding the possibility of any nonanalytic or unphysical behavior in the single-particle propagator. We argue that nonlocality in a normalized particle/hole theory is consistent with the Schr\"odinger equation; we already know that correlations with the Schr\"odinger equation (can) behave this way. A complete discussion of the entanglement, statistics, and dynamics is in Sec.~\ref{sec:analysis}. First, we finish developing our model and then compute spectra, our original motivation.

  \label{sec:sol_method}

\subsection{Nearly-single-reference concept}
  As we have said, we consider the multiconfigurational character of the problem essential, even for the most weakly-correlated systems. If we set $\epsilon = 0$, the approximate, single-reference $\overline{G}$ trivially matches the quantization condition. The broadening is always $\pm i \eta$, the value needed to match the quantization condition. In the high frequency limit $\overline{\omega} \rightarrow \infty$, far from any poles of $D$ at finite energies, the contribution of $\overline{\Sigma}^*$ goes to zero. As one searches these high values of $\overline{\omega}$ for solutions, the multiplicity of the Hartree-Fock quantized solution is arbitrarily high. The multiplicity saturates at these peaks. We refer to these solutions at $\overline{\omega} \rightarrow \infty$ as trivial solutions.

As discussed before, there are no trivial solutions in the exact $\mg$ because the residues are all $<1$, so we know these trivial solutions are incorrect. The exact $\mg$ guides our model. The residues are $<1$ because the true ground state (and all eigenstates) is \textit{multiconfigurational}, or a superposition of many different configurations. We point out that, even for the simplest, ordinary downfolding of $\mh$ with $\epsilon = 0$, there are many \textit{non}trivial solutions at finite values of $\overline{\omega}$ where the structure of $\mathrm{Re} \, \overline{\Sigma}^*$ gives a meaningful shift away from the Hartree-Fock energy. These are the solutions we are interested in and consider physical.

Applying the additional broadening $\pm i \epsilon$ removes the trivial solutions. In the high $\overline{\omega}$ limit, the broadening of $\overline{G}$ is $\pm i (\eta + \epsilon)$, which does not match the quantization condition. The trivial solutions are removed, and the search for quantized solutions can properly cover the entire $\overline{\omega}$-axis. We also consider the additional broadening by $i \epsilon$ to be quite physical, as it describes the actual broadening of spectral peaks by interactions. Since we have thrown out the amplitudes, we manually broaden with $i \epsilon$ to recover a realistic behavior and simulate the multiconfigurational system. The effect of the quantization condition is to sharpen this broadened pole into a properly quantized spectral peak. Additionally, $\epsilon$ determines the broadening of the virtual space correlation function. With nonzero $\epsilon$, $D$ is complex valued on the real $\overline{\omega}$-axis. With the finite lifetime broadening, a finite imaginary contribution is necessary to meet the quantization condition, and it is possible to pick up this imaginary part from the virtual space since $D$ is now complex valued.

If the broadening on \textit{both} sides of the eigenvalue problem used for the downfolding is set to $\mp i (\eta+\epsilon)$, then the trivial solutions are not removed. The imaginary parts in $D$ cancel, $\overline{\Sigma}^*$ is real valued, and the trivial solutions remain. Indeed, the purpose of adding $\epsilon$ to only the LHS of Eq.~\ref{ph_hamiltonian} is to additionally broaden the spectrum without changing the imaginary shift where the poles ``should" be located. That remains $\mp i \eta$ on the RHS of Eq.~\ref{ph_hamiltonian}. The LHS is a model correlation function, but it does not change the position of individually resolved poles, represented by the RHS. This lifetime-like broadening on one side of the equation gives meaningful structure to the imaginary part of $\overline{G}$, the portion which matters for the quantization condition.

This downfolding procedure which applies an extra $\epsilon$ broadening to only one side is far from perfect. The idea for the solution method, however, just does not fit in a single-reference framework with an easy downfolding of the Hermitian $\mh$. Of course, our prescription still identifies the downfolding frequency as a free parameter. It also accomplishes two more points. First, it forces the imaginary part of $\overline{\Sigma}^*$ to take a meaningful structure. This is absolutely \textit{mandatory} for our overall concept based on the inverse problem and to imitate the exact theory. Second, by establishing a downfolding procedure which can be applied consistently across different systems, we can associate the broadening of the virtual space correlation function $D$ with the lifetime of the excitations. While not perfect, this consistent treatment establishes a useful characteristic scale, $\epsilon$, for the broadening of the virtual space that depends on the system under consideration.

A final adjustment is necessary to complete the model. In our single-reference $\overline{G}$, particle and hole channels are not coupled to each other. This is incorrect and another failure of our single-reference construction. Furthermore, the sign of $\mathrm{Im} \, \overline{\Sigma}^*$ is incorrect to sharpen the overbroadened poles into a quantized peak. Within any channel (particle or hole), additional interactions with the virtual space can only broaden the pole further. It is actually not possible to meet our quantization condition only by applying the universal $\epsilon$ broadening. To compensate, we add an additional shift $i \gamma$ to the denominator of our approximate $\overline{G}$. $i \gamma$ takes the same sign in both channels, breaking particle/hole symmetry. $\overline{G}$ is now
\begin{equation}
\overline{G}(\omega) = \frac{1}{\omega - ( \lambda_P + \mathcal{F}  +  \overline{\Sigma}^*(\overline{\omega}) ) \pm i \left( \eta + \epsilon \right) + i \gamma  }  \label{final_g_bar}
\end{equation}
which can be more conveniently written as
\begin{equation}
\overline{G} = \frac{1}{  \omega  - (\lambda_{P} + \mathcal{F} + \overline{\Sigma}^*(\overline{\omega}) ) \pm i \eta + i m } 
\end{equation}
for $m \equiv \gamma \pm \epsilon$. If $\mathrm{Im} \, \overline{\Sigma}^*$ is of the same sign as $m$, it is possible for an imaginary contribution from $\overline{\Sigma}^*$ to match the quantization condition. Their contributions would cancel and the broadening is then $i \eta$. The symmetry breaking shift $i \gamma$ indicates the character of the background or true ground state. For the exact $\mg$, it is partially the role of the multiconfigurational ground state to remove the trivial solutions. $i \gamma$ plays this role by removing the trivial solutions even for the case $\epsilon = 0$.

Additionally, $i \gamma$ represents coupling to the opposite channel that would appear in the exact theory. In this simplified picture, coupling to the opposite channel is represented by a constant shift. Depending on the sign of $\gamma$, it is easier to find quantized solutions for either particles or holes. This makes sense as an indicator of the system's filling. For a mostly filled system, for example, there are more hole channels of appreciable strength than particles. Instead of being symmetric, this bias places more poles of $D$ on one side of the $\overline{\omega}$-axis than the other. The effect of this asymmetry is that the inverse problem exists primarily on one side of the $\overline{\omega}$-axis. As with $\epsilon$, we consider $\gamma$ to be physically meaningful.

A meaningful structure to $\mathrm{Im} \, \overline{G}$ is the key to a nontrivial inverse problem and the behavior we expect to be physical and imitate the exact theory. Our overall concept is: \textit{a nonzero imaginary contribution from the virtual space is necessary to match the quantization condition. Poles in the virtual space are broadened and shifted off of the real $\overline{\omega}$-axis.} Again, our construction is guided by the structure of the exact $\mg$. No poles of $\mg$ are \textit{quantized}, and to quantize one requires a nonzero contribution from somewhere else. Our parameters $\epsilon$ and $\gamma$ are adjustments to fit the single-reference theory into these conditions.

These adjustments to the time-ordered Schr\"odinger propagator form our nearly-single-reference (NSR) model. We expect this to be a physically meaningful and valid concept overall, but especially so for systems with well-defined spectral peaks analogous to long-lived quasiparticles. Of course, one of the main points of this work is that the theory is general and does not \textit{depend} on a quasiparticle picture. There is no redefinition of the field operators for a locally conserved current or any related quasiparticle interpretation of the spectrum. Nonetheless, many systems are of the long-lived quasiparticle type which does align well with our NSR model. We emphasize, however, that we generate an ensemble of collapse statistics for a bare particle instead of taking a quasiparticle interpretation of $G$ in some form (lifetimes, renormalized ``physical" field operators, local time evolution, quasiparticle equation, etc.).

A possible failure of the NSR model is the lack of $\omega$ dependence to the downfolding. Very naively, based on the Lehmann representation of $\mathcal{G}$ and the block structure of the L\"owdin downfolding, the exact downfolding could be of the form
\begin{eqnarray}
\overline{G}(\omega,\overline{\omega}) &=& G(\omega) +  \mathrm{correction}  \\
&=& G(\omega) + f(\omega) h(\omega,\overline{\omega}) f^{\dagger} (\omega)   \label{exact_gbar}
\end{eqnarray}
where the correction is a function of both $\omega$ and $\overline{\omega}$. This is a significant difference compared to our single-reference theory, in which $\overline{\Sigma}^*$ is a function of only $\overline{\omega}$. If the exact downfolding is indeed $\omega$ dependent, it may be that more of less solutions exist at certain frequencies than others. Roughly, more solutions may exist at high frequencies than low frequencies (or vice versa). Such an $\omega$-induced bias to the inverse problem would give additional shape and structure to the multiplicity that is beyond our NSR model. An $\omega$ dependence could also allow one to quantize a spectrum with an actual shape, instead of just one with a single, overbroadened pole which our current theory can handle. We also expect the lifetime broadening to play the role of setting the characteristic scale in the complex $\overline{\omega}$-plane in the exact theory since the Lehmann amplitudes are included in the exact $\mg$.

There are technical challenges to reaching such a downfolding. $\mathcal{G}$ is simply the definition of a correlation function and not an eigenvalue problem. We need, in principle, to downfold the entire correlation function with all of its multiconfigurational character. An exact downfolding and solution certainly seems quite complicated. It is only by approximating the actual correlation function with the Schr\"odinger propagator that we reach a simple, approximate external downfolding based on the $\epsilon$-shifted Hermitian operator ($\mh$) in the denominator of $U_S^T(\omega)$. Our NSR concept based on $U_S^T(\omega)$ is quite a special and simplified case, though still meaningful, in order to focus on the solution method and physics in a transparent way.

The same solution method based on our inverse problem applies to the exact downfolding. The general quantization condition is that one chooses the free downfolding parameter so that the exactly downfolded $\mathcal{G}$ is a single, normalized (unit residue), and $\eta$-broadened simple pole on the $\omega$-axis. Even if the L\"owdin downfolding cannot be applied \textit{per se} to the exact $\mg$, our idea is general enough to implement with any downfolding or compressive method applied to $\mg$ with a free parameter related to the choice of projection. In the exact theory, there should be no cutoff on $\overline{\omega}$, no trivial solutions because all residues of $\mathcal{G}$ are $<1$, and no need for the phenomenological adjustments $\epsilon$ and $\gamma$.

  \label{sec:nsr}

\section{Numerical study}
  As a simple proof-of-concept test of the theory, we consider the exact many-body Hamiltonian for a two-level system and search the NSR model for $\delta$ function solutions. The quantization condition is
\begin{widetext}
\begin{equation}
\delta_{\eta}( \omega - \omega_t ) = = \frac{1}{\pi} \,  \mathrm{Abs}  \Bigg[ \mathrm{Im} \Bigg[ \,  \cfrac{ 1 }{ (\omega - \bigg( P \mh P + P \mh Q \,  \cfrac{1}{ \overline{\omega} - Q \mh Q + i \epsilon } \, Q \mh P \bigg) + i \eta + i m } \, \Bigg] \Bigg ]   \label{collapse_condition}
\end{equation}
\end{widetext}
where the $==$ sign indicates that one searches the $\overline{\omega}$-axis for values which satisfy the equality. We compute the multiplicity for the diagonal particle addition matrix element of our two-level system. The center of the $\delta_{\eta}$ function on the $\omega$-axis is $\omega_t$. By construction, the quantization condition is met when the imaginary part of the second term in the denominator is equal to $m$. The solution method is now equivalent to the root finding problem
\begin{eqnarray}
\mathrm{Im} \,  \overline{\Sigma}^*   &==& m \;   \\
\mathrm{Im} \left( P \mh Q \, \frac{1}{ \overline{\omega} - Q \mh Q + i \epsilon } \, Q \mh P \right) - m &==& 0 \; . \label{root_finding}
\end{eqnarray}
We solve Eq.~\ref{root_finding} within a substantial numerical tolerance. When a solution is found, one must also store the real part of $\overline{\Sigma}^*$. The position of the $\delta$ function on the $\omega$-axis is $\omega_t = P \mh P + \mathrm{Re} \, \overline{\Sigma}^*$.

For testing purposes, we compare the multiplicity to the approximation with $\overline{\omega} = \omega$. We denote this approximation $A^*$,
\begin{equation}
A^*(\omega) = \frac{1}{\pi} | \mathrm{Im} \, \overline{G}(\omega) |  \big|_{\overline{\omega} = \omega} \; .
\end{equation}
While it is not ideal to compare two approximate theories to each other ($A^*$ to $W$) without a baseline reference ($A$, the spectral function of $G$), the comparison still has value. First, calculating $W$ provides a proof-of-concept for our solution method based on the inverse problem of searching for $\delta$ functions. By itself, this provides useful evidence that the idea is plausible. Second, we do expect the comparison between $W$ and $A^*$ to be meaningful for weakly-correlated systems. For weak correlation, we expect both methods to produce quasiparticle peaks in agreement with each other. We avoid a direct comparison between our approximate $W$ and the exact spectral function of $G$, $A(\omega)$. This comparison is somewhat unbalanced since $A$ properly includes all multiconfigurational character of the eigenstates but our NSR model does not. A future study of strongly-correlated systems should compare the exact $A(\omega)$ to the multiplicity computed from the exactly downfolded $\mg$, but such a study is beyond the scope of this work.

\begin{figure}[htb]
\begin{centering}
\includegraphics[width=\columnwidth]{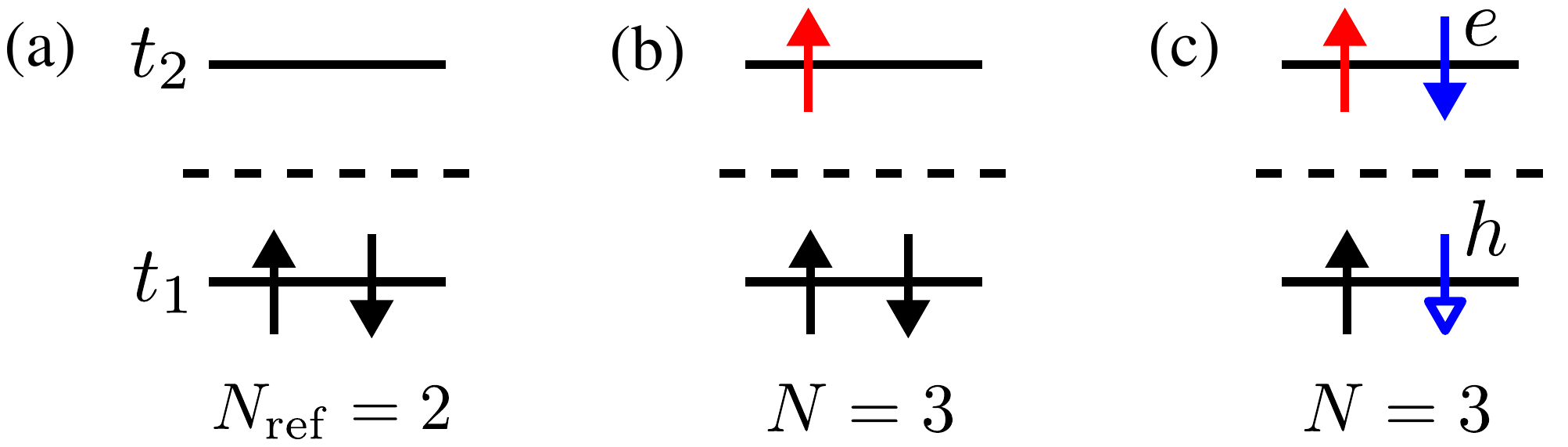}
\caption{Schematic of the two-level system in our numerical study. The added particle, shown in (b), can be coupled to a virtual electron-hole pair, shown in (c). \label{two_level}}
\end{centering}
\end{figure}

\begin{figure}[htb]
\begin{centering}
\includegraphics[width=\columnwidth]{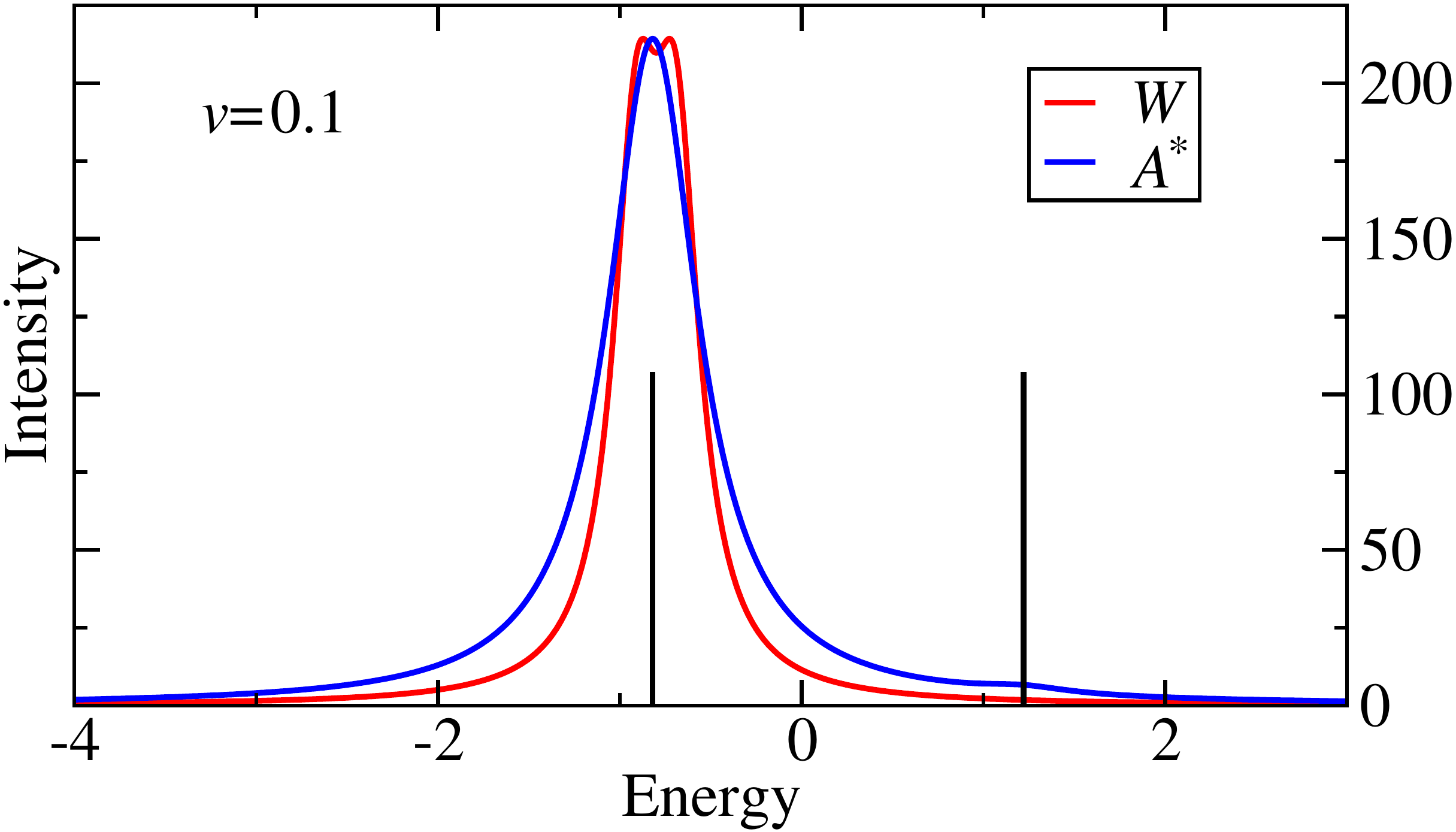}
\includegraphics[width=\columnwidth]{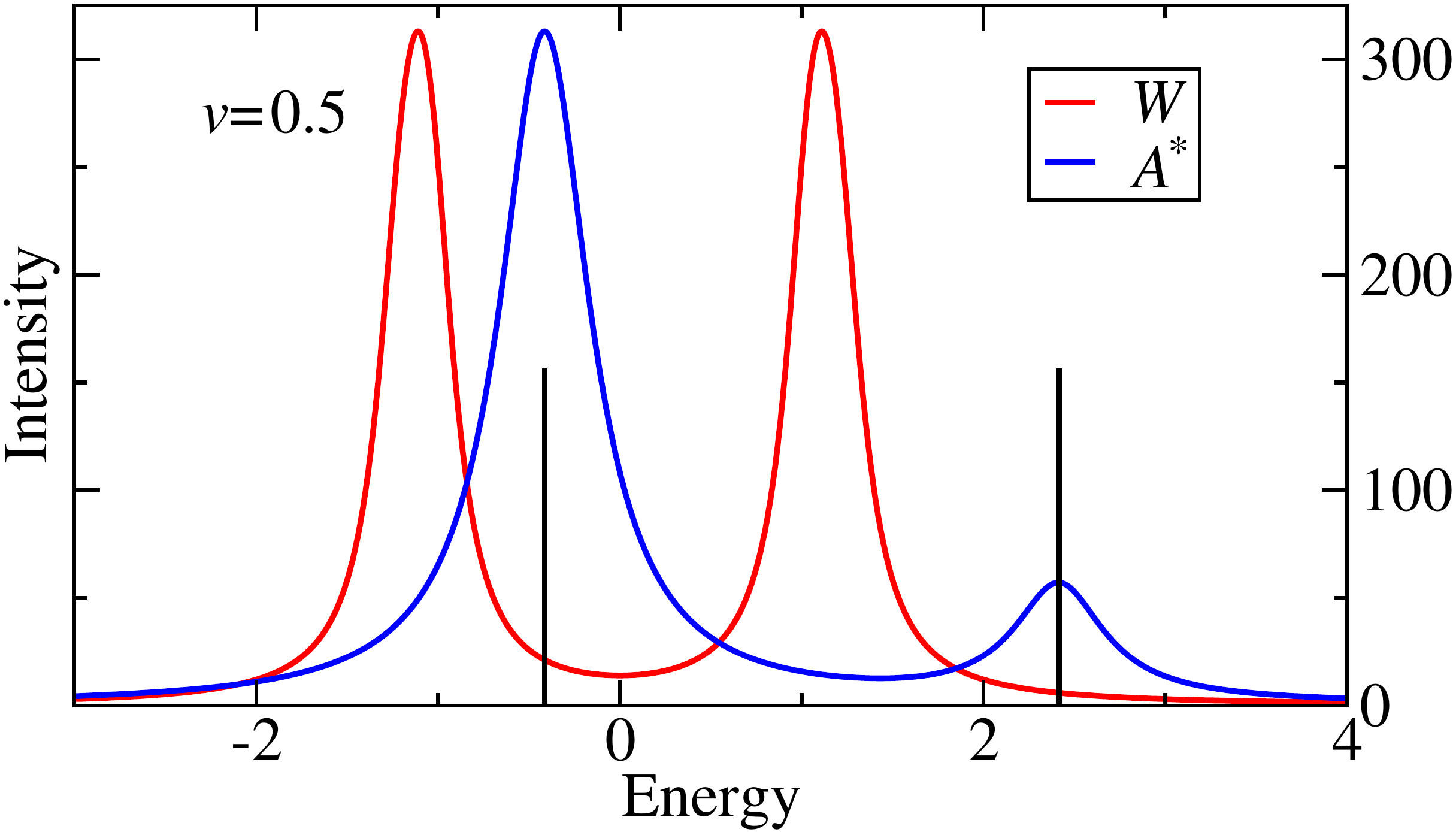}
\includegraphics[width=\columnwidth]{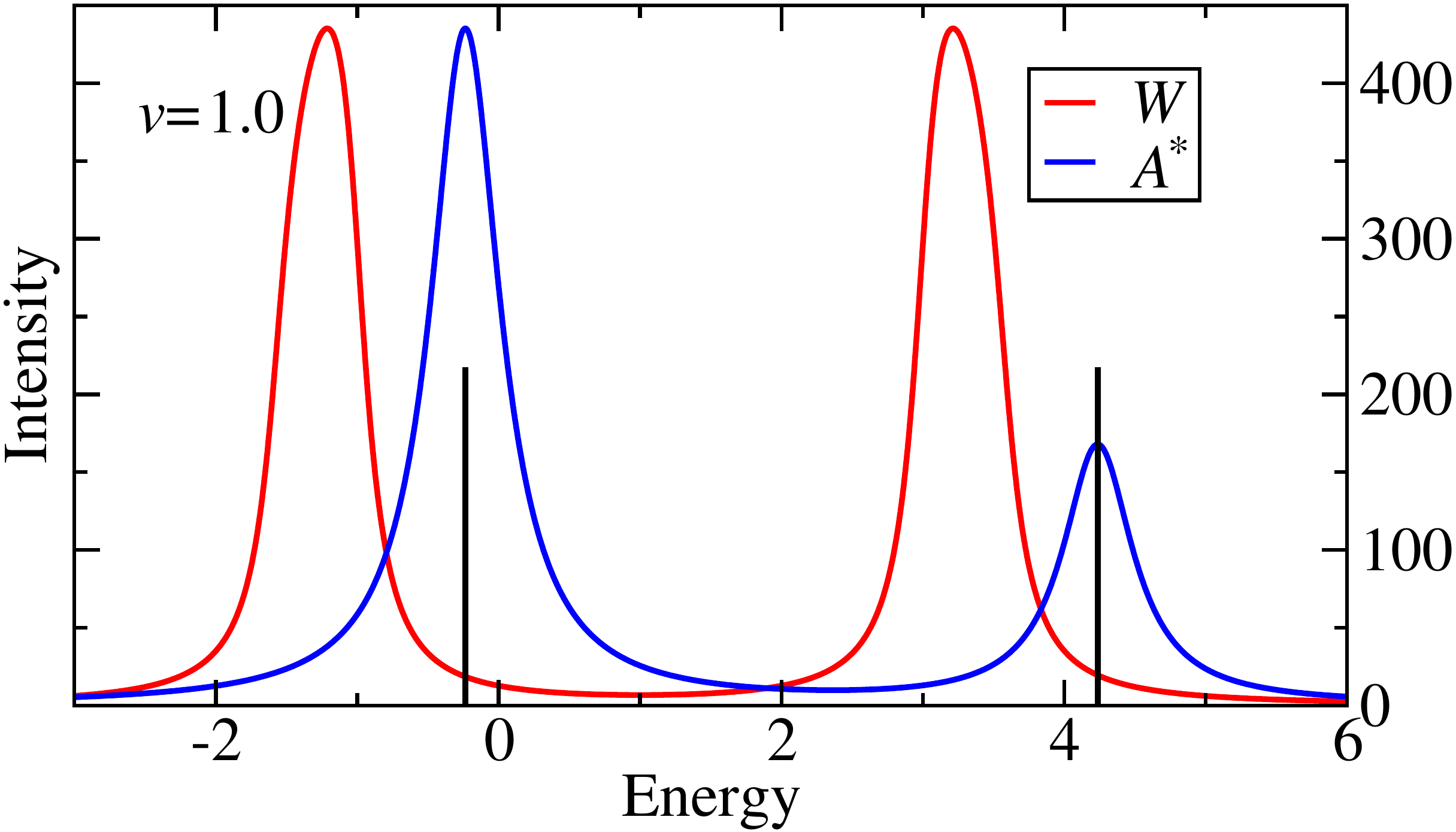}
\caption{Approximate spectral function $A^*$ and multiplicity $W$ as a function of interaction strength $v$. $t_1 = -1.0$, $t_2 = 1.0$. For comparison, the maximum height of $A^*$ is set to match the maximum of $W$. The scale on the right side shows the absolute value of $W$ for the chosen set of numerical parameters. Vertical black lines indicate the eigenvalues of $\mh$. \label{semi_spectra}}
\end{centering}
\end{figure}

We consider the particle addition problem with $N=3$ ($N=2$ is the reference configuration). For our calculations on a two-level system at half filling, there is only one particle addition matrix element. We compute the diagonal multiplicity for this one matrix element. We consider a metal and an insulator and, for simplicity, set the interaction matrix elements to a real constant, $v_{ijkl} = v$. The multiplicity compared against the $A^*$ approximation for the insulating case is shown in Fig.~\ref{semi_spectra} for three different values of interaction strength.

For weak interaction strength, shown in Fig.~\ref{semi_spectra}a, the multiplicity and $A^*$ spectrum almost perfectly agree. They both produce a single, sharp peak analogous to a long-lived quasiparticle solution. Despite the simplicity of the calculation, we consider the multiplicity spectrum in Fig.~\ref{semi_spectra}a to be a success. The two spectra are computed by different methods, and their underlying physics and meaning are very different from each other. 

\begin{figure}[htb]
\begin{centering}
\includegraphics[width=\columnwidth]{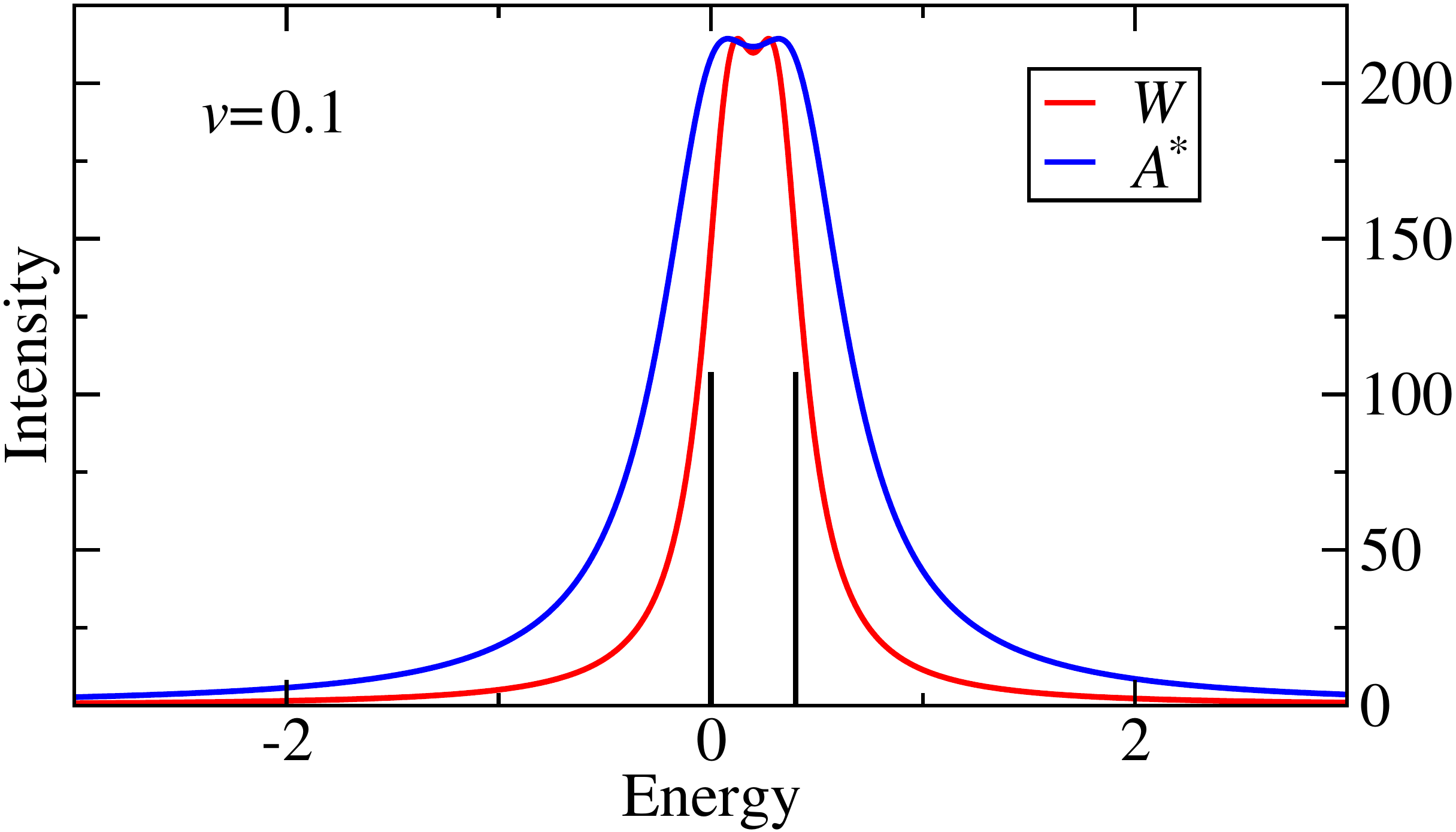}
\includegraphics[width=\columnwidth]{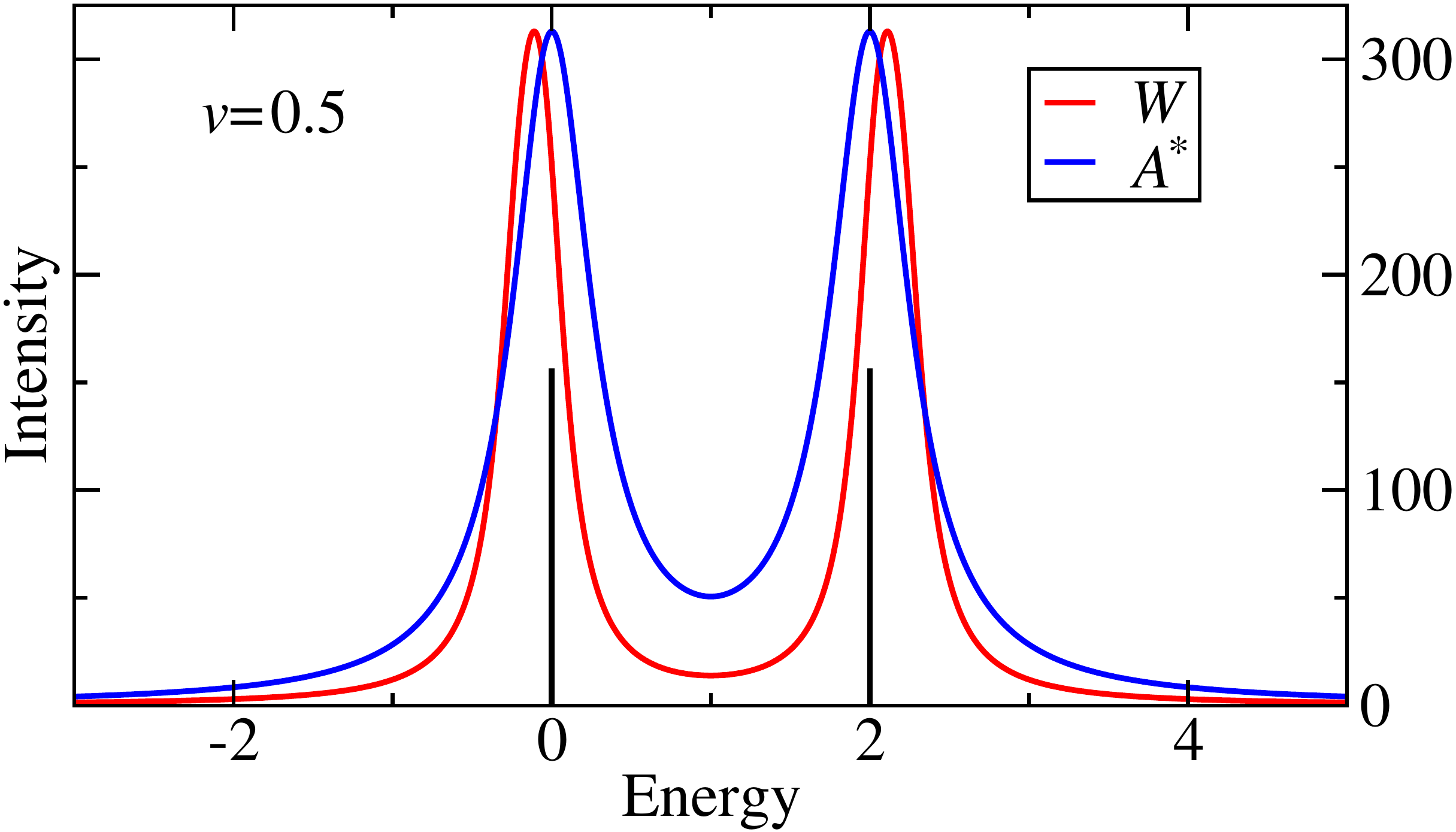}
\includegraphics[width=\columnwidth]{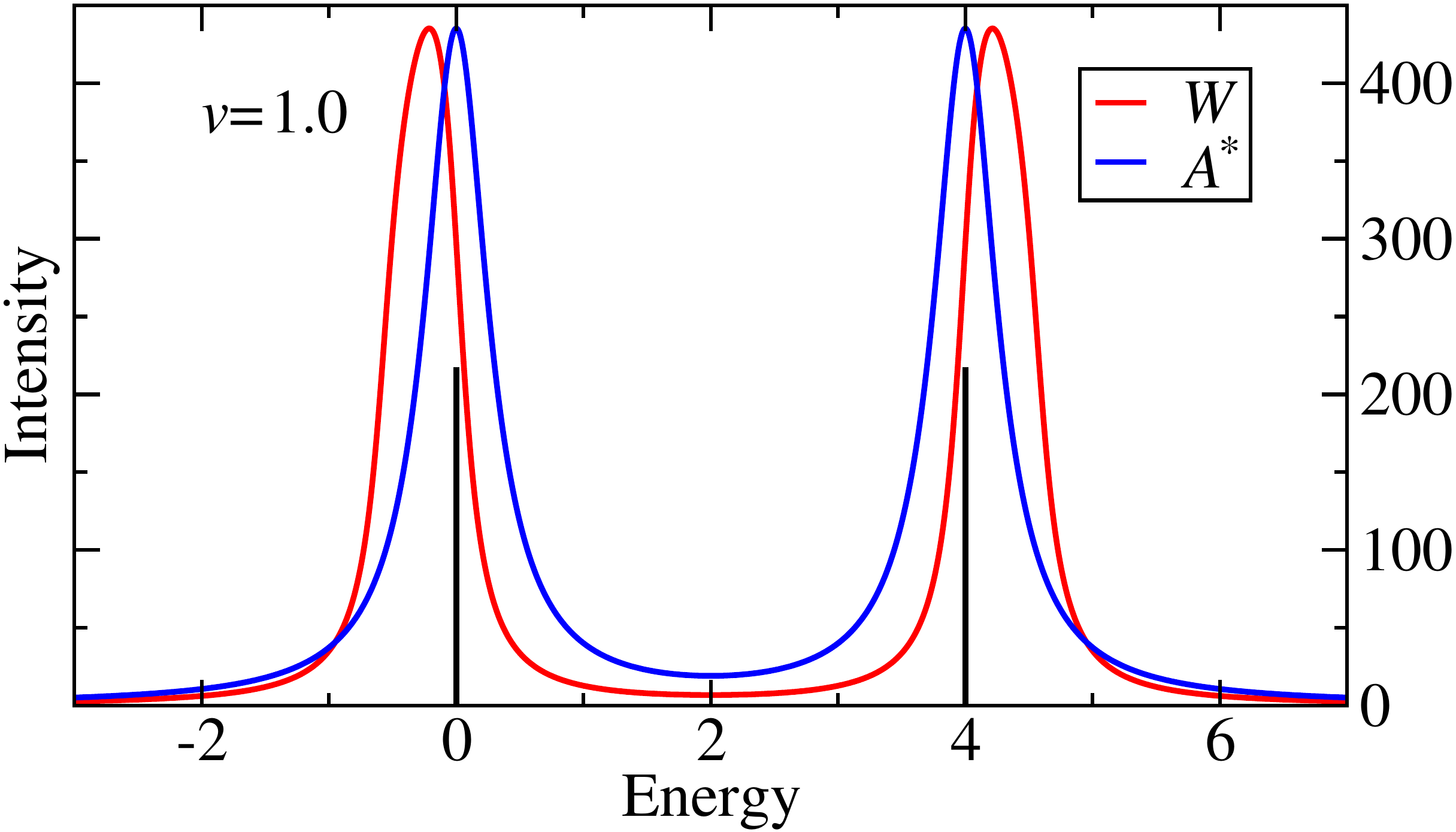}
\caption{Same as Fig.~\ref{semi_spectra} but with $t_1 = t_2 = 0$. \label{metal_spectra}}
\end{centering}
\end{figure}
For medium interaction strength, $v=0.5$, the multiplicity splits into two clear peaks. The $A^*$ approximation features a main peak and a weaker satellite feature. The two equal peaks in $W$ are because of the even symmetry around the center of $\delta(\overline{\omega})$ in $\mathrm{Im} \, \overline{\Sigma}^*$; the splitting vanishes for small or zero $\epsilon$. Satellite features in the multiplicity with different peak heights are also possible, but only if we allow for varying coupling strength between different elements of $P$ and $Q$. For $v=1$, the peak spacing is greater, but the multiplicity is qualitatively unchanged. Our main conclusion is that $W$ produces split peaks for high interaction strength. A more detailed comparison of the peak positions in $W$ and $A^*$ is not meaningful because our current numerical study is only approximate in the strongly-correlated regime. This is precisely the regime where we expect $W$ to significantly deviate from $A(\omega)$, whenever that comparison can be made.

The evolution of $W$ as a function of interaction strength is also shown for a metal in Fig.~\ref{metal_spectra} where we have set $t_1 = t_2 = 0$. The progression of the spectra is largely the same as in Fig.~\ref{semi_spectra}. For weak interaction strength, a single well-defined peak forms. At higher interaction strength, the peak splits into two bands. There is better quantitative agreement between $W$ and $A^*$ for the metal than for the semiconductor. As with the semiconducting case, quantitative features of $W$ should be compared against $A$ when the full multiconfigurational character of the problem is included in $W$.

The most important message of our numerical study is that, for weak correlation, the quasiparticle solutions produced with $W$ and $A^*$ are essentially indistinguishable. The two methods contain different physics and have very different interpretations, but their spectra are similar for weak correlation. It is possible to numerically solve a heavily simplified version of our proposed inverse problem and compute the multiplicity spectrum, and the results are reasonable. For strong correlation, $W$ produces split peaks and satellite features which are typical of experimental spectra in strongly-correlated materials. In general, these additional peaks appear as intersections of $\overline{\Sigma}^*$ with the solution region near additional poles of $D$ on the $\overline{\omega}$-axis.

  \label{sec:numerics}  
  
\section{Analysis, extensions, and interpretation}
  In this section, we outline concepts that are extensions to the theory presented in Sec.~\ref{sec:theory} or reinterpretations of the quantum many-body problem. Many of the ideas must be developed in parallel to reach our main conclusions. We will mostly stay within our nearly-single-reference (NSR) concept but occasionally address its failures or a more exact theory.

  \label{sec:analysis}
  
\subsection{Internal and external dynamics}
  Our discussion in the manuscript so far is aligned with describing the dynamics of a closed system as seen by an outside observer. The details of what happens with the collapse and how are somewhat unimportant from this vantage point, separated from the system undergoing collapse. We also want to develop a microscopic understanding of the collapse as it is experienced \textit{within} the closed system. This is a more complicated analysis that continues through the rest of the manuscript. Understanding this perspective is partly motivated by relativistic ideas, but also by the need for a self-consistent description of the dynamics within a closed system. The observer detecting the collapse is now part of, and entangled with, the quantum system.

To preface the discussion, we again emphasize that there is not yet a complete theoretical connection between the Green's function and many-particle wave function pictures. We have discussed this point extensively: as a superposition of configurations, the many-body wave function is not observable; the exact $G$ does not conserve particle number; local continuity is inconsistent with the Schr\"odinger equation. While the $\overline{G}$ theory and solution method in Sec.~\ref{sec:theory} could be a solution, the theory still needs to be interpreted. Fully connecting the single- and many-particle pictures with each other could require radical new concepts and interpretations of the physics.

We rewrite the nonlocal Dyson equation here for discussion (we group all terms into $\overline{\Sigma}^*$):
\begin{equation}
\overline{G}(\omega) = G^0(\omega) + G^0(\omega) \, \overline{\Sigma}^*(\overline{\omega}) \, \overline{G}(\omega)  \label{nonlocal_dyson}
\end{equation}
While Eq.~\ref{nonlocal_dyson} does describe collapse of the system onto a quantized solution, the interpretation of this equation from the single-particle perspective is difficult. The collapsed timeline does not pass through the spacetime coordinate of the wave function. The state is instantaneously shared between the two points without ever traversing the spacetime interval in between. As we have described, this nicely models quantum entanglement and collapse, but makes little sense from the perspective of an observer embedded in the system.

We consider Eq.~\ref{nonlocal_dyson} to be a version of the Einstein-Podolsky-Rosen (EPR) paradox \cite{einstein_pr_47}. Our theory is not based on a relativistic starting point and, as it is, the nonlocality is not a problem. If we require the correlation to propagate at a finite speed or that the information is transmitted across a finite spacetime interval, however, then there is a contradiction. These ideas are not included in our theory, but a relativistic extension requires them. The contradiction is already visible for the simplest nonrelativistic, interacting Hamiltonian on which our theory is based.

We move towards an extension of the $\overline{G}$ theory to address these issues. With a clear and quantifiable mechanism for entropy creation upon collapse, an information theoretical description of the dynamics seems natural. We develop our theory in this direction. As the system evolves according to $U(t,t')$ from its initial condition, it carries many possible outcomes. The number of observable outcomes it holds is the total multiplicity, $W^{\mathrm{tot}}$, which is the sum of individual multiplicities for every possible collapse. The multiplicity depends on three parameters: initial ($x$) and final ($y$) points and frequency $\omega$. We sum over all three to compute the total multiplicity.
\begin{equation}
W^{\mathrm{tot}} = \sum_{x,y,\omega} W_{xy} (\omega)
\end{equation}
As pointed out in Sec.~\ref{sec:theory}, even the \textit{initial} single-particle state is not determined until the collapse, hence the sum over $x$. In general, solutions exist at many different frequencies $\omega$. Because we cannot find a way of choosing which frequency is realized with collapse, we also assume the frequency of the solution is selected stochastically and include it in the sum for $W^{\mathrm{tot}}$. This choice can always be revisited, and redefining $W^{\mathrm{tot}}$ does not affect our overall conceptual framework.

These many possible outcomes are information which must be carried or stored by the system as it evolves. It is information contained in the wave function. Our overall picture of the quantum system is as a finite informational resource. For an informational resource which organizes information based on generalized bits which can take $b$ different values, the information capacity required to hold all possible outcomes is $S^{\mathrm{tot}} = \mathrm{\log}_b \, W^{\mathrm{tot}}$. The information formed from the wave function is always defined with respect to quantized correlations.

When the system collapses, the system gains information by reducing the number of possible outcomes. Each downfolding solution is a separate microstate which can be connected to a macrostate, an observed spectral function. Upon collapse, a number of microstates can be eliminated. Only the microstates which match the one realized macrostate need to be retained, which all together take information capacity $S_{xy}(\omega) = \mathrm{log}_b \, W_{xy} (\omega)$ (collapse entropy). The difference between $S^{\mathrm{tot}}$ and $S_{xy}(\omega)$ is gained information. By eliminating possible outcomes, the collapse allows the system to gain information of magnitude 
\begin{equation}
I_{xy}(\omega) = S^{\mathrm{tot}} - S_{xy}(\omega) \; .  \label{info_diff}
\end{equation}
$I_{xy}(\omega)$ is the Shannon information,\cite{shannon} or self-information, for the $x, y, \omega$ two-point event and equivalent to
\begin{equation}
I_{xy}(\omega) = \mathrm{log}_b \, p_{xy}^{-1} (\omega) \label{shannon_info}
\end{equation}
where $p_{xy}(\omega)$ is the probability of the $x,y,\omega$ event. The information can be gained by an observer which was part of the system's original time evolution. Collapse is only possible onto these internal points, or observers, entangled in the original time evolution. This, in turn, determines the observers which can gain information when their event is realized. An external observer who was not part of the time evolution of the many-body state cannot gain information from collapse of the system's time evolution operator. The information gain in Eq.~\ref{shannon_info} assumes the internal observer initially had no information about their system, a point we return to in Sec.~\ref{sec:interpretation}.

We propose that the gained information manifests itself in some way. There are two parts to our construction for the physical realization of $I$: the capacity for the information and the information itself. If the wave function carries information, then we identify spacetime as information capacity (We will often exclude ``space" from ``spacetime" in this work, though this must be included in the fully relativistic theory. The simplified picture we use is enough to develop the ideas.). We associate coherence in the wave function with information. In $\overline{G}$, it is coherence which determines the quantization condition and multiplicity of each solution, which also determines the information gained with collapse. The idea of coherence only makes sense in a finite time or space interval. Without any spacetime, there is no notion of coherence and can be no information.

There is therefore an opportunity in the theory to introduce a new spacetime coordinate that carries the self-information. It could also help solve the EPR paradox by describing the dynamics as it is felt within the closed system \textit{on a separate spacetime coordinate}. With collapse, it is clear how information is gained $-$ but there is not yet any channel to hold it. Our argument relies on the idea that information can only exist in some medium. We claim that the notion of information does not make sense without first defining the substance in which it is held. We introduce the idea of internal and external times. The external time is the one considered so far in this work. It describes the evolution of the many-body wave function. We denote it $t$ or $dt$. The internal time is the one experienced by the points of collapse which exist inside the system. We denote it $d\tau$. (From here on, internal and external do not refer to the internal and external times of perturbation theory discussed in Sec.~\ref{sec:theory}.)

\begin{figure*}[htb]
\begin{centering}
\includegraphics[width=2.0\columnwidth]{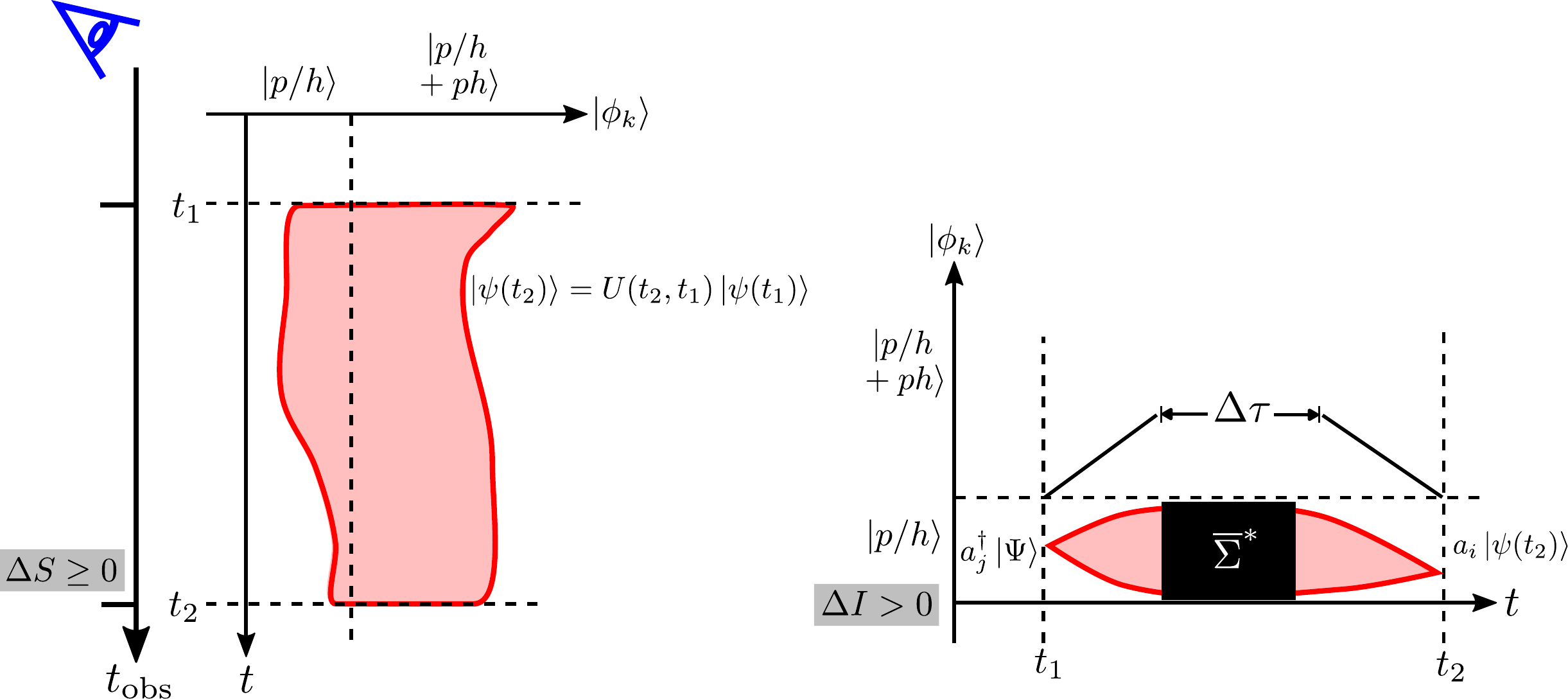}
\caption{Observer, external, and internal times are denoted $t_{\mathrm{obs}}$, $t$, and $\tau$, respectively. Initially, $t_{\mathrm{obs}}$ and $t$ run parallel to each other. When the collapse occurs, the observer sees an instantaneous rewriting of the system's history. The instantaneous nature of the event is represented by the rotation of the $t$-axis to be orthogonal to $t_{\mathrm{obs}}$ on the right side of the figure. With the collapse, an internal spacetime bridge, indicated by $\Delta \tau$ and which is imperceptible to the observer, forms between the end points. \label{nonlocal_timelines}}
\end{centering}
\end{figure*}

Fig.~\ref{nonlocal_timelines} demonstrates the concept. There are three time coordinates in the figure. The external observer has their own time $t_{\mathrm{obs}}$. The many-body wave function evolves on the $dt$ scale. We propose that upon collapse, the informational resource creates a third time coordinate, labeled $d \tau$, that connects the two end points of the event on a type of spacetime bridge or link. Initially, the time coordinates of the observer and wave function can run parallel to each other, as shown on the left side of Fig.~\ref{nonlocal_timelines}. With the collapse, the external observer sees an instantaneous rewriting of the system's history. This is represented by the orthogonality of $t$ to $t_{\mathrm{obs}}$ on the right side of the figure. It is not until this instant that the causal, quantized event on $dt$ exists. At the same instant on $t_{\mathrm{obs}}$, an internal spacetime bridge is formed to connect initial and final points.

The $d\tau$ spacetime coordinate exists only between the two end points and is imperceptible to the external observer. It may have a different duration than the interval $\Delta t$, a point we return to later. With each collapse of the system, the resource creates a new spacetime bridge in order to accommodate the information gained with collapse. Unlike the external coordinate in which the wave function propagates, the internal coordinate is created only when the time evolution operator collapses. The internal time serves a fundamentally different purpose. There are two spacetimes which can exist simultaneously: one in which static information exists, and one through which information is gained. Very importantly, the two-point structure of the event actually \textit{allows} for a notion of distance between the points and a bridge which connects them. We are not concerned with the properties of the bridge (space-like, time-like, etc.), only proposing that it exists and that it has a clear purpose $-$ to carry the self-information $I$. Understanding its details is beyond the scope of this work. If collapse in some region of external spacetime is for some reason not possible, then no internal spacetime can form.

Simply packaging the self-information into a carrier without introducing a new spacetime coordinate is not a solution to the EPR paradox in Eq.~\ref{nonlocal_dyson}. Because of the nonlocality in the external dynamics, it is not possible for the carrier to pass through the $dt$ coordinate of the wave function. This makes sense: the information \textit{gained} is \textit{not} the information in the wave function. A new channel is necessary. Our analysis is based on the nonlocal structure and physics of Eq.~\ref{nonlocal_dyson}, but the informational considerations are significant. The implication of our finding is that information cannot be gained through the same medium in which it is stored.

The internal spacetime should have a finite duration by some measure in order to have a finite information capacity. The capacity of the link depends on the amount of information it must carry. The duration of the internal time step is
\begin{equation}
d\tau_{xy}(\omega) \sim d \tau_0 \,I_{xy}(\omega)  \label{dilation}
\end{equation}
in units of some fundamental time step $d \tau_0$. On the fundamental time step $d \tau_0$, the phase of some object can take $b$ different values. The dilation in Eq.~\ref{dilation} describes the duration of the time bridge as seen by an external observer, if that were possible. The bridge should always have a finite size in order to hold information. If the end points coincided on $d \tau$, then this interval of zero width would have zero information capacity. The bridge only forms, however, when information is \textit{gained}, $I_{xy}(\omega) > 0$.

We define the proper internal time by the completion of the event so that an internal observer does not experience the dilation in Eq.~\ref{dilation}. A specific, benchmark two-point event constitutes a uniform $d\tau$ step as it is experienced by an internal observer and no matter the dilation to $d\tau$. Internally, the benchmark event always seems to take the same amount of time. The passage of the internal proper time is defined by each quantized event or information gain, no matter how long it takes on some absolute scale. These are the only perceivable events, and their occurrence \textit{defines} the passage of internal time. The proper internal time step is therefore constant.

The second part of our construction is the information itself. It is in the internal spacetime that we propose a theory describing the propagation of some quantized information carrier appears. At the instant of collapse, the informational resource creates a spacetime bridge. At the same time, an information carrier is suspended in the bridge. The details are far beyond the scope of this work, but its purpose is to carry $I_{xy} (\omega)$ and the correlation between the two points. At the instant of the collapse, these correlations are already determined, even though it would appear internally as if they are transmitted by the carrier. An internally embedded observer would have no access to the correlation from the external dynamics. As part of our concept of a \textit{finite} informational resource, this signal can only be sent at a finite speed. Our preferred picture is that the gained information is suspended in the bridge since the information is a joint property of both points, but it may take some rate to encode the information. This rate is also finite. By construction, the end points of the internal spacetime are fixed by the collapse $-$ a single field operator at the initial and final points creates/annihilates a normalized particle on the true ground state, no matter its level of multireference correlation. The description of any correlation carrier on the internal time could be relatively simple since the virtual processes determining the correlations and multiplicity are actually in the external dynamics.

If it is a fundamental frequency, or phase change, that determines the information capacity rather than a time, the internal frequency must contract according to
\begin{equation}
d \omega_{\tau,xy\omega}= \frac{1}{ d\tau_{xy}(\omega) } = \frac{d \omega_0}{I_{xy}(\omega)} \; . \label{frequency_dilate}
\end{equation}
The finite informational resource could have such a fundamental frequency, $d\omega_0$, that sets the information capacity based on a maximum rate of phase change, roughly analogous to a finite bandwidth. We see the internal frequency $d \omega_{\tau}$ as a coarse graining around this fundamental frequency. By coarse graining around such a fundamental frequency, the most rapid phase changes would be forbidden and the information capacity limited. As the information capacity of the bridge increases, its frequency scale contracts according to Eq.~\ref{frequency_dilate}. 

In order to fit with our concept of a \textit{finite} informational resource, the total information ($S^{\mathrm{tot}}$) and information gain ($I_{xy}(\omega)$) should both be finite. It could be that the underlying resource is discrete. A discrete frequency grid could keep the multiplicity countable and finite. A finite limit on the information gain also places a finite limit on the length of any spacetime bridge. We also consider the convergence factor $\eta$ an indicator that the underlying resource is finite.

The picture we have outlined here is very complicated. A real system would frequently collapse and create different spacetime bridges. Each of these links is \textit{its own} spacetime. In practice, we will primarily treat the internal spacetime as if these many different bridges are combined to form a uniform background spacetime (though still internal). This uniform internal coordinate can perhaps be treated, for example, as a proper spacetime coordinate with the correct normalization. In this picture with the length of each individual bridge fixed, different information carriers can be allowed to propagate at different rates to compensate.

To complete the information gain process, the gained information is eventually stored in an updated time evolution operator and initial condition. After the collapse, the system begins its evolution again with the newly gained information. The information in the wave function is contained in both the time evolution operator and the system's initial condition, both of which are reset after the collapse. The information is not gained and lost, it is gained and stored. We assume the resource is able to hold the gained information in the new wave function and time evolution operator.

To summarize, we can define the information gained with each collapse based on the Shannon information. We model the quantum system as a finite informational resource and identify spacetime as information capacity. With each collapse, the resource creates a spacetime bridge of size $I_{xy}(\omega)$, and the information gained is suspended in this bridge. The system gains information with every collapse. After each collapse, the system's time evolution operator and initial condition are updated to reflect the gained information. Quantization is the protocol for counting microstates and gaining information.

Even without details about the form of the information carrier or how the gained information is stored, we can still substantially develop this phenomenology and make several qualitative predictions.

  \label{sec:internal_external}
  
\subsection{Induced collapse}
  
Can we use the $\overline{G}$ theory to describe the dynamics of a ``real" system far away in the universe? The difficulty with this proposal is that the collapse in the $\overline{G}$ formalism is induced by the measurement. It is not until the measurement and collapse that a normalized particle enters the theory. If we want to describe the dynamics of a system far away in the universe without an observer, we could let the wave function evolve across all possible configurations without ever collapsing. We do not consider this very satisfying, however, largely because it does not match our perception of the universe as a single configuration. We would like to describe normalized particles instead of the high-dimensional wave function. To utilize $\overline{G}$ for such a theory, some type of natural collapse without a human observer is necessary.

In the model we started building in Sec.~\ref{sec:internal_external}, informational considerations are key. With the collapse, an internal observer gains information, and the resource resets the system's time evolution operator to restart its unitary time evolution. The event \textit{changes} information. The stochastic nature of this process and the way it changes information suggest that it can be interpreted as an informational reset.

If we interpret the collapse as such a reset, it must be preceded by a reason to change information. We propose that the need for a reset is a conflict of information carried by competing wave functions. A conflict of information appears when two overlapping wave functions do not agree on all correlations between points, or all possible observable outcomes. Imagine, as a simplified example, that the many-body wave function suddenly encounters a particle with which it is not yet entangled. In the downfolded picture, poles in $D$ that would appear from this occupied state are not yet included in the system. These poles lead to additional quantized solutions and are related to extra information that would be contained in the fully entangled wave function. Even before the downfolding occurs, the information and correlations contained in the current wave function, not including the new particle, are incorrect. The finite informational resource is monitoring the system, interpreting it in terms of quantized solutions, and detects the bad information. To resolve this conflict, the resource dumps the current information, becomes entangled with the additional particle, and resets its time evolution. In this picture, the resource has some instruction to resolve any and all conflicts. Collapse onto a quantized solution serves as a type of reset to try to reach this goal. By collapsing onto a single outcome, information is gained, relations between points and occupation numbers are adjusted, and the conflict could be resolved.

A simple numerical check on two different wave functions is a possible test for such conflicts of information. For many-body wave functions far from each other, their overlap is very low and their numerical values do not conflict. As they evolve on the $dt$ scale and come into contact, however, a sort of orthogonality check could be meaningful. The information is encoded in the superposition of phases that exists at each point. If the two wave functions overlap within some numerical tolerance, their information agrees and their evolution continues. If not, a mixed state forms and the system collapses to try to resolve the conflict. We expect the collapses to come in pairs since a conflict to one wave function must also be a conflict to the other. At the simplest level, one can compare the numbers of solutions carried by each wave function, $W^{\mathrm{tot}}$, to find conflicts, though this simplified check cannot account for conflicts about specific correlations when their total number is identical.

We expect decoherence between the system and its environment to play a major role in creating conflicts and inducing collapse. The information contained in the many-body wave function is directly related to its coherence. The quantization condition, which determines the information, is a condition on only $\mathrm{Im} \, \overline{G}$. This is quite a sensitive condition that requires special coherence among all the single-particle states contributing to $\overline{\Sigma}^*$.

If the many-body wave function overlaps with a different wave function from its environment, their interference could alter the coherence at any point and the information contained in $\overline{\Sigma}^*$. This creates an informational conflict. The informational resource detects the conflict and resets the time evolution after incorporating the newly added degrees of freedom. The information then spreads as the wave function evolves in external time, and the process repeats. The more the system interacts with its environment, more informational conflicts occur and the system becomes further entangled. As the system becomes more entangled and more states contribute to the many-body wave function, the delicate coherence required to preserve information contained in the wave function becomes even more difficult to maintain $-$ as the system grows, there are more and more phases which must align properly to meet the quantization condition. We therefore expect a size extensivity to the rate of decoherence and collapse, with larger systems being more prone to decoherence and more frequent collapse. This potentially increasing loss of coherence as the system overlaps with its environment could lead to frequent informational conflicts and collapses.

Determining whether or not the system undergoes a collapse with each event could be essential to understanding the dynamics. In our theory, collapse is the mechanism for entropy creation and evolves the system in one direction, as is known to be the case for real systems. This concept is not new $-$ collapse upon measurement has always been understood to be irreversible. However, with an \textit{ab-initio} model for collapse and a microscopic mechanism for entropy creation, we can now produce quantifiable statistics about the collapse dynamics.

We assume that the particle collapses onto a quantized solution with every informational conflict. We take a final step in completing our model, which is to assume that only these collapses create the observable statistics of both the external and internal dynamics. Although we expect decoherence to play a major role, we do not claim to know exactly how collapse is induced. Nonetheless, we can still continue with the picture that it is only these normalized events which are observable, and they occur naturally on their own. This difference in philosophy $-$ calculating individually collapsed, quantized events instead of just the probability amplitude \textit{of} a quantized event $-$ underlies our entire theory.

With this adjustment, the human observer's role in collapsing the system is removed. For a real system, collapse happens frequently. If the entanglement and multiplicity are functions of space, collapse into regions of space with high multiplicity is more likely than into regions with low multiplicity. As the system stochastically evolves between quantized solutions, the internal spacetime also changes as the observable particle's position and multiplicity are updated. Each two-point collapse defines two observable spatial points, both internal and external, and an internal time step $d \tau$.

  \label{sec:induced_collapse}
  
\subsection{Quantization force}
  We want to understand how the form of the inverse problem affects the shape of the multiplicity. Consider a cluster of particles at the origin as the ground state. Assume that the cluster is a superposition of negative and positive charge densities so that no electrostatic force exists at a test point $\mathbf{r}$. We want to know if there is any preference for an event an infinitesimal step towards or away from the cluster \textit{only} due to the quantization condition. To understand this infinitesimal, or semilocal, step, we consider the local (diagonal) multiplicity and its spatial gradient. Much of observed physics is local, which partly motivates this analysis. The gradient of the local multiplicity, if it is nonzero, gives an entropic description of how the particle infinitesimally evolves in space with each semilocal collapse. For this reason, and for the remainder of this work, we suppress the $\omega$ dependence of the multiplicity and consider a multiplicity which depends only on space.

We again consider our NSR system to search for general trends in the local multiplicity. This is our starting point for a minimal interacting system which isolates the effect of quantization as much as possible. In the NSR approximation, the local, or diagonal, element of $\overline{G}$ is
\begin{equation}
\overline{G}_{jj} = \frac{1}{  \omega  - (\lambda_{P,jj} + \mathcal{F}_{jj} + \overline{\Sigma}^*_{jj}(\overline{\omega}) ) \pm i \eta + i m }  \; .
\label{nsr_diagonal}
\end{equation}
In order to match our quantization condition, a finite imaginary contribution $im$ from $\overline{\Sigma}^*$ is necessary. In Eq.~\ref{nsr_diagonal}, the imaginary part needed from $\overline{\Sigma}^*$ for the quantization condition is fixed. The number of solutions, therefore, depends on the portion of the $\overline{\omega}$-axis which meets the condition $\mathrm{Im} \, \overline{\Sigma}^* == m$. The peak position for $\overline{A}(\omega)$ on the $\omega$-axis is determined by $\mathrm{Re}\, \overline{\Sigma}^*$. Similar to the condition on the imaginary part, the number of solutions at any given frequency depends on the portion of the $\overline{\omega}$-axis which meets some fixed value for $\mathrm{Re} \, \overline{\Sigma}^*$.

\begin{figure}[htb]
\begin{centering}
\includegraphics[width=0.75\columnwidth]{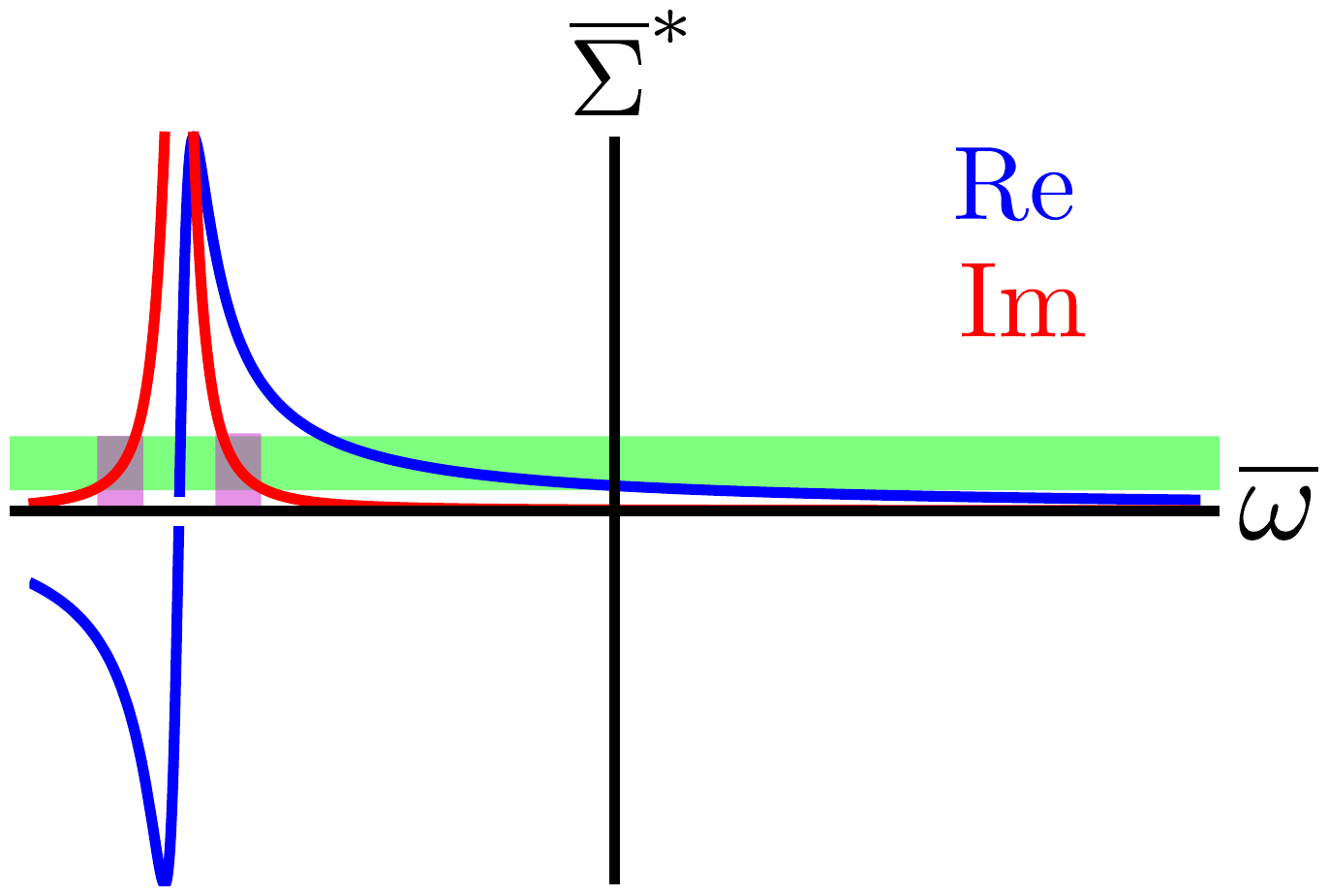}
\caption{Assume the single event spectral function is a Dirac-$\delta$ function when $\mathrm{Im} \, \overline{\Sigma}^*$ takes the value in the green region. The multiplicity is determined by the portion of the $\overline{\omega}$ axis meeting this condition, shown in purple for the $\mathrm{Im} \, \overline{\Sigma}^* = y$ solution region.  \label{branch_cut_2}}
\end{centering}
\end{figure}

Fig.~\ref{branch_cut_2} shows these effects graphically. For a fixed value of $\mathrm{Im} \, \overline{\Sigma}^*$ needed to match the quantization condition, indicated by the green band, the solution region on the $\overline{\omega}$-axis is given by the purple regions. The width of the purple regions on the $\overline{\omega}$-axis determines the multiplicity of the solution.
As indicated in Fig.~\ref{branch_cut_2}, $\overline{\Sigma}^*(\overline{\omega})$ itself has its own pole structure. $\overline{\Sigma}^*(\overline{\omega})$ depends on another correlation function, $D(\overline{\omega})$, as $\overline{\Sigma}^*(\overline{\omega}) = v \, D(\overline{\omega}) \, v$. $D$ has broadened poles at the multiparticle excitations of the system, each with their own amplitude.

For this discussion, we make the assumption that the three-particle excitations defining $D$ are long-lived and do not decay into beyond-three-particle excitations. We assume that $D$ is diagonal $-$ each three-particle excitation stays in its own channel. We make a final assumption for our cluster example, which is that a single dominant pole in $D$ contributes to the event of interest. The approximate $\overline{G}$ is
\begin{equation}
\overline{G}_{jj} = \frac{1}{  \omega  - (\lambda_{P,jj} + \mathcal{F}_{jj} +  v \frac{1}{\overline{\omega} - ( \lambda_Q + \mathcal{E}_Q ) \pm i \epsilon}  v ) \pm i \eta + i m }  \; . \label{approx_gbar}
\end{equation}
With these assumptions, we have replaced the matrix multiplication $v \, D \, v$ by a simple multiplication. 

The width of the solution region on the $\overline{\omega}$-axis depends on taking density of states-like (DOS-like) slices through the broadened pole in $D$. The imaginary part of this term is a $\delta$ function with strength set by $v^*v$,
\begin{eqnarray}
\overline{\Sigma}^*(\overline{\omega}) &=& v \, \frac{1}{\overline{\omega} \pm i\epsilon} v^*  \label{im_sigma} \\
\mathrm{Im} \, \overline{\Sigma}^*(\overline{\omega}) &=& \mp v^* v \, \delta_{\epsilon}(\overline{\omega}) == m \; ,  \label{re_sigma}
\end{eqnarray}
where $==$ indicates the quantization condition. The number of solutions to match a given value of $\mathrm{Im} \, \overline{\Sigma}^*$ therefore depends on the interaction strengths and the shape of $\delta_{\epsilon}(\overline{\omega})$. The flatness of the broadened $\delta$ function in $D$, or how rapidly it changes in the regime which gives quantized solutions, determines the multiplicity. In the single pole approximation of Eqs.~\ref{im_sigma} and \ref{re_sigma}, we can ignore the position of the pole on the $\overline{\omega}$-axis which sets the peak position. We assume the quantization condition is met at some frequency and only need to consider the nearby shape of the single pole on the $\overline{\omega}$-axis that dominates the physics of this one solution.

\begin{figure}[htb]
\begin{centering}
\includegraphics[width=\columnwidth]{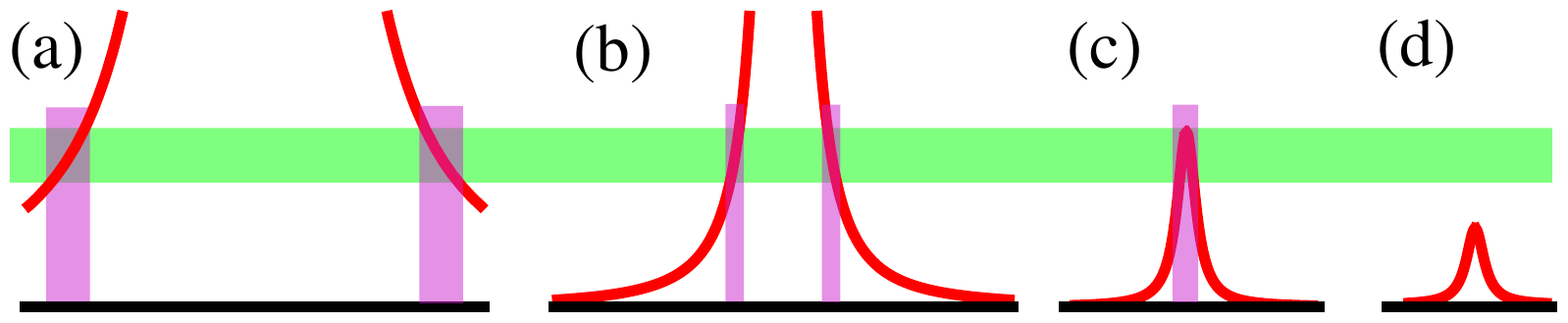}
\caption{Different coupling strengths $v$ affect the strength of the broadened $\delta$ function in $\mathrm{Im} \, \overline{\Sigma}^*$. As $v$ changes, the overlap of the green band with the red curve changes nontrivially. The multiplicity of the inverse problem mimics this shape. \label{amplitudes}}
\end{centering}
\end{figure}

We focus on varying $v$, the coupling strength between the particle and virtual spaces, and its effect on the multiplicity. Fig.~\ref{amplitudes} shows the imaginary parts of several poles with different coupling strengths. Their intersection with the green band, shown by the shaded purple regions, indicates the multiplicity of the solution. At weak coupling (panel (d)), there are no quantized solutions. The coupling is so weak that the contribution from $\overline{\Sigma}^*$ is effectively zero. As the coupling increases, the solution region passes through the entire broadened $\delta$ function. Eventually, as in panel (a), the broad, flat part of the $\delta$ function in $D$ gives a quantized solution and leads to very high multiplicity (this is not shown particularly well in the figure).

\begin{figure}[htb]
\begin{centering}
\includegraphics[width=\columnwidth]{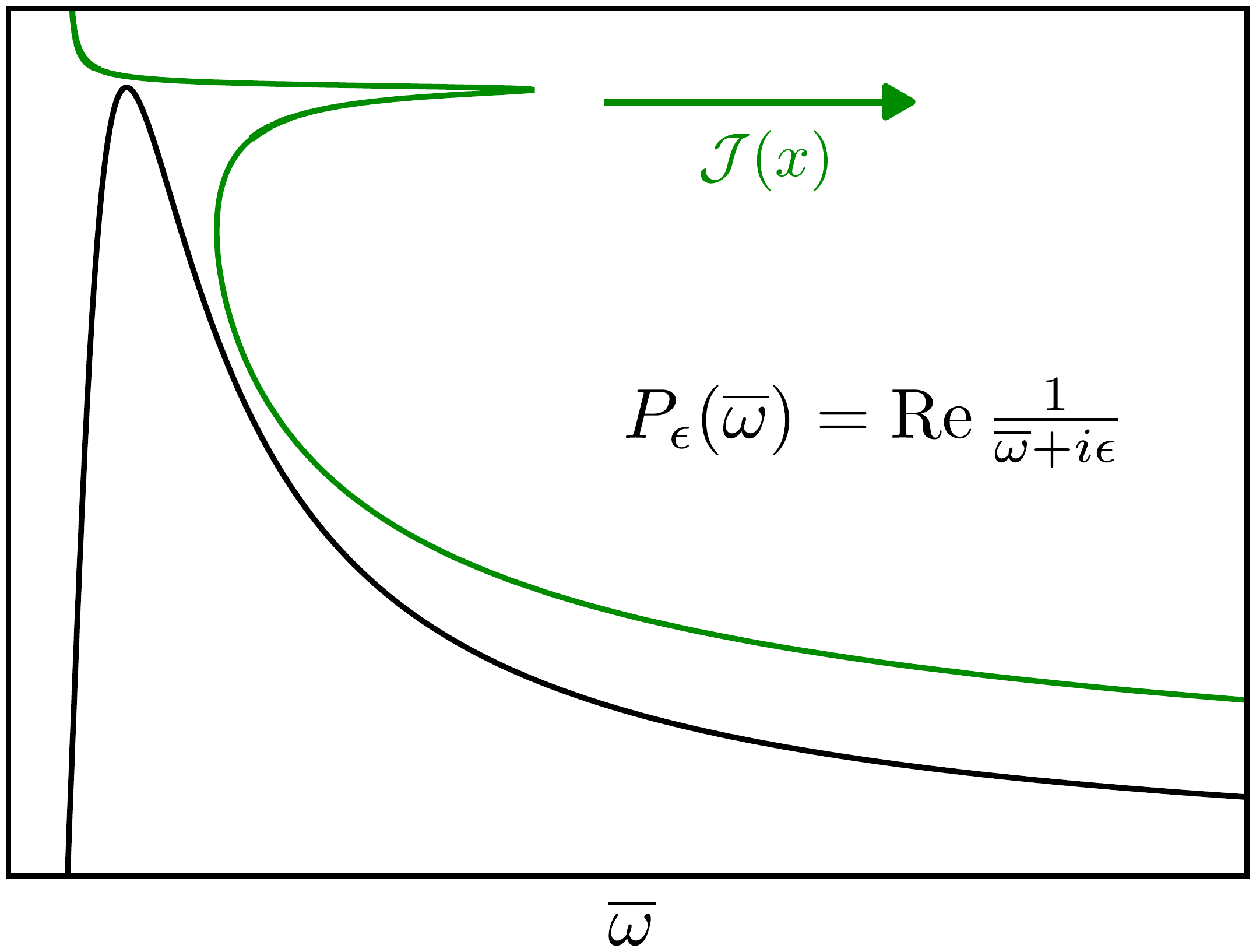}
\includegraphics[width=\columnwidth]{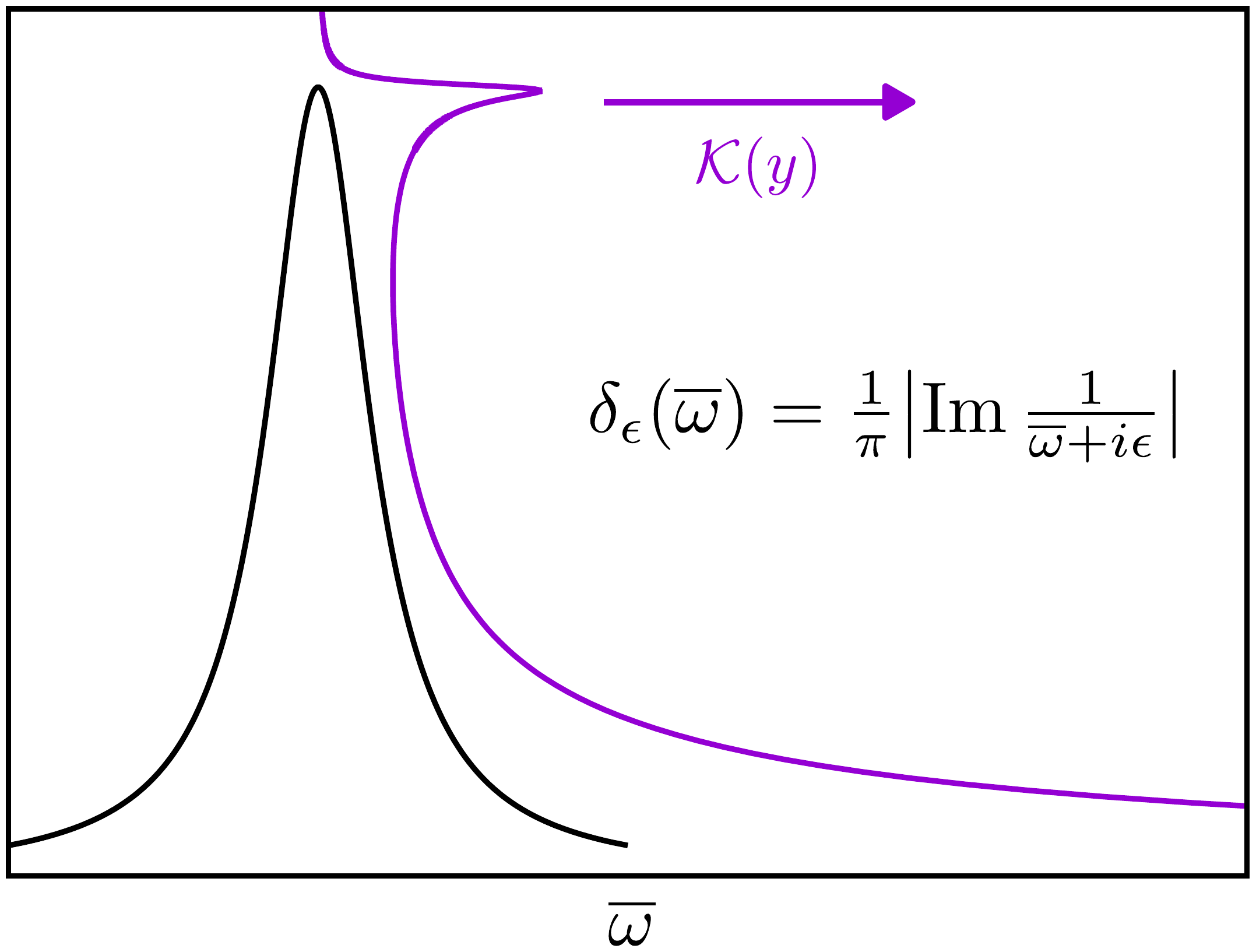}
\caption{The real and imaginary parts of a broadened pole and their DOS-like functions $\mathcal{J}$ and $\mathcal{K}$. The figure shows actual data for numerically exact functions. The curves have been scaled differently to facilitate plotting and should not be directly compared to each other. Only their shapes are meaningful. \label{j_curve}}
\end{centering}
\end{figure}

Therefore, the shape of the broadened pole in $D$ gives a noticeable structure to the multiplicity. The $\delta$ function has a broad, flat region, a finite maximum, and an inflection point which determine the onset of different regimes in the multiplicity. The shape is quite detailed. For this reason, we introduce two functions, $\mathcal{J}$ and $\mathcal{K}$, to help the analysis.
\begin{eqnarray}
P_{\epsilon}(\overline{\omega}) &=& \mathrm{Re} \, \frac{1}{\overline{\omega}+i\epsilon}  \\
\mathcal{J}(x) &=& \int d\overline{\omega} \, \delta(x - P_{\epsilon}(\overline{\omega})) \\
\delta_{\epsilon}(\overline{\omega}) &=& \frac{1}{\pi} \bigg| \mathrm{Im} \, \frac{1}{\overline{\omega} + i \epsilon} \bigg|  \\
\mathcal{K}(y) &=& \int d\overline{\omega} \, \delta( y - \pi \delta_{\epsilon}(\overline{\omega}))
\end{eqnarray}
Our use of the outer $\delta$ functions in these integrands is only a matter of convenience. The functions $\mathcal{J}$ and $\mathcal{K}$ count the number of solutions on the $\overline{\omega}$-axis which meet certain conditions, $\mathrm{Re} \, \overline{\Sigma}^* = x$ or $\mathrm{Im} \, \overline{\Sigma}^* = y$, respectively. They are defined in analogy to computing a DOS by integrating a $\delta$ function over some parameter. Numerically computed $\mathcal{J}$ and $\mathcal{K}$ are shown in Fig.~\ref{j_curve}.

We first focus on the strong coupling regime. We consider the multiplicity at a certain frequency determined by $\mathrm{Re} \, \overline{\Sigma}^* = x$. At strong coupling, the fixed value $x$ is met by the broad, flat part of the broadened pole in $D$ (see Eq.~\ref{approx_gbar}). Both real and imaginary parts of $D$ in the solution region are far from the pole in $D$. In this regime, $\mathrm{Im} \, D$ is very flat and changes slower than $\mathrm{Re} \, D$ (this is simply an empirical observation). For this reason, we treat $\mathrm{Im} \, D$ as fixed $-$ it is roughly constant and is assumed to match the quantization condition. We do not need to consider its effect on the multiplicity for nearby, infinitesimal changes to $v$ because it changes more slowly than the real part. Equivalently, we can think of counting nontrivial solutions to the inverse problem without any modifications ($\epsilon = \gamma = m = 0$), although this changes the shape of $\mathcal{J}$ and $\mathcal{K}$ when $\epsilon = 0$. The multiplicity at a fixed frequency is then determined by how rapidly $\mathrm{Re} \, \overline{\Sigma}^*$ changes close to a certain point where $\mathrm{Re} \, \overline{\Sigma}^* = x$. As the coupling strength increases, the solution region is pushed into a broader, flatter region of the pole in $D$, increasing the multiplicity. This DOS-like description of the solution region is quantified by the function $\mathcal{J}(x)$.

\begin{figure}[htb]
\begin{centering}
\includegraphics[width=\columnwidth]{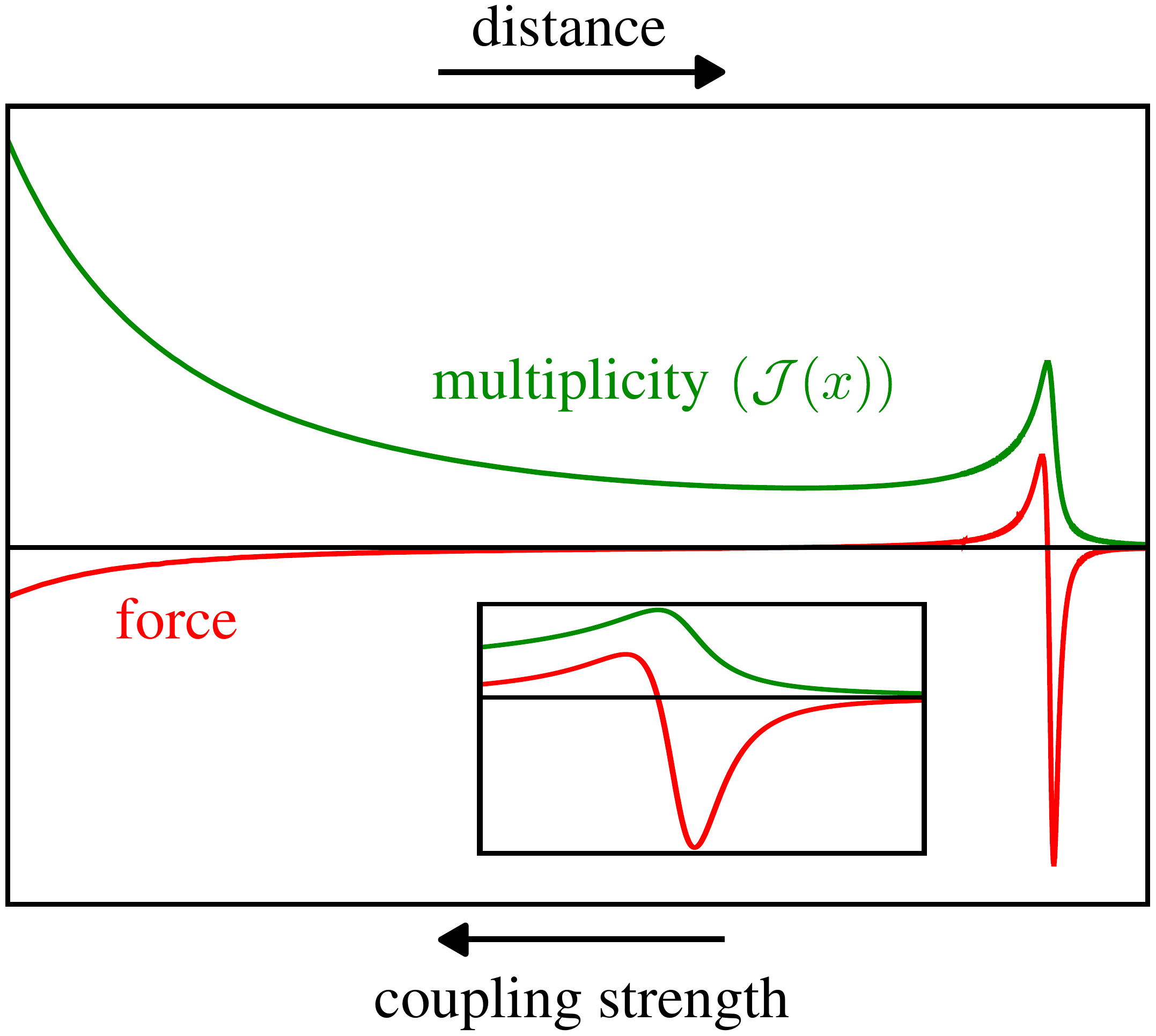}
\includegraphics[width=\columnwidth]{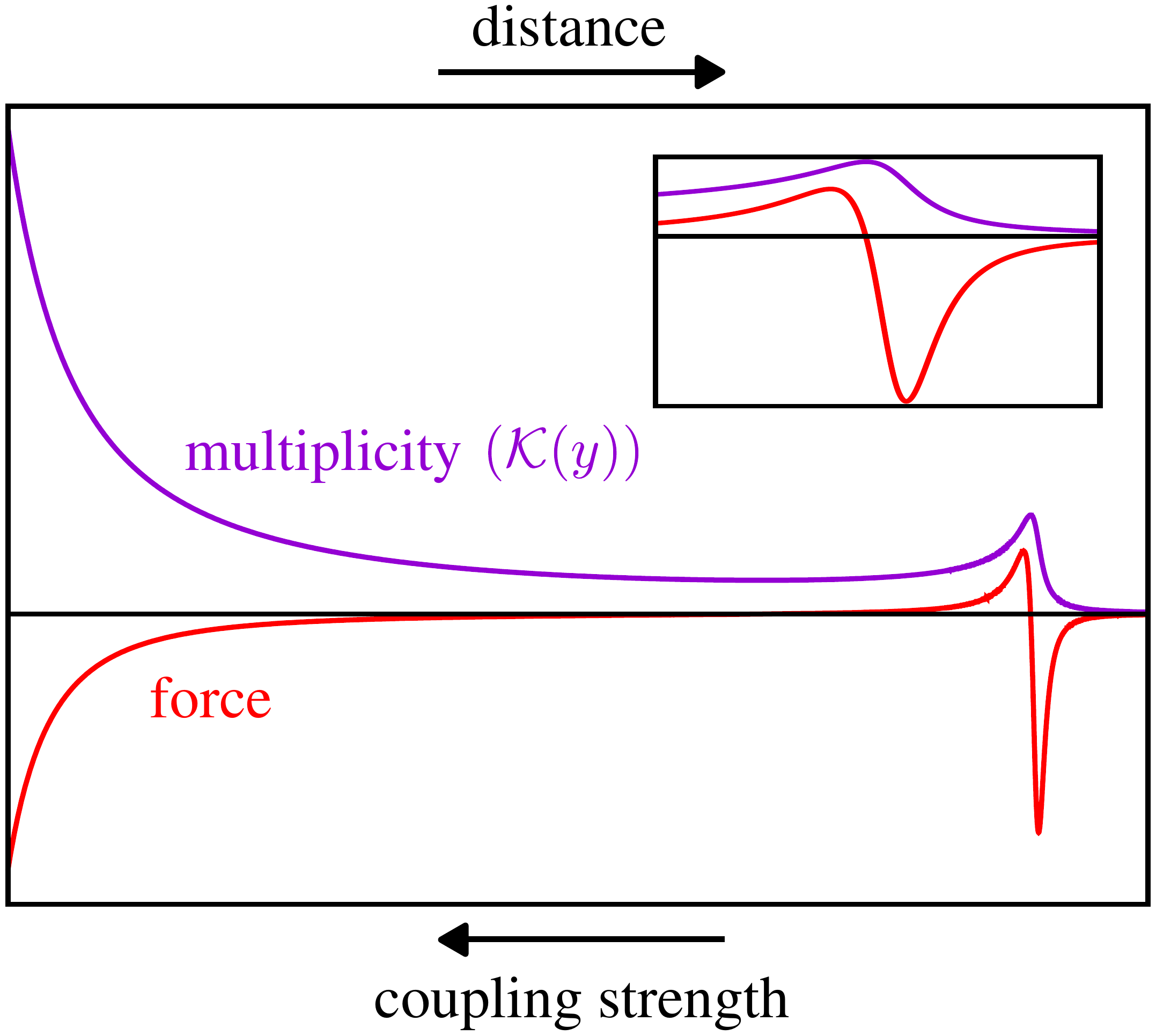}
\caption{$\mathcal{J}$ is a model for the multiplicity at a fixed external frequency, while $\mathcal{K}$ is the multiplicity of quantized solutions at all frequencies.  The force is attractive to the cluster when the red curve is below the x-axis, shown by the black horizontal line, and the repulsive force is above. The inset is an enhancement of the region where the force rapidly changes sign. Arrows indicate the directions of increasing distance or coupling strength. The curves have been scaled differently to facilitate plotting and should not be directly compared to each other. Only their shapes are meaningful. \label{eff_force}}
\end{centering}
\end{figure}

We can therefore use $\mathcal{J}(x)$ as a model for the multiplicity at strong coupling, as shown in Fig.~\ref{eff_force}. The coupling strength $v$ depends on the distance between the test point and the $D$ excitation. $D$ excitations require an occupied state and are localized to the cluster. $D$ does not depend on the point of creation/annihilation and is unchanged as we vary the distance to the test point. For a $1/r$ interaction, we can estimate the effect of varying the distance by treating $v$ as a parameter. The direction of decreasing distance is analogous to increasing coupling strength; increasing distance corresponds to decreasing the coupling strength. As described above, the multiplicity increases $-$ faster and faster $-$ as the coupling strength increases or the distance decreases. The gradient of $\mathcal{J}$ is an estimate of the local entropic force, $F =  \nabla \mathcal{J}$. At strong coupling, the force is attractive (see Fig.~\ref{eff_force}). To accentuate that this entropic force is only due to counting quantized solutions, we refer to it as the quantization force.

The quantization force is a new effect that is absent from the Born rule interpretation of the single-particle Green's function or wave function. Fig.~\ref{eff_force} shows that more quantized solutions exist infinitesimally closer to the cluster than away from it. We have not changed the size of the Hilbert space but instead counted the multiplicity of \textit{quantized} solutions. In our demonstration of the quantization force, both the particle number and poles of $D$ are fixed. Unlike the two-point problem in Sec.~\ref{sec:internal_external}, the statistics of the local problem considered in this subsection are based on a more simple single-point configurational entropy. The quantization force exists only because a quantized collapse \textit{occurs} and is universal to any quantum many-body system undergoing collapse. This entropic force is not felt by the many-body wave function. 

For a more complicated system, the force is mixed with other effects like electrostatics or strong correlation in $\overline{\Sigma}^*$. It could therefore be very difficult to isolate in an actual interacting system but become noticeable on the correct scale. Measurements of an interacting quantum many-body system with a potential ($\phi = -W$) shaped like either $\mathcal{J}$ or $\mathcal{K}$ (or some combination), perhaps including new or unexplained bound states, could provide evidence that collapse is, indeed, occurring in that system if there is not any prior \textit{ab-initio} explanation for a potential with that form. While simple in principle, the quantization condition could dramatically alter the shape and symmetry of the observable physics compared to the microscopic description.

The quantization force could lead to additional, entropically bound states which appear to be of a different type than electronically bound states. In fact, even if we consider a fixed value for the imaginary part of $\overline{\Sigma}^*$, our conclusions are the same. Tracking the DOS of $\mathrm{Im} \, \overline{\Sigma}^* = m$ gives an estimate of the total number of solutions without considering at which frequency they appear. With either choice, the gross features of $\mathcal{J}(x)$ and $\mathcal{K}(y)$ are quite similar (though scaled differently). Both functions clearly produce an attractive force at strong coupling. Beyond a certain point, this force could overwhelm all other effects.

The situation at weak coupling is much more intricate. As $v$ weakens, there is a wide, flat regime of the multiplicity with very weak force. The extent of this regime depends on the value of $\epsilon$. As $v$ weakens further, any fixed condition on $\overline{\Sigma}^*$ passes through the rapidly changing part of the pole in $D$. This regime includes an inflection point in both $P_{\epsilon}(\overline{\omega})$ and $\delta_{\epsilon}(\overline{\omega})$ and a local maximum in the multiplicity. This local maximum creates a narrow window of repulsive entropic force driving the particle away from the cluster before another attractive regime sets in, as shown in the insets of Fig.~\ref{eff_force}. Going further away from the cluster, the coupling strength is so weak that no solutions exist, as in Fig.~\ref{amplitudes}d. This suggests that there is a finite boundary to what is observable. The semilocal trajectory simply ends at the boundary. Again, these conclusions are largely unchanged if we consider either $\mathcal{J}(x)$ or $\mathcal{K}(y)$. We do not expect our simplified analysis to hold perfectly in this regime, where the detailed structure of the pole in $D$ and competition between $\mathrm{Re}\,\overline{\Sigma}^*$ and $\mathrm{Im} \, \overline{\Sigma}^*$ matter. We also have not parameterized the curves in Fig.~\ref{eff_force} to a $1/r$ interaction. However, we do draw two conclusions for the weak coupling scenario: (1) a finite boundary to the observable problem exists and (2) a local multiplicity maximum close to the observable boundary is very likely.

$\mathcal{J}$ and $\mathcal{K}$ are the inherent shape of the downfolding and solution method. Their shape could approximate the multiplicity even for a more exact downfolding, perhaps in the entire complex $\overline{\omega}$-plane. This shape applies as long as the multiplicity is determined by the intersection of the quantization condition with a ``projection surface" in the complex $\overline{\omega}$-plane defined by a dominant pole or set of poles, and the pole residue depends parametrically on coupling strength or distance. The numerical values needed to match the quantization condition are not necessarily fixed, as we assume in NSR, but as long as they vary slowly, $\mathcal{J}$ and $\mathcal{K}$ still apply.

Our conclusions remain largely intact even if we consider the case $\epsilon = \gamma = 0$, equivalent to $m=0$, when searching for nontrivial solutions of the inverse problem. For this case of the unaltered downfolding of the Schr\"odinger propagator, a meaningful shape to $\mathcal{J}$ still appears and creates an attractive quantization force. The narrow local multiplicity maximum disappears, however.

As alluded to above, $\epsilon$ is not necessarily constant but may depend on distance. This makes sense since $\epsilon$ is a measure of the interaction strength, which itself depends on distance. $\epsilon$ is set by the overlap between the bare field operator(s) acting on the ground state and the eigenstates of the system; very low overlap implies large spectral broadening and very large $\epsilon$, which may occur for a strongly-correlated multireference system. High overlap implies a small value of $\epsilon$, as in a weakly-correlated system. In this case, the local multiplicity maximum is very narrow. Based on these trends, $\epsilon$ likely increases at strong interaction strength or close range. Different values of $\epsilon$ change the curvature of the multiplicity surface, size of the internal spacetime, and quantization force. This is a \textit{separate} effect than just moving the particle along a multiplicity surface with fixed $\epsilon$. If we connect to the quasiparticle picture, there is an identifiable trend between lifetime and curvature of $W$: short lifetime (large $\epsilon$) implies substantial curvature, and a long lifetime (small $\epsilon$) increases the roughly flat regions of $\mathcal{J}$ and $\mathcal{K}$ (horizontal regions of Fig.~\ref{eff_force} with small force).

\begin{figure}[htb]
\begin{centering}
\includegraphics[width=0.85\columnwidth]{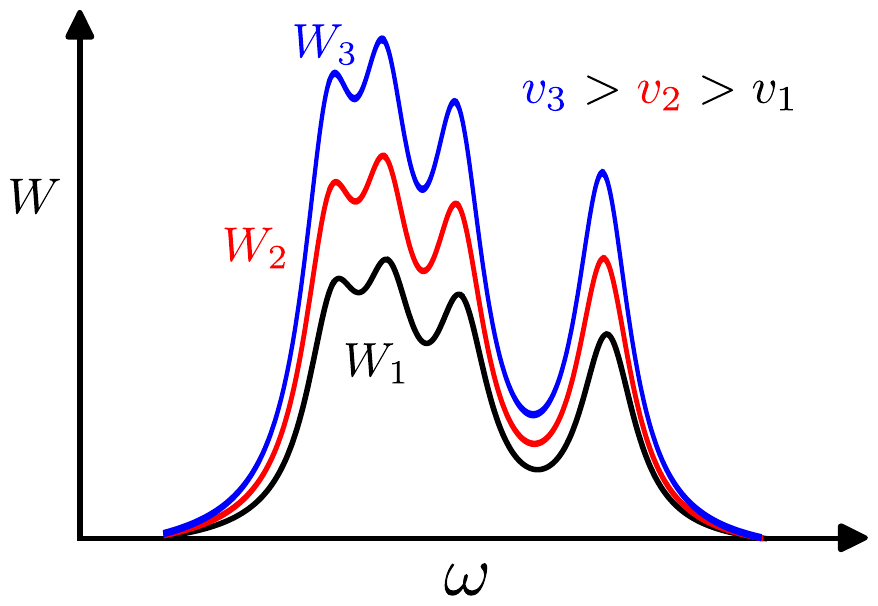} \caption{Change in the multiplicity as a function of coupling strength $v$. Greater $v$ is associated with shorter distance. In the attractive regime of the quantization force, $W$ can continue to increase as $v$ increases. $W$ follows no sum rule. \label{mult_shift}}
\end{centering}
\end{figure}

The multiplicity follows no sum rule. There is no limit on the number of quantized solutions that can exist, and the multiplicity can rise monotonically at every energy. This is possible because every individual $\delta$ function solution is already normalized. This scenario is demonstrated schematically in Fig.~\ref{mult_shift}. This effect can be traced to the lack of energy dependence in $\overline{\Sigma}^*$. In our NSR example, the structure of $\overline{\Sigma}^*$ determines the multiplicity, but it has no $\omega$ dependence. Even if $\overline{\Sigma}^*$ adds additional satellite solutions away from the main peak, they still parametrically depend on the coupling strength $v$ in the same way as the main peak. $\overline{\Sigma}^*$ can contribute solutions at every energy. Even if an $\omega$ dependence to the downfolding does exist for the exactly downfolded $\mg$, the multiplicity still has no sum rule. Monotonic changes to the entire multiplicity as a function of coupling strength (Fig.~\ref{mult_shift}) are still allowed in the exact theory.

The theory may also explain why so much of observed physics is local in space. In our NSR model, only a very small contribution from $\overline{\Sigma}^*$ is necessary to quantize any diagonal element of $\overline{A}$. In contrast, the offdiagonal elements depend entirely on coherence in $\overline{\Sigma}^*$ to produce the full imaginary part needed for the quantized pole. These offdiagonal, or spatially nonlocal, elements require a larger imaginary contribution from $\overline{\Sigma}^*$ to match the quantization condition than the diagonal elements. Any condition on $\mathrm{Im} \, \overline{\Sigma}^*$ is sensitive to the decoherence effect discussed in Sec.~\ref{sec:induced_collapse}, with the greater numerical condition being more sensitive to decoherence. Therefore, we expect the offdiagonal elements to be more sensitive to decoherence than the diagonal. If the global phases of all states are uncorrelated, the coherence required to produce a properly quantized pole becomes more difficult as more entangled states contribute to the wave function. As mentioned before, the rate of decoherence could be size extensive. As the system becomes more entangled with its environment, the size extensivity of decoherence could suppress offdiagonal elements of $W$ and cause local collapses to dominate the particle's observable trajectory. In this case, the particle follows an apparent single-point trajectory along the semilocal multiplicity gradient.

  \label{sec:quant_force}
  
\subsection{Interpretation}
  With the essential concepts of our framework in place, we give some overall perspective of the reformulation and discuss its consequences.

We summarize the theory so far. In our model, the system's external time evolution changes when the collapse occurs. The system first evolves according to the exact time evolution operator, $U(t,t')$, and exists as a superposition of configurations. When there is an informational conflict, the particle's history is downfolded into the nonlocal Dyson equation. With this collapse of the time evolution operator, the system gains the self-information $I_{xy}$. The equation describing how the system evolved from $t'$ to $t$ has changed. We interpret this as an actual change to the system's history on the $dt$ time scale and as the event which creates the internal spacetime. The new history depicts the creation, propagation, and annihilation of a normalized particle. This event is defined to be causal on $dt$, although it appears externally as an instantaneous, nonlocal change to the system's timeline. Internally, the self-information is carried by the spacetime bridge. As a real system evolves, its history is frequently rewritten to form quantized, causal events with each collapse. With each collapse, the many-body time evolution operator and initial condition are updated to hold the newly gained information. External time evolution starts again. The system is constantly monitored for conflicts and then rearranged to try to resolve them.

The proposal raises very strange possibilities. For one, events which are separated by a large interval $\Delta t = | t - t'|$ on the external time may have a smaller or larger separation on $d \tau$. In our picture, the duration of the time bridge is related to the multiplicity of the solution but does not depend on $\omega$ or the time interval $\Delta t$. This could look strange but violates nothing about our starting point. Highly specialized experimental set-ups in which both spacetimes are detetectable, either directly or indirectly, could look extremely strange. The fact that such blatant nonlocality is extremely rare in practice suggests that much of what we perceive occurs in an internal spacetime.

Our phenomenology relies on the existence and collapse of a wave function but supposes that it can never be directly observed $-$ only collapsed states can be observed. The wave function is a carrier of information about possible outcomes, but it is neither observable itself nor a probability amplitude. Observed reality is always a normalized two-point correlation. These are the only events allowed by the rules of our resource, and they are synonymous with changes to information. The ``internal observer" exists at one point of the two-point collapse. Different internal observers do not experience the same events, spacetimes, nor information gains. Because the multiplicity is a function of two points, an individual observer's information gain is always self-referencing. For example, if a selected internal observer is moved to a different point, but the partner point is held fixed, $p^{-1}$ changes. At the new point, the observer gains different information through a new internal spacetime with different size. The observed reality and internal spacetime are not determined until the collapse, and they are different for different observers.

Even if observable events are not the wave function, the effect of information contained in the wave function is certainly real. Both the wave function and collapsed states exist and they affect each other. The possible outcomes carried in the wave function determine what information can be gained through the internal spacetime, and the gained information updates the information contained in the wave function. There cannot be changes to information or conflicts without the information itself. In our picture, the combined effects of the wave function and collapsed events create one reality. 

In our model, the many-body wave function is constantly monitored and interpreted by the resource in terms of quantized events to form information. Our model suggests that this constant observation \textit{is} necessary to use information theory. To form information, one needs the bare state \textit{and} the interpreter. Without this interpreter always working, the bare state is meaningless and rules based on information cannot apply. Without the state, the interpreter can form no information, and without the interpreter, the state is meaningless. Together, they create something which can obey a set of informational rules.

Such a system would obviously give special purpose to the observer, as is already known to be the case in quantum mechanics, and a role played by the informational resource in our theory. This picture suggests that the quantum system has an inherent, invisible capability for information detection and processing that is always working $-$ the back-end informational resource. Somehow, the system just knows about its possible outcomes and their conflicts. Although it is a strange or uncomfortable idea, double-slit experiments, quantum erasers, and entanglement all behave like their subsystems know about the information they contain.

\begin{figure}[htb]
\begin{centering}
\includegraphics[width=0.75\columnwidth]{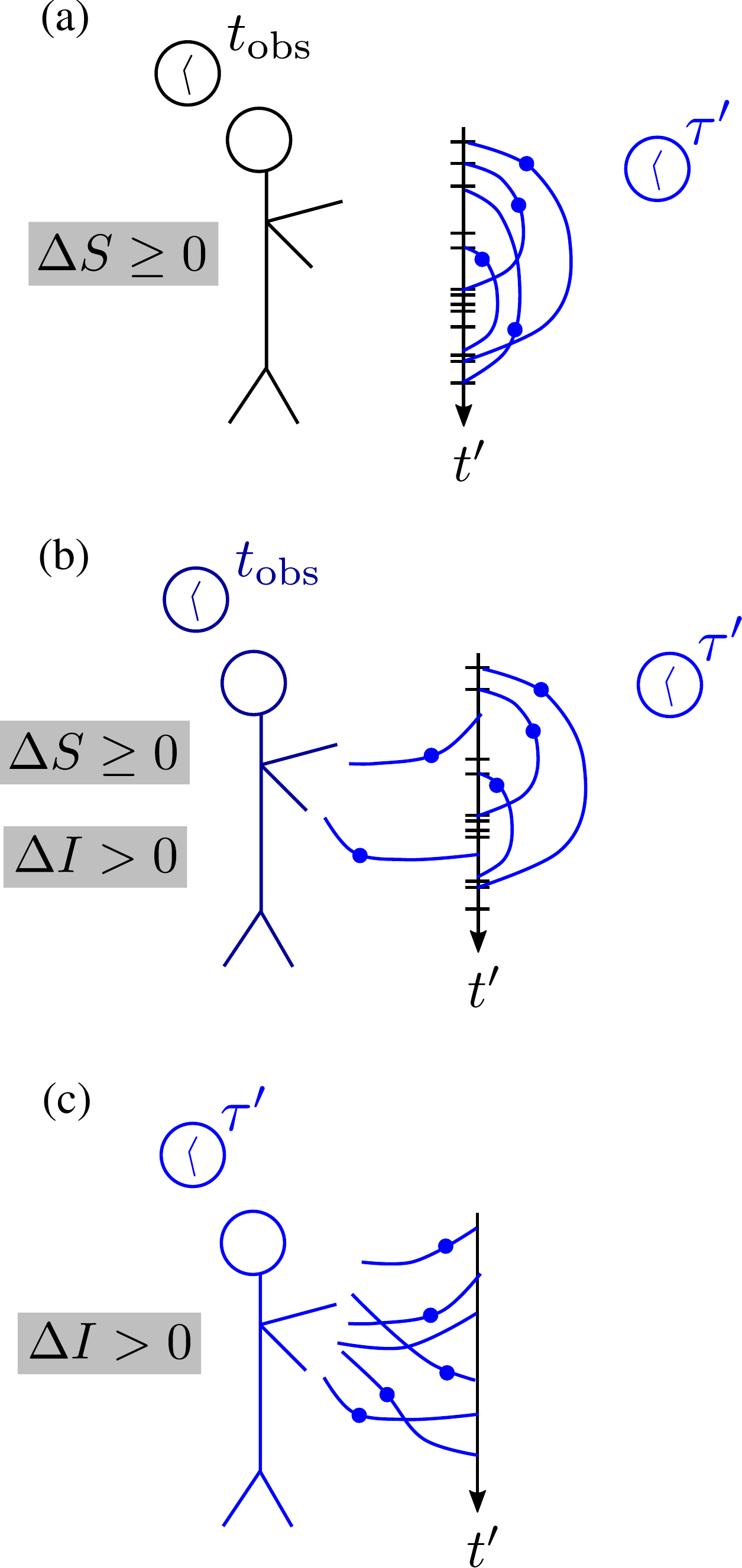}
\caption{An observer completely disentangled from their sample, as in panel (a), sees only a set of point-like, nonlocal events, represented by the horizontal lines on the $t'$ time scale. They are unaware of the sample's internal time coordinate $\tau'$ and have no access to it. Their ignorance of the system grows with every collapse. In the perfectly entangled experiment, shown in (c), the experimentalist is part of the same internal spacetime as the experiment, shown in blue. They gain self-information with every collapse. Real systems are a mixture of internal and external dynamics, falling somewhere in between these two limits (b). \label{time_scales}}
\end{centering}
\end{figure}

Our starting point was modeling collapse of the many-body wave function on the external time scale $dt$ as seen be an outside observer. In this case, the observer is completely disentangled from the sample under investigation and only sees the experiment's external time coordinate, labeled $t'$ in Fig.~\ref{time_scales}a. If the experimentalist is entangled with the sample and their measurements are through an internal spacetime, however, the actual experimental set-up is reversed. In this case, the experimentalist detects the event and gains information through internal dynamics. In this case, the observer and experiment are contained in the same internal spacetime.

The two cases of perfect entanglement or disentanglement with the sample are two extreme limits of the theory. Real systems fall somewhere between these two limits. The spectroscopist and their equipment are not perfectly entangled or disentangled with their sample but instead have some mixed level of entanglement. To better describe a mixed level of entanglement, we more carefully consider the meaning of the self-information and collapse entropy. An internal observer has access to the spacetime bridge and its information carrier. We associate this manifestation of the self-information with a statistical mechanical energy. For an internal observer, the forward direction through the irreversible events creating the internal spacetime is defined by the way of increasing information, $d I / {d \tau} > 0$.

Externally, the collapse looks different. Without access to the internal spacetime or self-information, there is no perceived propagation of self-information across the internal spacetime. Instead, the collapse appears to an external observer as a loss of information, or increase of ignorance. The magnitude of the ignorance gain is the collapse entropy, $S_{xy}$. The entropy only increases or remains constant on the external time scale, $dS / dt \ge 0$. From this vantage point, we associate the collapse entropy with a statistical mechanical entropy. Without the ability to detect any correlation carrier on the internal spacetime, it could be that external events are detectable only by the entropy they create, perhaps as heat or finite temperature, and the way these events increase ignorance over time.

The statistics describing the rearrangement of a real system's information would then be a combination of internal (energetic, information gain) and external (entropic, ignorance gain) effects. Energy and entropy are different aspects of the same phenomenon $-$ collapse. The collapse certainly appears different internally and externally; it is our own interpretation to associate these with energy and entropy. A macroscopic experiment is a statistical counting of individual particles which are entangled with the sample or not, leading to both energetic and entropic effects on a macroscopic scale. The net level of information gain or loss depends on details of the system and its environment. For a single event, the magnitudes of information gained internally plus the information lost externally equals the total information carried by the original wave function, $S^{\mathrm{tot}} = I_{xy}(\omega) + S_{xy}(\omega)$.

The two-point nature of the problem is unavoidable. There is no way around the fact that the multiplicity is a function of two points. The information in the wave function is about \textit{relations between points}, not single points. The single-point configurational entropy we use in Sec.~\ref{sec:quant_force} to describe the quantization force is a simplified picture. If we only know the position of the particle, we do not know the path it took to get there. Therefore, we do not know the entropy created with each collapse that occurred on its trajectory, or even how many times it collapsed. If the particle travels a very unlikely path, the system gains more information than it does on a very likely path. The current observed state alone is not enough to know the information gained by the system or the history of its internal spacetime up to that point. As we have outlined the theory, these collapse statistics about information gain create an entire history of the internal spacetime. The future internal spacetime does not yet exist and its shape and information gain are not predetermined.


In the ground state, no single-particle state is either definitely occupied or definitely empty. A single field operator acting on the ground state does not create a normalized particle. Yet, the field operators represent our way of manipulating the physical degrees of freedom. It is only by treating \textit{pairs} of field operators that we can find solutions which match observable physics. These pairs are the relevant degrees of freedom, not single points. A single-point theory, or a single-reference theory in which a field operator creates a normalized particle, would need a completely different mechanism for information gain (if it can have any such mechanism). The interacting problem has a completely different structure than the noninteracting one. In a deterministic system, for example, all information is known at any one time and therefore no information can be gained. Without irreversible events, there is no perception of a preferred direction. In our theory, the multiconfigurational character of the system, time evolution into a higher dimensional state than can be observed, and nonlocality appear essential to its two-point structure, collapse, and the ability to gain information. Without the high-dimensional state created by the interactions, there is no choice of the projection and no meaningful information or structure to the multiplicity.

We then claim that every meaningful measurement is actually a two-point process, and any apparent single-point measurement is embedded in an internal spacetime and just half of a pair. For any apparent and meaningful single-point measurement, the other half of the two-point process has already occurred in the external spacetime, even if we are unaware of it and no matter their separation. Normalized measurements must always come in particle pairs, and their two-point correlation is determined instantly on the external time. Our overall perception of space itself depends on measurements, observations, and information gain. Our experience is then most aligned with the spacetime through which we gain information, the internal spacetime.

Our interpretation of the collapse entropy suggests that the quantum system obeys a set of rules for information gain and loss. By taking the perspective that informational aspects of the system are fundamental and other properties follow, one can attempt to formulate a set of informational rules or axioms describing the dynamics. We have already pointed out that an observer must be part of the system's time evolution to gain information with collapse. We postulate that any observer able to gain information about a system is entangled with it. With perfect knowledge of the microstate, an EPR experimentalist is not able to gain information about the system. We propose that any observer who cannot gain information about a system is external, or disentangled, from the system of interest. This criterion can be met in at least a couple of ways. An observer who knows all of a system's possible outcomes has complete information. They cannot possibly gain information and therefore cannot be a part of its time evolution, which would automatically permit them to gain information with collapse. On the contrary, an internal observer can gain information about their own system via collapse, which means their information must be incomplete. Complete information on a system forbids entanglement with it, and entanglement with a system forbids complete information. This case of knowing the exact many-body wave function is very special and not necessary to be considered external. An external observer with very little information about a system and who is not able to gain more information still qualifies as external.

As mentioned in Sec.~\ref{sec:internal_external}, the Shannon information gain assumes that the internal observer initially had no information about their system. Collapse is \textit{nonlocal} in time, which indeed suggests that it is not possible for an internal observer to hold any information about their current system during its \textit{original} time evolution. If nonlocality is necessary to gain information, it is consistent that an internal observer before the collapse is completely ignorant of their current state (though they may have information about their past). In this case, defining the gained information as the Shannon information makes sense. This interpretation agrees with the picture outlined in the above paragraph that an internal observer's information must be incomplete. There may be no allowed way for a single point, or observer, to know about its relations with other points without nonlocality. This makes some intuitive sense. If an observer can ``see" another point and know their relation with it, they are not actually an individual, internal point themselves. This would contradict the meaning of the internal points and the way information is formed. The information is a \textit{many-body} effect, which is necessarily inaccessible to one internal body. Only an external observer can see the point-to-point relations. At the same time, their ignorance does allow an internal observer to \textit{gain} information. This is, of course, not rigorous, but the overall picture is that an internal observer just cannot know their own current wave function or relational information.

We conclude that the ability to gain information based on the two-point multiplicity and collapse of the time evolution operator can be used to classify an observer's level of entanglement. We favor this criterion as the most fundamental choice for defining entanglement.

An EPR entangled pair demonstrates some of our concepts. Externally, the two-point correlation is determined instantly and nonlocally with collapse. As we have outlined our theory, an internal observer would receive the information gained across a finite amount of internal spacetime. This could appear as two separate measurements internally, even though the two-point collapse determines the correlation instantaneously. An external experimentalist who prepared the EPR pair already knew how they were correlated in the many-body state. They knew everything that could happen since their information about the system was perfect, and they gain no information with the collapse. With collapse onto a single outcome, the experimentalist actually loses information proportional to the collapse entropy because they cannot know which microstate was realized. To an internal observer without initial information, the collapse creates the Shannon information. For an external observer with initial information, the collapse creates ignorance. The original information in the wave function is irretrievably lost.

The example becomes strange when considering the informational accounting on a global scale. Treat the EPR experimentalist and experiment as two separate systems, external to each other, as we have already done. When they collide and the system collapses, the entangled pair gains self-information and the experimentalist loses information. Our question is: does \textit{the} information about the entangled pair increase or decrease? The experimentalist rightly says that information about the entangled pair decreased; the entangled pair correctly insists that information about itself increased. There is not an objective and consistent statement about \textit{global} information gain or loss with the collapse, even if \textit{parts} of the total system undergo a definite information gain or loss. Here, we have taken a global perspective, which assumes that both the experiment and experimentalist belong to a single, larger quantum system which holds information about both and has fixed ``hardware." This could be an especially useful, or possibly even necessary, construct as a way for detecting conflicts. A single, global resource with information about all of its parts could easily detect conflicts between its subsystems since they exist within its own total information, which it can easily access and check. It functions as a global observer. With EPR, does the global observer's information about the entangled pair increase or decrease? Somehow, it supports both. There is a structure that allows for parts of the system to behave very differently than the total. This strange concept is based in understanding how a single entity (the global resource) can have conflicts or contradictions with itself, the effect which promotes information gain, and the way that an outside body can hold information about a subsystem(s) that is inaccessible \textit{within} the subsystem.

In our model, the informational resource constantly monitors the system to detect and resolve any informational conflicts. It has some directive to form an information preserving state without conflicts $-$ a many-body pure state, or universal wave function in which every particle is consistently entangled. If such a state could ever be reached, the theory says that internal space and time would vanish. Everything would be described by a single, fundamental, and objective wave function that contains all information and has a reversible, deterministic evolution: all possible outcomes are held forever. The external time still runs and the wave function exists, but there are no events. There are no more changes to information and the internal spacetime disappears entirely. Whether or not this is actually possible requires much closer investigation. Conversely, it suggests that internal spacetime began from some initial conflict of information.

Because the internal spacetime and direction through it are defined by information gain, the system could appear on the internal time to spontaneously arrange itself in a way to gather information. Suppose a system collapses upon encountering a conflict of information. The reset may resolve the conflict and prevent future collapses. If not, and the system encounters the same conflict again, it will collapse again. The system continues the search to find a permitted arrangement so that the conflict disappears. This stochastic process of conditioning the system to a specific conflict could last several internal time steps. After the conflict is finally resolved, there are no additional internal time steps. The system has been rearranged and its multiplicity restructured according to the information it gained, which will affect its collapse response to future conflicts. In this way, the system becomes conditioned to its environment. The system's entire history of conflict, collapse, and information gain determines its current informational conflicts and possible outcomes. The information it holds about what could happen agrees with the possible outcomes of the environment.

No matter how much external time passes before a new conflict is encountered, it would appear internally as if the system \textit{immediately} set out to solve the next conflict. It is not until the next conflict and collapse that information is gained and the internal time begins running again. In this way, it could seem as if the system actively seeks out conflicts and resolves them by some unseen direction. There could actually be huge intervals of external time between events when essentially nothing is happening, but they would be impossible to detect. Only events which change information are perceivable. On the internal spacetime, the many-body system could appear to self-assemble, gather information, respond to its environment, and resolve conflicts all on its own. It collects information about itself with every collapse. The system continuously restructures its internal spacetime to gain information and ``match" the environment. There could be a natural peak in the system's observable events and complexity, which we associate with a net level of information gain, as it grows and increases its boundary with the environment, maximizing the number of events, but before it reaches a universal wave function and its boundary and all observable events disappear. As it gathers and stores information, the system could become highly ordered to maintain coherence and information.

The details of such a system and how it fits into the resource are extremely complicated. Within our phenomenology, we do consider it possible for such a self-organized and information gathering system to appear, as we described. However, if conflicts can only be detected by a global observer, as we suggested previously, then the information gathering system actually belongs to a larger, total system. It appears necessary for the system to have at least \textit{some} information about its environment to make an informational \textit{comparison} to detect conflicts, which requires an observer global to both the system and its environment. Conflicts which promote its information gain can only be detected by an observer global to the system undergoing collapse \textit{and} the environment inducing collapse. The information gathering system is not truly independent from its environment since they share an observer. The information gathering subsystem is a compartmentalized portion of the global information which is somehow detached from the resource so that it undergoes information gain, not loss, and can also hold consistent information about its environment.

Instead of focusing on the subsystem, a \textit{total} system of mixed entangled and disentangled parts could, in principle, experience a similar net gain of \textit{inconsistent} information as it observes its own subsystems. The global nature of the resource, which is the component that actually forms and holds information about the current state, could also be the key to understanding how one subsystem (the experimentalist) can have objective information \textit{about} another subsystem (the EPR entangled pair) with which it is not entangled. With this in mind, the global observer of a mixed and inconsistent system has the potential advantage of gaining information \textit{within} a subsystem while at the same time having information \textit{about} that subsystem.

The quantization condition and information gain depend on $\mathrm{Im} \, \overline{G}$, a condition that is very sensitive to coherence of the single-particle states. However, the global phase of each single-particle state is arbitrary. A real orbital can arbitrarily be made imaginary, completely changing the coherence for the quantization condition. The theory appears gauge dependent. Here, ``gauge" simply refers to the set of phase shifts attached to each single-particle state. Because the information contained in the wave function is sensitive to the gauge, adjustments to the gauge could be a very efficient way of storing gained information. In a complicated system with many particles, there is enormous flexibility among the gauge degrees of freedom for holding information about quantized solutions. With each collapse, the resource could rotate the phases of the collapsed points to hold their newly gained information. For an internal observer, their phase rotation gives them some information about the many-body system they are part of, though we associate this information with past correlations and not the current information in the updated many-body state. Offdiagonal elements of $\overline{G}$ have greater gauge dependence than a diagonal element, potentially increasing the decoherence of offdiagonal elements and decreasing their multiplicity.

  \label{sec:interpretation}
  
\subsection{Curvature of the multiplicity surface}
  We return to describing the dynamics of a particle collapsing along a semilocal trajectory on the multiplicity surface. With each local creation and annihilation, the amount of self-information gained depends on the shape of the semilocal multiplicity. If we assume the particle always evolves perfectly along the multiplicity gradient (no rare or spatially nonlocal events occur), the information gained with each collapse still depends on the multiplicity of all \textit{other} possible infinitesimal steps. These other possible but unrealized outcomes are necessary to compute the probability of the realized step and its self-information. This determines the information gained and size of the internal spacetime as the particle evolves. Ideally, one would sum the number of solutions for all possible infinitesimal steps around a given point to compute $W^{\mathrm{tot}}$ and then $I_{xy}$. One could repeat this procedure along the particle's entire trajectory to map the internal spacetime. The number of solutions in any direction away from a given point is related to the gradient in that direction. To calculate the probability and self-information of a realized local step, one could compare the gradients of the multiplicity in all directions instead of explicitly summing the solutions. The change in multiplicity gradient around the central point, or curvature, contains the same information about the internal time dilation and information gain as an explicit sum for $W^{\mathrm{tot}}$, as we now demonstrate.

\begin{figure}[htb]
\begin{centering}
\includegraphics[width=0.8\columnwidth]{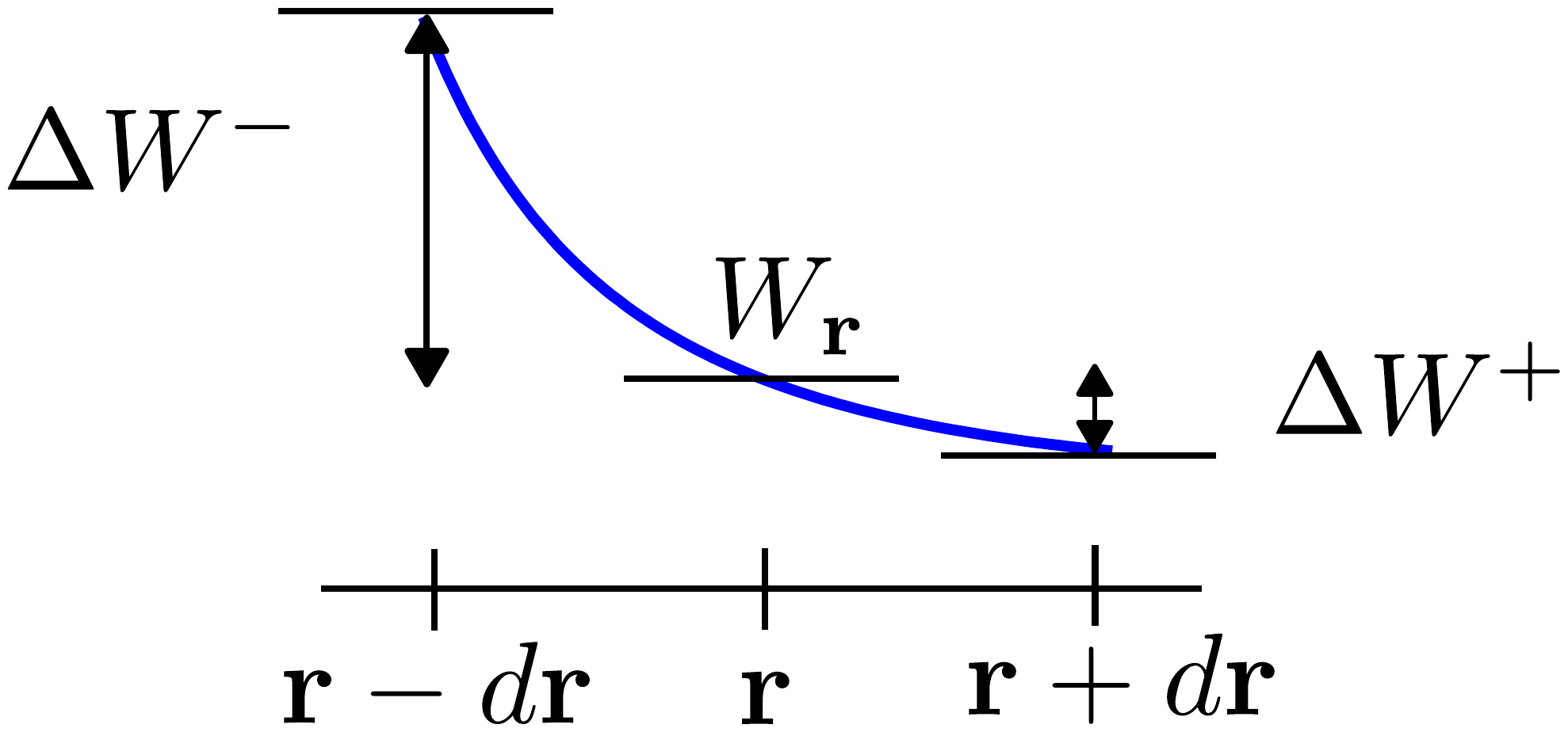}
\caption{Three-point model for the local multiplicity, $W_{\mathbf{r}}$, and two semilocal steps at $\mathbf{r}-d\mathbf{r}$ and $\mathbf{r}+d\mathbf{r}$. The model is set up for a concave up multiplicity, shown in blue, and defined for $\Delta W^- > 0$ and $\Delta W^+ > 0$. \label{concavity}}
\end{centering}
\end{figure}

To simplify the problem, we assume a test particle on the multiplicity surface can only stay in place, take an infinitesimal step towards the cluster ($d\mathbf{r}^-$), or take an infinitesimal step away ($d \mathbf{r}^+$). We compare the diagonal multiplicities at these three points ($W$, $W^-$, and $W^+$) to each other to calculate the information gained with a local creation and annihilation. For the model shown in Fig.~\ref{concavity}, the inverse probability for the in-place local step is
\begin{equation}
p^{-1} = 3 + \frac{\Delta W^- - \Delta W^+}{W} \; . \label{inverse_prob}
\end{equation}
The inverse probability of the local step, and therefore the amount of self-information gained, is greater for $\Delta W^- - \Delta W^+ > 0$ than $\Delta W^- - \Delta W^+ < 0$. This condition on the forward and backward gradients of $W$ is a statement about the concavity of $W$. For a constant gradient of $W$, the forward and backward gradients in Eq.~\ref{inverse_prob} cancel. The entropic force depends on the gradient of $W$, but the internal time dilation depends on a second derivative-like curvature. We use $\mathcal{J}$ in Fig.~\ref{eff_force} as a guide to the multiplicity for our cluster example. In the attractive regime of the quantization force, $\mathcal{J}$ is concave up. Furthermore, the concavity is not constant. We verify numerically that the curvature increases with increasing coupling strength. A local event closer to a cluster at the origin gains more information than a local event farther away. The internal spacetime close to the origin dilates more than the internal spacetime far away.

Crudely, the internal time dilation for the semilocal multiplicity with the shape in Fig.~\ref{concavity} is something like
\begin{eqnarray}
 d \tau_{\mathbf{r}\mathbf{r}} &=& d\tau_0 \, I_{\mathbf{r}\mathbf{r}}  \\
 d \tau_{\mathbf{r}\mathbf{r}} &=& d\tau_0  \, \mathrm{log}_b  \left( C + \frac{1}{W} \frac{d^2 W}{d\mathbf{r}^2} r_0^2  \right)  \; . \label{curvature}
\end{eqnarray}
for the constant $C$ that depends on dimensionality. For the three-point model in Fig.~\ref{concavity}, $C=3$. The second term in the logarithm of Eq.~\ref{curvature} is some type of dimensionless curvature about the central point, roughly the second derivative, that depends on a coherence length ($r_0$). Beyond $r_0$, the number of quantized solutions dramatically drops due to decoherence and sensitivity to the initial condition set by the previous collapse. There is likely a memory effect in the multiplicity due to the initial condition (updated Lehmann amplitudes) and update of the time evolution operator which makes it most likely that the system repeats the same collapse (or spatially nearby).

The second derivative of $W$ plays a special role in determining $p^{-1}$, as demonstrated in Fig.~\ref{concavity}. Generalizations to dimensions greater than one should follow. A constant multiplicity has zero force and a constant $d \tau$ on the entire multiplicity surface whose magnitude depends on the dimensionality. The collapse entropy behaves differently: it depends on the magnitude of $W_{xy}$ and changes with the constant background value of $W$. A linear multiplicity in our three-point model has nonzero entropic force but constant $d \tau$ that again depends on the dimensionality constant $C$. The dilation for a nonvanishing but constant second derivative of $W$ depends on the magnitude of the curvature. The dilation depends on the sign of the curvature: the internal time at a multiplicity minimum, for example, is more dilated than at a maximum with the same curvature. Finally, for a multiplicity with a second derivative that depends on position, which is the case for $\mathcal{J}$, the magnitude of the internal time step varies on the multiplicity surface. We find it quite interesting and essentially accidental that, using $\mathcal{J}$ for the multiplicity at strong coupling, the gradient of the multiplicity evolves the particle in the direction of \textit{greater} information gain with each step (internal time dilation is greater closer to the cluster).

\begin{figure}[htb]
\begin{centering}
\includegraphics[width=\columnwidth]{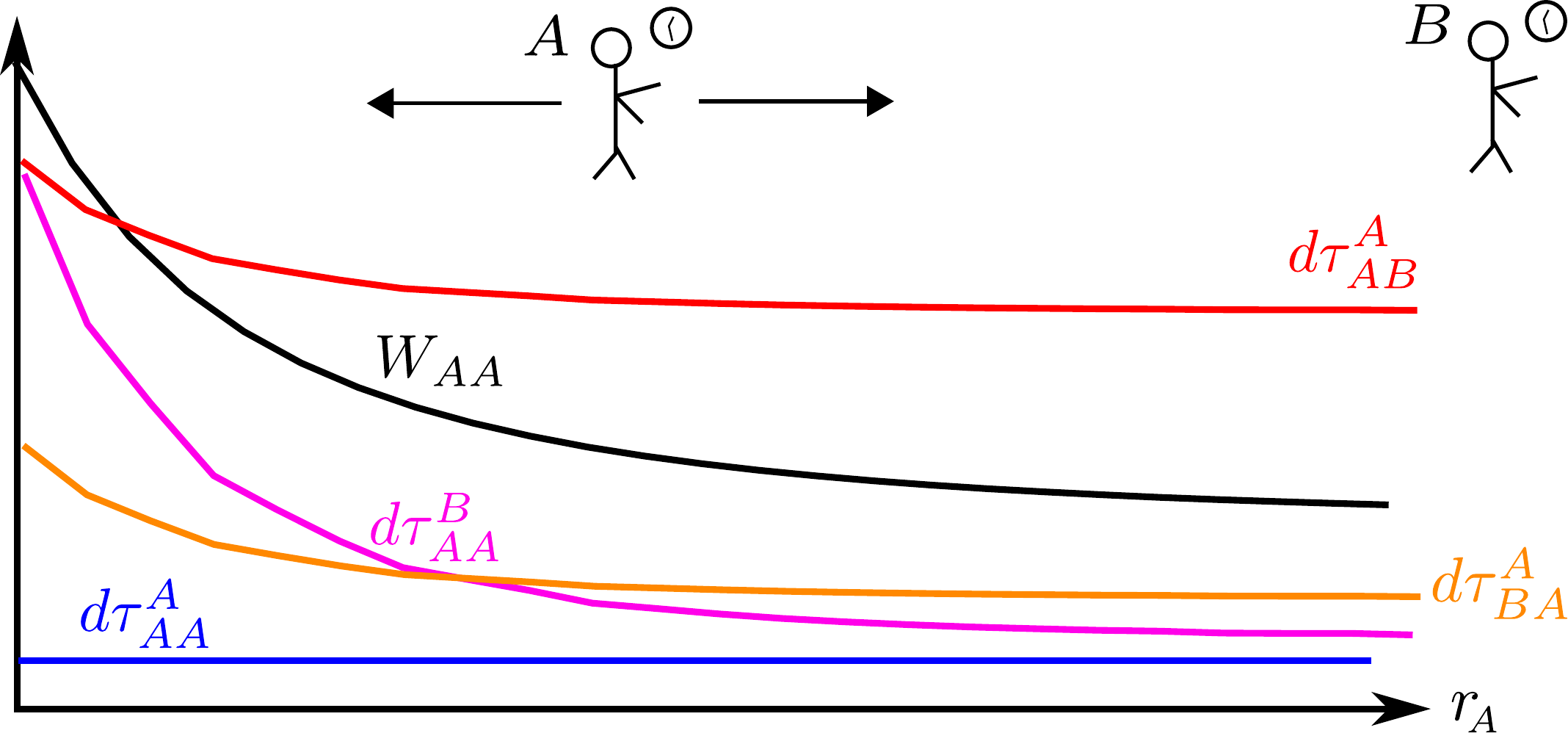}
\caption{Colored lines show the dilation of the indicated internal spacetime coordinates as a function of Alice's distance from the origin. The curves are only a qualitative guide and should be interpreted loosely. The red curve is with Alice's initial condition and orange with Bob's. We neglect the local multiplicity maximum at weak coupling. All internal times should meet at the right side where Alice and Bob are at the same point; the figure is too crowded to show this.  \label{space_entanglement_strong_2} }
\end{centering}
\end{figure}

To demonstrate these effects, we place two observers, Alice and Bob, on the multiplicity surface around our cluster of entangled particles. As mentioned in Sec.~\ref{sec:internal_external}, the internal spacetime is defined by correlations between two points. We simply claim that a proper, local spacetime coordinate can be defined by normalizing the internal spacetime to a benchmark event. These proper internal time coordinates are the ones experienced by internal observers Alice and Bob, who each exist locally at their own point. We introduce the notation $d \tau_{ij}^k$ to denote the duration of the time bridge created for a two-point event initiated at $i$, terminated at $j$, as measured on the proper and local time of $k$. Here, $i$, $j$, and $k$ can take values $A$ (Alice) or $B$ (Bob). The proper times measured by both, $d \tau_{AA}^A$ and $d \tau_{BB}^B$, are constant and set to 1, $d \tau_{AA}^A = d \tau_{BB}^B = 1$. We place Alice closer to the cluster than Bob. By the concavity argument above, Alice's internal time step is longer, or more dilated, than Bob's when they are both measured by the same clock. This effect, $d \tau_{AA}^B > d \tau_{BB}^B$, is shown in Fig.~\ref{space_entanglement_strong_2}. Consequently, Bob experiences more events in a fixed amount of absolute time and ages faster than Alice.

The spatially nonlocal problem is more challenging. These spatially nonlocal spacetimes represented by $d \tau_{AB}^B$ and $d \tau_{BA}^A$ are also shown in Fig.~\ref{space_entanglement_strong_2}. Bob and Alice represent different initial conditions and different $W^{\mathrm{tot}}$. We assume that local solutions dominate the multiplicity; at strong coupling, Alice has greater total multiplicity $W^{\mathrm{tot}}$ than Bob. By the same arguments about the shape of the multiplicity, we expect an observer at strong coupling to carry more quantized solutions than one at weak coupling. If the multiplicity of the spatially nonlocal A-B event, $W_{AB}$, is equal to the multiplicity of the opposite process, $W_{BA} = W_{AB}$, the self-information gained with either one depends on $W^{\mathrm{tot}}$. Therefore, the self-information gained by the A-B process with Alice's initial condition (greater $W^{\mathrm{tot}}$ than Bob) will be higher than for Bob's initial condition. Because the total multiplicities are sensitive to the observer's position, the Alice-Bob and Bob-Alice internal spacetimes are not symmetric, even if their particular two-point process is symmetric.

Fig.~\ref{space_entanglement_strong_2} is a major simplification of a very complicated problem. It is possible, in principle, to compute all of these quantities from the multiplicity for quantized solutions. Even without specifying the detailed structure of each spacetime bridge, we can still compute its length. Our conclusions are at least contained and self-consistent within our NSR model and informational phenomenology. Our emphasis is on the existence of different curves in the figure and their qualitative properties. As we have outlined the theory, the general structure shown in Fig.~\ref{space_entanglement_strong_2} is possible. Most importantly, and in the approximation that nearby, diagonal solutions dominate the multiplicity, the dilation of the internal time can be related to the curvature of $W$. The relation between the curvature of $W$ and internal time is general, when semilocal solutions dominate, and independent of the specific shape of $W$ or the validity of our $\mathcal{J}$ approximation.

  \label{sec:curvature}
  
\subsection{Zero quantized solutions}
\begin{figure}[htb]
\begin{centering}
\includegraphics[width=0.75\columnwidth]{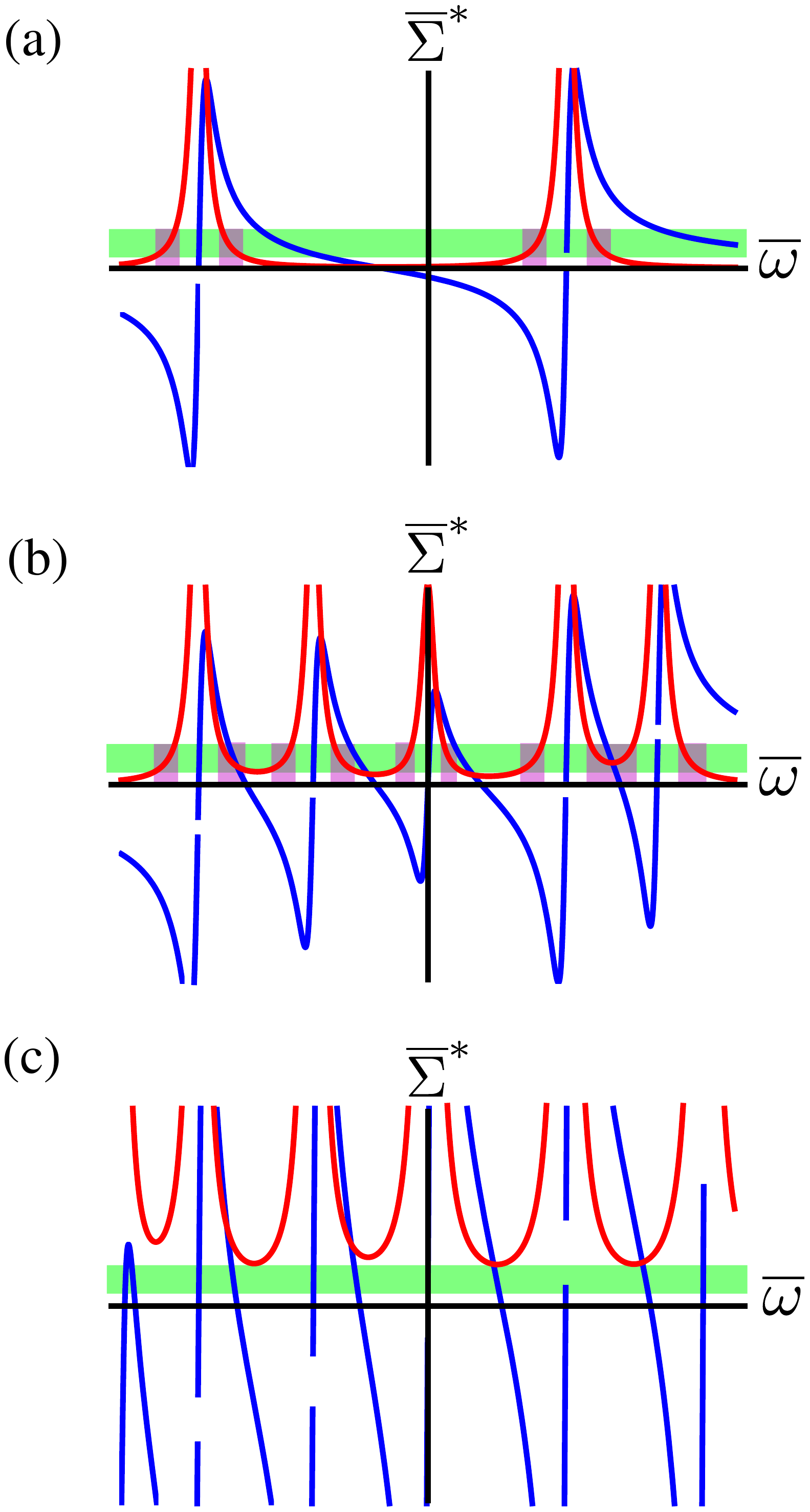}
\caption{As in Fig.~\ref{branch_cut_2} but with greater entanglement. The couplings to each pole can become so great and the level spacing so small that no quantized solutions exist (c). This condition depends sensitively on the broadening and strength of each pole. The case in (c) is highly entangled but with zero multiplicity. \label{branch_cut}}
\end{centering}
\end{figure}

  It is possible, in principle, that the inverse problem has no solutions. We first outline a specific way in which this could occur and then consider the more general case and the properties of such a system. $D$ is a size extensive quantity. As the number of entangled particles in a system grows, the number of poles in $D$ increases. This increases the region on the $\overline{\omega}$-axis which could give a quantized solution. The size extensivity is characterized by modified $\mathcal{J}$ and $\mathcal{K}$ functions, for example
\begin{eqnarray}
m_{\epsilon}(\overline{\omega}) &=& \sum_{k \in D} \pi \delta_{\epsilon}(\overline{\omega} - \omega_k) v_k^*v_k \\
\mathcal{M}(y) &=& \int d\overline{\omega} \, \delta ( y - m_{\epsilon}(\overline{\omega}) ) ) \label{mod_k}
\end{eqnarray}
where the sum runs over the poles of $D$. Here, it is necessary to include the coupling strengths $v$ in the calculation of $\mathcal{M}$ instead of treating $v$ as a parameter as we did in Sec.~\ref{sec:quant_force}. Computing modified $\mathcal{J}$ and $\mathcal{K}$ based on the actual $\overline{\Sigma}^*$ describes the true shape of the statistics better than our simplified single pole model in Sec.~\ref{sec:quant_force}, especially when the quantization force cannot be easily isolated. In practice, this is an extremely complicated task. The multi-pole structure of $\overline{\Sigma}^*$ is shown in Fig.~\ref{branch_cut}. The multiplicity is now determined by the intersection of the quantization condition with the many poles of $\overline{\Sigma}^*$. We generally expect an increase in the number of poles to lead to more $D$ crossings of the $\overline{\omega}$-axis and more solutions, enhancing effects like the quantization force and internal time dilation.

However, there could be a critical system size, entanglement level, or interaction strength beyond which the number of solutions begins to \textit{decrease}. If the level spacing in $D$ becomes very narrow, separate poles in $D$ can overlap and decrease the number of solutions. The interaction strength $v$ can also push the poles of $D$ out of the solution region. This possible regime of decreasing multiplicity could be a region of repulsive quantization force at close range or strong interaction strength that is not shown in Fig.~\ref{eff_force}. Very strong interaction strengths, or very short distances, may not give any quantized solutions at all. Additionally, if the particle is entangled with so many other particles that the poles of $D$ merge into a branch cut, it may not be possible to find any quantized solutions. The $\delta$ function condition in Eq.~\ref{im_g} requires finite behavior in $\mathrm{Im} \, \overline{\Sigma}^*$ to find a solution (finite value of $im$). If the poles merge together and $\overline{\Sigma}^*$ is singular on the entire $\overline{\omega}$ axis, quantized solutions may be impossible. This progression of $\overline{\Sigma}^*$ and the multiplicity from low to high level density is shown graphically in Fig.~\ref{branch_cut}.

It is a highly nontrivial problem to compute the multiplicity of a real system through such a transition. It depends in a complicated way on particle number, interaction strengths, and broadening. Nonetheless, we can remain consistent in our NSR model, which predicts that a system with a very dense level spacing may not have quantized solutions.

\begin{figure}[htb]
\begin{centering}
\includegraphics[width=\columnwidth]{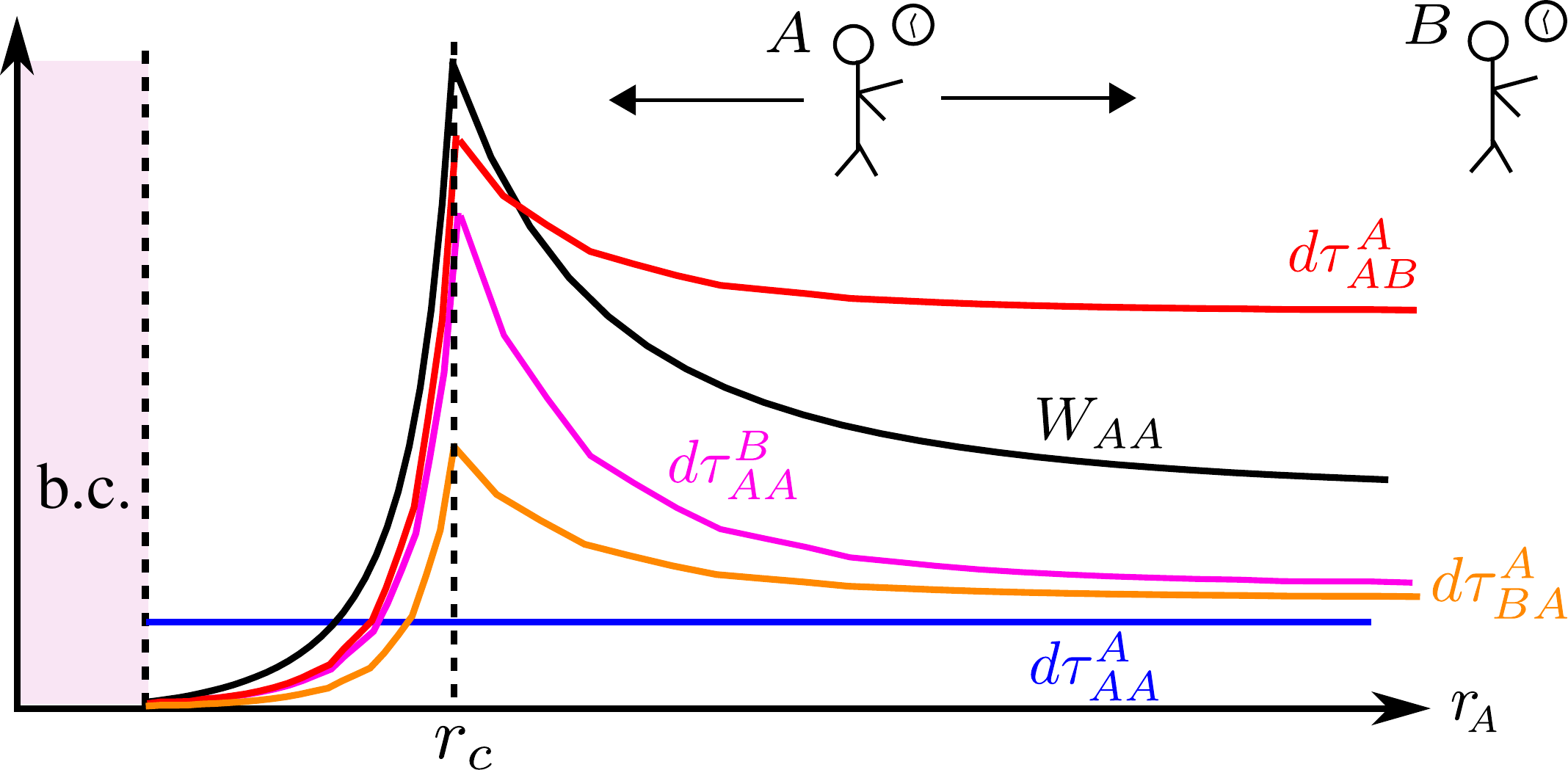}
\caption{Colored lines show the dilation of the indicated internal spacetime coordinates as a function of Alice's distance from the origin. The curves are only a qualitative guide and should be interpreted loosely. Below the distance labeled $r_c$, the multiplicity turns over and the number of solutions begins to decrease. Eventually, all internal spacetime coordinates vanish at the boundary of the branch cut (b.c.). \label{space_entanglement_strong}}
\end{centering}
\end{figure}

This behavior also depends on understanding how new particles are entangled and added to $D$. Generally, we expect more entangled particles and greater level density closer to the cluster. With each collapse on its trajectory, the particle becomes more entangled with the particles in the cluster. For this reason, these branch cuts most likely exist at the end of a particle's trajectory towards the center of a highly entangled cluster. Fig.~\ref{space_entanglement_strong} is a modified guide to the internal spacetimes showing the onset of this regime with zero quantized solutions. Below the cutoff radius $r_c$, the multiplicity turns over and shrinks as the poles of $D$ begin to merge. At even shorter distance, the poles merge and no quantized solutions exist.

The example in Fig.~\ref{space_entanglement_strong} is just one specific case for which no quantized solutions exist. There may be other examples in more ordinary systems that do not depend on such a high level density. It could be a complicated numerical fact that no solutions to the inverse problem exist, without any transparent explanation. In these cases, understanding the existence or nonexistence of solutions requires knowing the very detailed structure of the exact $\overline{G}$ from an exact downfolding of $\mg$. The Lehmann amplitudes, which mix different excitation levels in a complicated way, could be very important in these more conventional systems without very high level density. A zero multiplicity phase could also be facilitated by extremely weak coupling. If, hypothetically, $\overline{G}$ requires a finite imaginary contribution $im$ to match the quantization condition, but the coupling to $D$ is so weak that $\mathrm{Im} \, \overline{\Sigma}^*$ is never of magnitude $m$, no quantized solutions exist. This is analogous to searching for solutions beyond the boundary of the observable problem, as discussed in Sec.~\ref{sec:quant_force}, where the coupling strength is too weak to give quantized solutions.

A system without quantized solutions could be a hallmark of strongly-correlated physics. It would have no internal spacetime, and it would be impossible for an internal observer to probe it for information. Without any collapse on its interior with zero multiplicity, the system's entropy creation and other statistics exist only on its boundary.

The branch cut and the entire zero multiplicity region could suddenly appear when the quantization force accumulates enough particles at close range that the level density exceeds a certain threshold. The multiplicity could suddenly change shape from Fig.~\ref{eff_force} to Fig.~\ref{space_entanglement_strong} once it reachs a critical entanglement level. The branch cut would likely form at the end of the system's progression through stages of complexity and information gain since it requires a high level of entanglement, as opposed to several separate, disentangled parts, each with low entanglement and level density, that exist at the beginning of such a progression. Even when the branch cut exists at close range, an entangled particle far away with weaker coupling strength could still have quantized solutions and be attracted to the zero multiplicity region by the quantization force. The branch cut region can contribute to the quantization force at long range \textit{even though} observable particles are absent in its interior $-$ it is not really a branch cut at weak coupling.\footnote{In fact, this property applies to more than just a region without quantized solutions. A particle feels the quantization force of a cluster with which it is entangled even if no observable events occur \textit{within} the cluster (even if solutions do exist there) and the particle always collapses locally. The entangled particles creating the multiplicity surface do not need to be observable.} In this single-point picture, a particle on the multiplicity surface would stochastically evolve towards the region until it contributes to the branch cut itself and effectively disappears.

A possible precursor to this zero multiplicity phase is extremely low-energy excitations. On the way to the phase with $W^{\mathrm{tot}}=0$ in some parameter space, the system would likely first pass through a regime with extremely low $S^{\mathrm{tot}}$. $S^{\mathrm{tot}}$ limits the maximum amount of information gain (see Eq.~\ref{info_diff}). Small $S^{\mathrm{tot}}$ gives any realized event a relatively high probability and low information gain. If the information gain is related to the energy of the excitation, these events would be very low-energy excitations on the phase's boundary, which could mean either a real space surface or more abstract boundary in a phase diagram. 

As an aside, our model also permits events which create entropy but gain no information. A system with $S_{xy} = S^{\mathrm{tot}}$ \textit{has} quantized solutions and creates entropy with collapse onto $x,y$, \textit{but} it has no internal spacetime or information gain. This exceptionally strange behavior is because there is only one possible outcome, but this single outcome has multiplicity $>1$. For this event, an external observer gains ignorance but no internal spacetime forms. Such systems could appear to an internal observer to expel information carrying fields $-$ there is no information to carry $-$ and their events would be internally reversible. These events could also precede the onset of a phase without quantized solutions: before $S^{\mathrm{tot}}$ drops to zero, the multiplicity would naturally localize, and its last finite (though small) value would be for a single event. Events at a very sharp local maximum in $W$, as appears in $\mathcal{J}$ at weak coupling and with small $\epsilon$, could also behave this way (create entropy but gain no information and lack an internal spacetime). This particular point is also \textit{stable}. Events are also possible, in principle, which gain information but create no ignorance ($S_{xy} = 0$, $S^{\mathrm{tot}} \neq 0$).

As already alluded to, an important point of understanding is if a system without quantized solutions contains any information. In our theory, the amount of information is always defined with respect to the number of quantized two-point solutions that the wave function carries. Without any quantized solutions, we say that such a system contains no information. The wave function exists here, but its interpretation by the resource gives no information. Without any information or events, there is no reality. Additionally, even if the wave function is in some way incoherent in this region, we also wonder if conflicts of information exist in this region as we have sketched them. We consider the conflicts to also be related to differences in information about quantized solutions. Without any quantized solutions, there can be no conflicts about quantized solutions.

Case specific or model details aside, we emphasize the overall concept: systems without quantized solutions are generally allowed and naturally included in our theory. This concept is in stark contrast to the standard theory, in which unitary time evolution of the wave function covers all of space. With a Born rule interpretation of the wave function, there is a probability amplitude for observing a particle everywhere. This quantum mechanical effect (lack of solutions) stems from our strict interpretation and enforcement of quantization. If such unobservable regions of space can actually be detected, combining these detections with our theory could provide support that a collapse \textit{does}, in fact, occur with every event that contributes to the observable statistics. This is a feature of internal spacetime dynamics, as we have outlined them, which require the collapse of a wave function in an external spacetime to create (or not) the internal spacetime. To an internal observer, these regions mark the end of spacetime. Systems with a history of these features could indicate that collapse occurred naturally in the past and before any explicit human observer probed the system.

  \label{sec:zero_sols}
  
\section{Conclusion}

The most rigorous result of this work is the appearance of two frequencies in the nonlocal Dyson equation for $\overline{G}$. This appears to be a new fundamental freedom of quantum many-body theory (projection freedom) that we take full advantage of. Our motivation for a theory beyond the single-particle $G$ was clear: norm conservation. A point of emphasis in this work is to use only the Schr\"odinger equation. In addition to discovering a second frequency in the downfolding, we used physical arguments to motivate the nonlocality and understand the physical meaning of the downfolding procedure. Downfolding necessarily gives a choice of the projection, and we keep only the projections that match what is observed. The theory and solution method are an alternative to the Born rule that is still entirely consistent with the Schr\"odinger equation.

Our numerical calculations for a two-level system show good agreement between the multiplicity and an approximate local spectral function for long-lived quasiparticle solutions. The calculation is heavily simplified but still meaningful. This is an encouraging result considering the significantly different nature of the two calculations. Future benchmarking against either the spectral function of $G$ or experimental spectra requires an exact downfolding of the Lehmann representation of $\mg$ to give the exact $\overline{G}$.

In the third section, we interpreted and extended the result. Our reformulation can be considered at different levels of abstraction. At the highest level, there is the overall concept of the many-body problem as an inverse problem, which allows us to adopt a certain protocol (quantization). This has immediate consequences that are independent of finer details: unphysical behavior can be discarded, and there may be systems without solutions. At the same high level of abstraction is the idea that collapse creates both information and ignorance depending on the perspective of the observer. This naturally leads to the information theoretic reasoning that the gained information requires its own channel, the internal spacetime. Our core physical principle is that the reduction of the probabilistic state creates a new spacetime to hold the gained information. It is a long-standing idea, which we exploit and develop here, that objective collapse is an essential component of quantum dynamics. At the deepest level, there are hints of a deep underlying mathematical and informational structure to our proposed resource.

From these foundational principles, it follows that the particle follows the multiplicity gradient according to an entropic force based on counting quantized solutions. The system evolves irreversibly with a forward direction defined by information gain/loss. We showed that the dilation of the internal spacetime as the particle collapses along its semilocal trajectory can be related with a simple model to the curvature of the multiplicity surface. The quantization condition creates a size, shape, and history for the two-point internal spacetime. Our quantization condition suggests that the system is extremely sensitive to decoherence in a size extensive way, which could both induce collapse and localize the particle. At the most practical level, we proposed a model for the inverse problem and shape of the multiplicity surface based on an adjusted single-reference theory.

Many of the principles which we invoke or their consequences are actually longstanding ideas which are supported by the $\overline{G}$ theory. We recognize that our informational model has elements in common with various different modern proposals in physics,\cite{rovelli,smolin,penrose_1996,diosi_1989,swingle,maldacena_susskind,raamsdonk,Bong2020,frauchiger_2018} which we take as a positive sign. The idea of gravity induced collapse\cite{penrose_1996,diosi_1989,bassi} and the ER=EPR conjecture\cite{maldacena_susskind} both initially helped us interpret our finding. Instead of gravity induced collapse, our idea is \textit{collapse induced gravity}. The ideas in play are major foundational concepts in general relativity.\cite{einstein_gr} With this in mind, our quantization force perhaps plays the role of gravity and regions without quantized solutions could be related to singular points in spacetime. We do not have expertise in general relativity, but we consider the fundamental statistical mechanics behind these effects in our theory motivation to explore this link further.

There are many avenues for further research. Many of our ideas need to be refined and formalized. The particle spin degree of freedom needs to be included and could explain the character of the internal spacetime. Spin could also play the role of an internal many-body-like superposition which holds information, perhaps about relations between spin up and down, for instance. With these additions, a final, closed set of equations for the collapse dynamics of the particle, internal spacetime, and their connection to information in the wave function would be within reach. On the applications side, we are most interested in studying strongly-correlated materials, which was our original motivation. The theory could provide an unprecedented level of understanding to the spectra of non-Fermi liquids.

\section{Acknowledgments}

This work was supported by the Academy of Finland through grant no.~316347.

  \label{sec:conclusion}
  
\appendix*

\section{Informational rules}
  This serves as an informal summary of our current model for the informational resource and a starting point for a more rigorous set of informational axioms from which, ideally, quantum mechanics could be reverse engineered. The set is a mixture of rules extracted from the physics of the $\overline{G}$ theory and rules built in to our finite informational resource model.

\begin{figure}[htb]
\begin{centering}
\includegraphics[width=\columnwidth]{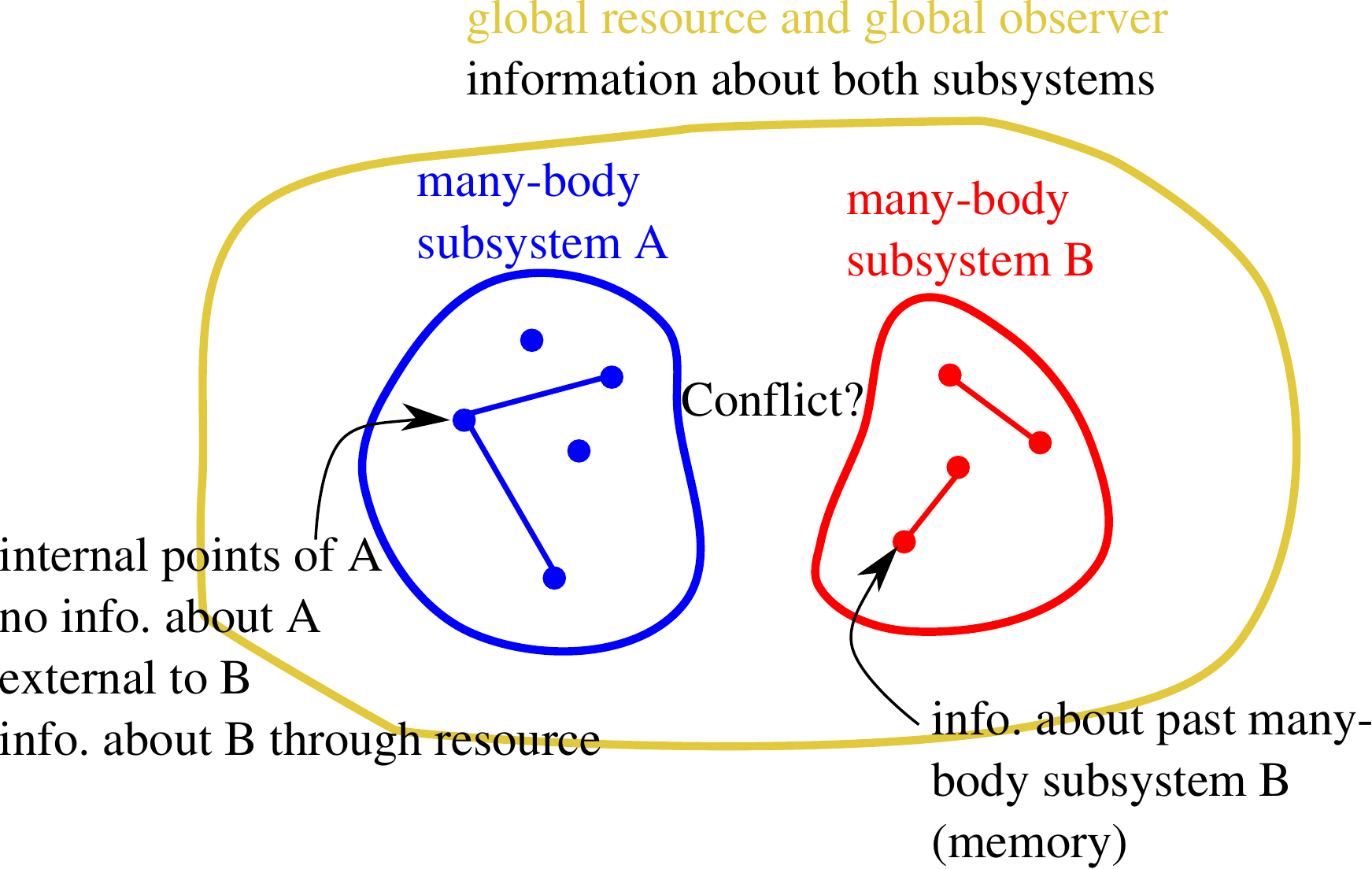}
\caption{Schematic showing the total system broken into two disentangled subsystems and their monitoring by the global resource. To summarize the idea: conflicts can be detected only by comparison. This requires an observer with information about both subsystems. The global resource's information is inconsistent because the subsystems are mixed. Furthermore, with collapse of a subsystem, the global observer supports both information gain and loss. The entire process on the global scale is inconsistent and self-referencing, but for either local subsystem it is consistent. Because they are totally separate, or disentangled, the subsystems can have consistent information about each other, but not themselves. We attribute this structure largely to the many-body character of the information. A nested structure, with two disconnected circles placed inside every circle, could also be explored. \label{subsystems_resource}}
\end{centering}
\end{figure}

\begin{enumerate}
\item The external time evolution of the wave function obeys the Schr\"odinger equation.
\item Counting quantized, two-point correlations obtainable by downfolding the time evolution is the protocol to define microstates. This counting sets the information in the wave function.
\item The resource constantly monitors and interprets the system to form information about quantized solutions.
\item The instruction of the resource is to resolve all conflicts of information.
\item Collapse is a reset protocol to dump conflicting information.

\item An internal observer gains the Shannon information with collapse onto their point.
\item Gained information is subjective and relative to both points of the event. An internal observer's information gain is relative to themselves.

\item An observer who can gain information about a system must be part of the system's time evolution, defining an internal, or entangled, observer.
\item An internal observer's information is incomplete.

\item Information is gained through an internal spacetime.
\item Information can only be gained at a finite rate on some absolute scale.
\item Stored or static information is held in the external spacetime.
\item An observer who cannot gain information about a system cannot be part of the system's time evolution, defining an external, or disentangled, observer.
\item An external observer loses an amount of information with collapse proportional to the collapse entropy.
\item Information loss occurs instantly.
\item An observer with perfect $-$ which could mean complete, nonsubjective, or non-self-referencing $-$ information about a system must be external to that system.
\item Well-posed but unsolvable quantization conditions which yield no information are allowed.

\item The global observer's information is inconsistent.

\item A proper coordinate system for any internal spacetime can be defined by normalizing to a single benchmark two-point correlation or, equivalently, information gain.

\end{enumerate}

  \label{sec:axioms}

\bibliographystyle{apsrev4-1}
\bibliography{./subtex/references}

\end{document}